\documentclass {article}
\usepackage{graphicx}

\begin{document}

\title{The holostar - a self-consistent model for a compact self-gravitating object}

\author{Michael Petri\thanks{email: mpetri@bfs.de} \\Bundesamt f\"{u}r Strahlenschutz (BfS), Salzgitter, Germany}

\date{June 16, 2003}

\maketitle

\begin{abstract}
Anisotropic compact stars were introduced as a new class of
solutions to Einstein's classical field equations of general
relativity in several recent papers. Some of these solutions
possess a two-dimensional membrane of tangential pressure situated
at the boundary between the matter-filled interior and the outer
vacuum space-time. In this paper the so called holographic
solution, in short "holostar", is discussed. The holostar is
characterized by the property, that the stress-energy content of
its membrane is equal to the gravitational mass of the holostar.

The holostar exhibits properties similar to the properties of
black holes. The exterior space-time of the holostar is identical
to that of a Schwarzschild black hole, due to Birkhoff's theorem.
The holostar possesses an internal temperature, proportional to
$1/\sqrt{r}$, from which the Hawking temperature law for
spherically symmetric black holes can be derived up to a constant
factor. The number of zero rest-mass particles within any
concentric region of the holostar's interior is proportional to
the proper area of its boundary, implying that the holostar is
compatible with the holographic principle and the Bekenstein
entropy-area bound. In contrast to a black hole, the
holostar-metric is static throughout the whole space-time. There
are no trapped surfaces, no singularity and no event horizon.
Information is not lost. The weak and strong energy conditions are
fulfilled everywhere, except for a Planck-size region at the
center. Therefore the holostar can serve as an alternative model
for a compact self-gravitating object.

Although the holostar is a static solution, it behaves dynamically
with respect to the interior motion of its constituent particles.
Geodesic motion of massive particles in a large holostar exhibits
properties quite similar to what is found in the observable
universe: Any material observer moving geodesically will observe
an isotropic outward directed Hubble-flow of massive particles
from his local frame of reference. The radial motion is
accelerated, with the proper acceleration falling off over time.
The acceleration is due to the interior metric, there is no
cosmological constant. The total matter-density $\rho$, viewed
from the extended Lorentz-frame of a geodesically moving observer,
decreases over proper time $\tau$ with $\rho \propto 1 / \tau^2$.
The radial coordinate position $r$ of the observer changes
proportional to $\tau$. The local Hubble value is given by $H
\simeq 1/ \tau$. The observer is immersed in a bath of zero
rest-mass particles (photons), whose temperature decreases with $T
\propto 1 / \sqrt{\tau}$, i.e. $\rho \propto T^4$. Geodesic motion
of photons within the holostar preserves the Planck-distribution.
The radial position of an observer can be determined via the local
mass-density, the local radiation-temperature or the local
Hubble-flow. Using the experimentally determined values for the
matter-density of the universe, the Hubble-value and the
CMBR-temperature, values of $r$ between $8.06$ and $9.18 \cdot
10^{60} r_{Pl}$ are calculated, i.e values close to the current
radius and age of the universe. Therefore the holographic solution
might serve as an alternative model for a universe with
anisotropic negative pressure, without need for a cosmological
constant.

The holographic solution also admits microscopic self-gravitating
objects with a surface area of roughly the Planck-area and zero
gravitating mass.

\end{abstract}

\section{\label{sec:intro}Introduction:}

In a series of recent papers  new interest has grown in the
problem of finding the most general solution to the spherically
symmetric equations of general relativity, including matter. Many
of these papers deal with anisotropic matter
states.\footnote{Relevant contributions in the recent past (most
likely not a complete list of relevant references) can be found in
the papers of \cite{Barve/Witten, Burinskii, Dev/00, Dev/03,
Dymnikova, Elizalde, Gair, Goncalves, Giambo, Hernandez, Herrera,
Herrera/2002, Ivanov, Mak, McManus/Coley, Mielke/BosonStars,
Petri/bh, Salgado, Singh/Witten} and the references given
therein.} Anisotropic matter - in a spherically symmetric context
- is a (new) state of matter, for which the principal pressure
components in the radial and tangential direction differ. Note
that an anisotropic pressure is fully compatible with spherical
symmetry, a fact that appears to have been overlooked by some of
the old papers. One of the causes for this newly awakened interest
could be the realization, that models with anisotropic pressure
appear to have the potential to soften up the conditions under
which spherically symmetric collapse necessarily proceeds to a
singularity.\footnote{See for example \cite{Singh/Witten}, who
noted that under certain conditions a finite region near the
center necessarily expands outward, if collapse begins from rest.}
Another motivation for the renewed interest might be the prospect
of the new physics that will have to be developed in order to
understand the peculiar properties of matter in a state of highly
anisotropic pressure and to determine the conditions according to
which such matter-states develop.

In this paper a particularly simple model for a spherically
symmetric, self gravitating system with a highly, in fact
maximally anisotropic pressure is studied. The model provides some
new insights into the physical phenomena of strongly gravitating
systems, such as black holes, the universe and - possibly - even
elementary particles.

The paper is divided into three sections. In the following
principal section some characteristic properties of the holostar
solution are derived. In the next section these properties are
compared to the properties of the most fundamental objects of
nature that are known so far, i.e. elementary particles, black
holes and the universe. The question, whether the holostar can
serve as an alternative, unified model for these fundamental
objects will be discussed. The paper closes with a discussion and
outlook.

\section{Some characteristic properties of the holographic solution}

In this section some characteristic properties of the holostar are
derived. As the exterior space-time of the holostar is identical
to the known Schwarzschild vacuum solution, only the interior
space-time will be covered in detail.

Despite the mathematically simple form of the interior metric, the
holostar's interior structure and dynamics turns out to be far
from trivial. A remarkable list of properties can be deduced from
the interior metric, indicating that the holostar has much in
common with a spherically symmetric black hole and with the
observable universe.

\subsection{The holostar metric}

The holographic solution is a particular case of a spherically
symmetric solution to the equations of general relativity with an
interior matter-density proportional to $1/r^2$. In units $c=G=1$
the metric and (interior) fields are given by \cite{Petri/bh} as:

\begin{equation} ds^2 = B(r) dt^2 - A(r) dr^2 - r^2 d\Omega
\end{equation}

\begin{equation} B(r) = \frac{1}{A(r)} = \frac{r_0}{r} (1-\theta(r-r_h))
+ \left( 1 - \frac{r_+}{r}\right) \theta(r-r_h)
\end{equation}

\begin{equation} \rho(r)= \frac{1}{8 \pi r^2} (1-\theta(r-r_h))
\end{equation}

\begin{equation} P_r(r)= - \rho(r)
\end{equation}

\begin{equation} P_\theta = \frac{1}{16 \pi r_h}
\delta(r-r_h)
\end{equation}

$r_h = r_+ + r_0$ is the position of the membrane, $r_+ = 2 M$ is
the gravitational radius of the holostar, $M$ its gravitating mass
and $r_0$ is a fundamental length. There is evidence
\cite{Petri/charge, Petri/thermo}, that $r_0$ is comparable to the
Planck-length up to a numerical factor of order unity ($r_0
\approx 2 r_{Pl}$).

The matter-fields, i.e. $\rho, P_r$ and $P_\theta$ can
be derived from the metric, once the metric of the whole
space-time is known.\footnote{See for example \cite{Petri/bh}}

\subsection{Proper volume and radial distance}

The proper radius, i.e. the proper length of a radial trajectory
from the center to radial coordinate position $r$, of the holostar
scales with $r^{3/2}$:

\begin{equation}
l(0, r) =  \int_{0}^{r}{\sqrt{A}dr}=\frac{2}{3} r \sqrt{\frac{r}{r_0}}
\end{equation}

The proper radial distance between the membrane at $r_h$ and the
gravitational radius at $r_+$ is given by:

\begin{equation}
l(r_+, r_h) =  \frac{2}{3} r_+ \sqrt{\frac{r_+}{r_0}} \big(
(1+\frac{r_0}{r_+})^{\frac{3}{2}}-1\big)\cong r_0
\sqrt{\frac{r_+}{r_0}} = \sqrt{r_+ r_0}
\end{equation}

The proper volume of the region enclosed by a sphere with proper
area $4 \pi r^2$ scales with $r^{7/2}$:

\begin{equation}
V(r) = \int_0^{r}{4 \pi r^2 \sqrt{\frac{r}{r_0}} dr} =\frac{6}{7}
\sqrt{\frac{r}{r_0}} V_{flat}
\end{equation}

$V_{flat} = (4 \pi /3) r^3$ is the volume of the respective sphere
in flat space.

Therefore both volume and radial distance in the interior
space-time region of the holostar are enhanced over the respective
volume or radius of a sphere in flat space by the square-root of
the ratio between $r_h = r_+ + r_0$ and the fundamental distance
defined by $r_0$.

The proper integral over the mass-density, i.e. the sum over the
total constituent matter, scales as the proper radius, i.e. as
$r_h^{3/2}$. There is evidence \cite{Petri/thermo} that the
interior matter should consist of at least one mass-less fermion
species.\footnote{It is quite remarkable and maybe not just a
coincidence, that $r^{3/2}$ is the dimension of a (supersymmetric)
fermion field in appropriate units.}

\subsection{Energy-conditions}

For a space-time with a diagonal stress-energy tensor, $T_\mu^\nu
= diag(\rho, -P_1, -P_2, -P_3)$, the energy conditions can be
stated in the following form:

\begin{itemize}
\item {weak energy condition: $\rho \geq 0$ and $\rho + P_i \geq 0$ }
\item {strong energy condition: $\rho + \sum{P_i} \geq 0$ and $\rho + P_i \geq 0$}
\item {dominant energy condition: $\rho \geq | P_i |$}
\end{itemize}

It is easy to see that the holostar fulfills the weak and strong
energy conditions at all space-time points, except at $r=0$, where
- formally - a negative point mass of roughly Planck-size is situated.
The dominant energy condition is fulfilled everywhere except at
$r=0$ and at the position of the membrane $r=r_h$.

According to quantum gravity, the notion of space-time points is
ill defined. Space-time looses its smooth manifold structure at
small distances. At its fundamental level the geometry of space
time should be regarded as discrete \cite{Penrose/spin-networks,
Penrose/twistor}. The minimum (quantized) area in quantum gravity
is non-zero and roughly equal to the Planck area
\cite{Ashtekar/area, Rovelli/Smolin}. Therefore the smallest
physically meaningful space-time region will be bounded by a
surface of roughly Planck-area. Measurements probing the interior
of a minimal space-time region make no sense. A minimal space-time
region should be regarded as devoid of any physical
(sub-)structure.

If the energy conditions are evaluated with respect to physically
meaningful space-time regions\footnote{With "Planck-sized region"
a compact space-time region of non-zero, roughly Planck-sized
volume bounded by an area of roughly the Planck-area is meant.},
i.e. by integrals over at least Planck-sized regions, the
following picture emerges: Due to the negative point mass at the
center of the holostar the weak energy condition is violated in
the sub Planck size region $r < r_0$. However, from the viewpoint
of quantum gravity this region should be regarded as inaccessible
for any meaningful physical measurement.

I therefore propose to discard the region $r < r_0$ from the
physical picture. This deliberate exclusion of a classically well
defined region might appear somewhat conceived. From the viewpoint
of quantum gravity it is quite natural. Furthermore, disregarding
the region $r < r_0$ is not inconsistent. In fact, the holographic
solution in itself very strongly suggests, that whatever is
"located" in the region $r<r_0$ is irrelevant to the (classical)
physics outside of this region, not only from a quantum, but also
from a purely classical perspective: If we "cut out" the region $r
< r_0$ from the holostar space-time and identify all space-time
points on the sphere $r = r_0$, we arrive at exactly the same
space-time in the physically relevant region $r >= r_0$, as if the
region $r < r_0$ had been included. The reason for this is, that
the gravitational mass of the region $r \leq r_0$ - evaluated
classically - is exactly zero. This result is due to the
(unphysical) negative point mass at the (unphysical) position $r =
0$ in the classical solution, which cancels the integral over the
mass-density $\propto 1/r^2$ in the interval $(0, r_0]$. Although
neither the negative point mass nor the infinite mass-density and
pressure at the center of the holostar are acceptable, they don't
have any classical effect outside the physically meaningful region
$r > r_0$.

Thus the holographic solution satisfies the weak (positive) energy
condition in any physically meaningful space-time region
throughout the whole space-time manifold. The same is true for the
strong energy condition.

Can one "mend" the violation of the dominant energy condition in
the membrane by a similar argument? Due to the considerable
surface pressure of the membrane the dominant energy condition is
clearly not satisfied within the membrane. Unfortunately there is
no way to fulfill the dominant energy condition by "smoothing"
over a Planck-size region\footnote{If the term "Planck-sized
region" would only refer to volume, allowing an arbitrary large
boundary area, we could construct a cone-shaped region extending
from the center of the holostar to the membrane, with arbitrary
small solid angle, but huge area and radial extension. In such a
cone-shaped region the integrated dominant energy condition would
not be violated, if the region extends from the membrane at least
half-way to the center.}, as is possible in case of the weak and
strong energy conditions. Therefore the violation of the dominant
energy condition by the membrane must be considered as a real
physical effect, i.e. a genuine property of the holostar.

Is the violation of the dominant energy condition incompatible
with the most basic physical laws? The dominant energy condition
can be interpreted as saying, that the speed of energy flow of
matter is always less than the speed of light. As the dominant
energy condition is violated in the membrane, one must expect some
"non-local" behavior of the membrane. Non-locality, however, is a
well known property of quantum phenomena. Non-local behavior of
quantum systems has been verified experimentally up to macroscopic
dimensions.\footnote{See for example the spin "entanglement"
experiments, "quantum teleportation" and quantum cryptography,
just to name a few phenomena, that depend on the non-local
behavior of quantum systems. Some of these phenomena, such as
quantum-cryptography, are even being put into technical use.} This
suggests that the membrane might be a macroscopic quantum
phenomenon. In \cite{Petri/thermo} it is proposed, that the
membrane should consist of a condensed boson gas at a temperature
far below the Bose-temperature of the membrane. In such a case the
membrane could be characterized as a single macroscopic quantum
state of bosons. One would expect collective non-local behavior
from such a state.

\subsection{"Stress-energy content" of the membrane}

One of the outstanding characteristics of the holostar is the
property, that the "stress-energy content" of the membrane is
equal to the gravitating mass $M$ of the holostar:

For any spherically symmetric solution to the equations of general
relativity the total gravitational mass $M(r)$ within a concentric
space-time region bounded by $r$ is given by the following mass function:

\begin{equation} \label{eq:massfunction}
M(r) = \int_0^r{\rho \widetilde{dV}} = \int_0^r{\rho 4 \pi r^2 dr}
\end{equation}

$M(r)$ is the integral of the mass-density $\rho$ over the
(improper) volume element $\widetilde{dV} = 4 \pi r^2 dr$, i.e.
over a spherical shell with radial coordinate extension $dr$
situated at radial coordinate value $r$. Note, that in a
space-time with $A B = 1$ the so called "improper" integral over
the interior mass-density appears just the right way to evaluate
the asymptotic gravitational mass $M$: The gravitational mass can
be thought to be the genuine sum (i.e. the proper integral) over
the constituent matter, corrected by the gravitational red-shift.
The proper volume element is given by $dV = 4 \pi r^2 \sqrt{A}
dr$. The red-shift factor for an asymptotic observer situated at
infinity, with $B(\infty) = 1$, is given by $\sqrt{B}$. Therefore
the proper integral over the constituent matter, red-shift
corrected with respect to an observer at spatial infinity, is
given by: $M = \int{\rho \sqrt{A B} 4 \pi r^2 dr}$. This is equal
to the improper integral in equation (\ref{eq:massfunction}),
whenever $A B = 1$.

If the energy-content of the membrane is calculated by the same
procedure, replacing $\rho$ with the two principal non-zero
pressure components $P_\theta = P_\varphi$ in the membrane, one
gets:

\begin{equation} \label{eq:Energy:Pt}
\int_0^\infty{(2 P_{\theta}) 4 \pi r^2 dr} = \frac{r_h}{2} = M +
\frac{r_0}{2}\cong M
\end{equation}

Note, that the  tangential pressure in the membrane of a holostar,
$P_{\theta} = 1 / (16 \pi (2M+r_0))$, is (almost) exactly equal to
the tangential pressure attributed to the event horizon of a
spherically symmetric black hole by the membrane paradigm
\cite{Price/Thorne/mem, Thorne/mem}, $P_{\theta} = 1 / (32 \pi
M)$.

The integral over the mass-density (or over some particular
pressure components) might not be considered as a very
satisfactory means to determine the total gravitational mass of a
self gravitating system. Neither the mass-density nor the principal
pressures have a coordinate independent meaning: They transform
like the components of a tensor, not as scalars.

In this respect it is quite remarkable, that the gravitating mass
of the holostar can be derived from the integral over the trace of
the stress-energy tensor $T = T_\mu^\mu$. In fact, the integral
over $T$ is exactly equal to $M$ if the negative point mass $M_0 =
-r_0/2$ at the center is included, or - which is the preferred
procedure - if the integration is performed from $r \in [r_0,
\infty]$, as was suggested in the previous section.

\begin{equation} \label{eq:Energy:T}
\int_0^\infty{T 4 \pi r^2 dr} = \int_0^\infty{(\rho - P_r - 2
P_{\theta}) 4 \pi r^2 dr} = \frac{r_h}{2} + M_0 = M
\end{equation}

Contrary to the mass-density and/or the pressure, $T$ is a
Lorentz-scalar and therefore has a definite coordinate-independent
meaning at any space-time point.

It is quite interesting that the gravitating mass of the holostar
also can be derived from the so called Tolman-mass $M_{Tol}$ of
the space-time:

\begin{equation} \label{eq:M:Tolman}
M_{Tol} = \int_0^\infty{(\rho + P_r + 2 P_{\theta}) \sqrt{-g}
d^3x} = \frac{r_h}{2} + M_0 = M
\end{equation}

The Tolman mass is often referred to as the "active gravitational
mass" of a system. The motion of particles in a wide class of
exact solutions to the equations of general relativity indicate
that the sum of the matter density and the three principle
pressures can be interpreted as the "true source" of the
gravitational field, or rather of the field's action, its
gravitational attraction.\footnote{Poisson's equation for the
local relative gravitational acceleration $g$ is given by $\nabla
\cdot \textbf{g} = -R_{00} = - 4 \pi G (\rho + \sum{P_i})$ in
local Minkowski-coordinates. Therefore $\rho$ and the three
principal pressures appear as source-terms in Poisson's equation
for $g$, i.e. can be considered as active gravitational mass
densities.}

Also note, that the rr- and tt-components of the Ricci-tensor are
zero everywhere, except for a $\delta$-functional at the position
of the membrane:

$$R_0^0 = R_1^1 = -\frac{1}{2 r_h}\delta = -\frac{1}{4 M + 2 r_0}\delta$$

\subsection{\label{sec:eq:motion}The equations of geodesic motion}

In this section the equations for the geodesic motion of particles
within the holostar are set up. Keep in mind that the results of
the analysis of pure geodesic motion have to be interpreted with
caution, as pure geodesic motion is unrealistic in the interior of
a holostar. In general it is not possible to neglect the
pressure-forces totally. In fact, as will be shown later, it is
quite improbable that the motion of massive particles will be
geodesic throughout the holostar's whole interior space-time.
Nevertheless, the analysis of pure geodesic motion, especially for
photons, is a valuable tool to discover the properties of the
interior space-time.

Disregarding pressure effects, the interior motion of massive
particles or photons can be described by an effective potential.
The geodesic equations of motion for a general spherically
symmetric space-time, expressed in terms of the "geometric"
constants of the motion $r_i$ and $\beta_i$ were given in
\cite{Petri/bh}:

\begin{equation} \label{eq:beta:r:2}
\beta_r^2(r) + V_{eff}(r) = 1
\end{equation}

\begin{equation} \label{eq:Veff:beta}
 V_{eff}(r) = \frac{B(r)}{B(r_i)}\left(1 -
\beta_i^2(1-\frac{r_i^2}{r^2})\right)
\end{equation}

\begin{equation} \label{eq:beta:t:2}
\beta_{\perp}^2(r) = \frac{B(r)}{B(r_i)}\frac{r_i^2}{r^2} \,
\beta_i^2
\end{equation}

$r_i$ is the interior turning point of the motion, $\beta_r$ is
the radial velocity, expressed as fraction to the local velocity
of light in the radial direction, $\beta_{\perp}$ is the
tangential velocity, expressed as fraction to the local velocity
of light in the tangential direction. As $\partial r,
\partial \varphi$ and $\partial \theta$ are orthogonal, $\beta^2 =
\beta_r^2 + \beta_{\perp}^2$. The quantities $\beta^2, \beta_r^2$
and $\beta_\perp^2$ all lie in the interval $[0,1]$.

$\beta_i = \beta(r_i) = \beta_{\perp}(r_i)$ is the velocity of the
particle at the turning point of the motion, $r_i$. $\beta_r(r_i)
= 0$, therefore $\beta_i$ is purely tangential at $r_i$. For
photons $\beta_i =1$. Pure radial motion for photons is only
possible, when $r_i=0$. In this case $V_{eff} = 0$. Therefore the
purely radial motion of photons can be considered as
"force-free".\footnote{The region $r < r_0$ should be considered
as unaccessible, which means that pure radial motions for photons
is impossible. Aiming the photons precisely at $r=0$ also
conflicts with the quantum mechanical uncertainty postulate. The
photons will therefore always be subject to an effective potential
with $r_i > r_0$, i.e. an effective potential that is not
constant. However, whenever the photon has moved an appreciable
distance from its turning point of the motion, i.e. whenever $r
\gg r_i$, the effective potential is nearly zero. Therefore
non-radial motion of photons can be regarded as nearly
"force-free", whenever $r \gg r_i$.}

In order to integrate the geodesic equations of motion, the
following relations are required:

\begin{equation} \label{eq:beta:r}
\beta_r(r) = \frac{dr}{dt} /
\sqrt{\frac{B}{A}}
\end{equation}

\begin{equation} \label{eq:beta:t}
\beta_{\perp}(r) = \frac{r d\varphi}{dt} / \sqrt{B}
\end{equation}

$t$ is the time measured by the asymptotic observer at spatial
infinity. If the equations of motion are to be solved with respect
to the proper time $\tau$ of the particle (this is only reasonable
for massive particles with $\beta <1$), the following relation is
useful:

\begin{equation} dt = d\tau \frac{\gamma_i \sqrt{B(r_i)}}{B}
\end{equation}

$\gamma^2 = 1/(1-\beta^2)$ is the special relativistic
$\gamma$-factor and $\gamma_i = \gamma(r_i)$, the local
$\gamma$-factor of the particle at the turning point of the
motion.

Within the holostar $B = r_0 / r$. Therefore the equations of
motion reduce to the following set of simple equations:

\begin{equation} \label{eq:beta:r2:int}
\beta_r^2(r) = 1 - \frac{r_i}{r}\left(1 -
\beta_i^2\left(1-\frac{r_i^2}{r^2}\right)\right) =
\left(\frac{dr}{dt} \frac{r}{r_0}\right)^2
\end{equation}

\begin{equation} \label{eq:beta:t2:int}
\beta_\perp^2(r) = \beta_i^2 \left(\frac{r_i}{r}\right)^3 =
\left(\frac{r d\varphi}{dt}\right)^2 \frac{r}{r_0}
\end{equation}

The equation for $\varphi$ can be expressed as a function of
radial position $r$, instead of time:

\begin{equation} \label{eq:phi}
\frac{d\varphi}{dr} = \frac{1}{r^2 \beta_r(r)}
\sqrt{\frac{r_i}{r_0}} r_i \beta_i
\end{equation}

Equation (\ref{eq:phi}) determines the orbit of the particle in
the spatial geometry. It is not difficult to integrate. For $r \gg
r_i$ the radial velocity $\beta_r(r)$ is nearly unity, independent
of the nature of the particle and of its velocity at the turning
point of the motion, $\beta_i$. In the region $r \gg r_i$ we find
$d\varphi \propto dr/r^2$, so that the angle remains nearly
unchanged. This implies, that the number of revolutions of an
interior particle around the holostar's center is limited. The
radial coordinate position $r_{1}$ from which an interior particle
can at most perform one more revolution is given by
$\varphi(\infty) - \varphi(r_{1}) < 2 \pi$. Expressing $r_1$ in
multiples of $r_i$ yields:

\begin{equation}
\frac{r_1}{r_i} > \beta_i \sqrt{\frac{r_i}{r_0}} \frac{1}{2 \pi}
\end{equation}

The total number of revolutions of an arbitrary particle, emitted
with tangential velocity component $\beta_i$ from radial
coordinate position $r_i$ is very accurately estimated by the
number of revolutions in an infinitely extended holostar:

\begin{equation} \label{eq:Nrev}
N_{rev} \simeq \frac{1}{2 \pi}
\int_{r_i}^\infty{\frac{d\varphi}{dr}} = \beta_i
\sqrt{\frac{r_i}{r_0}} \frac{1}{2 \pi} \int_{0}^1{\frac{dx}{
\sqrt{1 - x \left(1 - \beta_i^2(1-x^2)\right)}}}
\end{equation}

The definite integral in equation (\ref{eq:Nrev}) only depends on
$\beta_i$. Its value lies between $1.4022$ and $2$, the lower
value for $\beta_i = 1$, i.e. for photons\footnote{The exact value
of the definite integral for $\beta_i=1$ is given by
$\frac{\sqrt{\pi}}{3}
\frac{\Gamma(\frac{1}{3})}{\Gamma(\frac{5}{6})}$}, and the higher
value for $\beta_i = 0$, i.e. for pure radial motion of massive
particles.\footnote{In this case the number of revolutions is zero,
as $\beta_i = 0$} The integral is a monotonically decreasing
function of $\beta_i^2$. From equation (\ref{eq:Nrev}) it is quite
obvious that particles emitted from the central region of the
holostar will cover only a small angular portion of the interior
holostar space-time.

In the following sections we are mainly interested in the radial
part of the motion of the particles.

From equation (\ref{eq:Veff:beta}) it can be seen, that whenever
$B(r)$ is monotonically decreasing, the effective potential is
monotonically decreasing as well, independent of the constants of
the motion, $r_i$ and $\beta_i$. Within the holostar $B(r) =
r_0/r$, therefore the interior effective potential decreases
monotonically from the center at $r=0$ to the boundary at $r =
r_h$, which implies that the radial velocity of an outmoving
particle increases steadily. The motion appears accelerated from
the viewpoint of an exterior asymptotic observer. The perceived
acceleration decreases over time, as the effective potential
becomes very flat for $r \gg r_i$.

For all possible values of $r_i$ and $\beta_i$ the position of the
membrane is a local minimum of the effective potential. Therefore
no motion is possible that remains completely within the
holostar's interior. Any interior particle has two options: Either
it oscillates between the interior and the exterior space-time or
it passes over the angular momentum barrier situated in the
exterior space-time (or tunnels through) and escapes to infinity.

In order to discuss the general features of the radial motion it
is not necessary to solve the equations exactly. For most purposes
we can rely on approximations.

For photons the radial equation of motion is relatively simple:

\begin{equation} \label{eq:beta:r2:photon}
\beta_r(r) = \sqrt{1 - \left(\frac{r_i}{r}\right)^3} =
\frac{dr}{dt} \frac{r}{r_0}
\end{equation}

An exact integration requires elliptic functions. For $r \gg r_i$
the term $(r_i/r)^3$ under the root can be neglected, so that the
solution can be expressed in terms of elementary functions.

For massive particles the general equation (\ref{eq:beta:r2:int})
is very much simplified for pure radial motion, i.e. $\beta_i =
0$:

\begin{equation} \label{eq:beta:r2:m:radial}
\beta_r(r) = \sqrt{1 - \frac{r_i}{r}}=\frac{dr}{dt} \frac{r}{r_0}
\end{equation}

This equation can be integrated with elementary mathematical
functions. The general equation of motion (\ref{eq:beta:r2:int})
requires elliptic integrals. However, in the general case of the
motion of a massive particle ($\beta_i \neq 0$ and $\beta_i^2 <
1$) the equation for the radial velocity component
(\ref{eq:beta:r2:int}) can be approximated as follows for large
values of the radial coordinate coordinate, $r \gg r_i$:

\begin{equation} \label{eq:beta:r2:int_approx}
\beta_r^2(r) \simeq 1 - \frac{r_i}{r}\left(1 - \beta_i^2\right) = 1 -
\frac{r_i}{r} \frac{1}{\gamma_i^2}
\end{equation}

Whenever $r \gg r_i$ the solution to the general radial equation
of motion for a massive particle is very well approximated by the
much simpler, analytic solution for pure radial motion of a
massive particle given by equation (\ref{eq:beta:r2:m:radial}).
One only has to replace $r_i$ by $r_i / \gamma_i^2$. The radial
component of the motion of a particle emitted with arbitrary
$\beta_i$ from $r_i$ is nearly indistinguishable from the motion
of a massive particle that started out "at rest" in a purely
radial direction from a somewhat smaller "fictitious" radial
coordinate value $\widetilde{r_i} = r_i/\gamma_i^2$.

\subsection{Bound versus unbound motion}

In this section I discuss some qualitative features of the motion
of massive particles and photons in the holostar's gravitational
field.

An interior particle is bound, if the effective potential at the
interior turning point of the motion, $r_i$, is equal to the
effective exterior potential at an exterior turning point of the
motion, $r_e$. The effective potential has been normalized such,
that at any turning point of the motion $V_{eff}(r) = 1$. A
necessary (and sufficient) condition for bound motion then is,
that the equation

\begin{equation} \label{eq:cond:turningpoint}
V_{eff}(r_e)=1
\end{equation}

has a real solution $r_e \geq r_h$ in the exterior (Schwarzschild)
space-time.

\subsubsection{Bound motion of massive particles}

For pure radial motion and particles with non-zero rest-mass
equation (\ref{eq:cond:turningpoint}) is easy to solve. We find
the following relation between the exterior and interior turning
points of the motion:

\begin{equation} \label{eq:re:0}
r_{e} = \frac{r_+}{1-\frac{r_0}{r_i}}
\end{equation}

Equation (\ref{eq:re:0}) indicates, that for massive particles
bound orbits are only possible if $r_i > r_0$. If the massive
particles have angular momentum, the turning point of the motion,
$r_i$, must be somewhat larger than $r_0$, if the particles are to
be bound. Angular motion therefore increases the central "unbound"
region. The number of massive particles in the "unbound" central
region, however, is very small. In \cite{Petri/thermo} it will be
shown, that the number of (ultra-relativistic) constituent
particles within a concentric interior region of the holostar in
thermal equilibrium is proportional to the boundary surface of the
region with $N \approx {(r/r_0)}^2$.

Let us assume that a massive particle has an interior turning
point of the motion far away from the central region, i.e. $r_i
\gg r_0$. Equation (\ref{eq:re:0}) then implies, that for such a
particle the exterior turning point of the motion will lie only a
few Planck-distances outside the membrane. This can be seen as
follows:

If a massive particle is to venture an appreciable distance away
from the membrane, the factor $1- r_0 / r_i$ on the right hand
side of equation (\ref{eq:re:0}) must deviate appreciably from
unity. This is only possible if $r_0 \approx r_i $. In the case
$r_i \gg r_0$ equation (\ref{eq:re:0}) gives a value for the
exterior turning point of the motion, $r_e$, that is very close to
the gravitational radius of the holostar. Any particle emitted
from the region $r_i\gg r_0$ will barely get past the membrane.
Even particles whose turning point of the motion is as close as
two fundamental length units from the center of the holostar, i.e.
$r_i = 2 r_0$, have an exterior turning point of the motion
situated just one gravitational radius outside of the membrane at
$r_e = 2 r_+$.

The above analysis demonstrates, that only very few, if any,
massive particles can escape the holostar's gravitational field on
a classical geodesic trajectory. As has been remarked before, an
escape to infinity is only possible for massive particles (with
zero angular momentum) emanating from a sub Planck-size region of
the center ($r_i < r_0$). It is quite unlikely that this region
will contain more than one particle.

In the case of angular motion the picture becomes more
complicated. In general the equation $V(r_e) = 1$ is a cubic
equation in $r_e$:

\begin{equation} \label{eq:re}
B(r_i) = \frac{r_0}{r_i} = (1 -
\frac{r_+}{r_e})\left(1-\beta_i^2(1-\frac{r_i^2}{r_e^2})\right)
\end{equation}

It is possible to solve this equation by elementary methods. The
formula are quite complicated. The general picture is the
following: For any given $r_i$ the particle becomes "less
bound\footnote{i.e. $r_e$ becomes larger than the value given in
equation (\ref{eq:re:0})}", the higher the value of $\beta_i^2$ at
the interior turning point gets. Particles with interior turning
point of the motion close to the center are "less bound" than
particles with interior turning point close to the boundary. For
sufficiently small $r_i$ bound motion is not possible, whenever a
certain value of $\beta_i^2$ is exceeded.

For particles with $r_i \ll r_e$ equation (\ref{eq:re}) is
simplified, as $r_i^2/r_e^2$ can be neglected. We find

\begin{equation} \label{eq:re:small}
\frac{r_0}{r_i} \approx (1 - \frac{r_+}{r_e})(1-\beta_i^2) = (1 -
\frac{r_+}{r_e})\frac{1}{\gamma_i^2}
\end{equation}

which allows us to determine $r_e$ to a fairly good approximation:

\begin{equation} \label{eq:re:approx}
r_{e} \approx \frac{r_+}{1-\frac{r_0 \gamma_i^2}{r_i}}
\end{equation}

The motion is bound, as long as $r_i > \gamma_i^2 r_0$. This
inequality seems to indicate, that any massive particle is able to
escape the holostar in principle, as long as $\gamma_i$ is high
enough. This however, is not true:

The geodesic motion of massive particles and photons is described
by an effective potential, which is a function of $r_i$ and
$\beta_i$. From equation (\ref{eq:Veff:beta}) it can be seen, that
the exterior effective potential for fixed $r_i$ is a
monotonically decreasing function of $\beta_i^2$, i.e. $V_{eff}(r,
\beta_i^2) \geq V_{eff}(r, 1)$. This is true for any radial
position $r$. Therefore the exterior effective potential of any
particle with fixed $r_i$ and arbitrary $\beta_i$ is bounded from
below by the effective potential of a photon. This implies that
whenever a photon possesses an exterior turning point of the
motion, any particle with $\beta_i^2 < 1$ must have an exterior
turning point as well, somewhat closer to the holostar's membrane.
Therefore escape to infinity from any interior position $r_i$ for
an arbitrary rest-mass particle is only possible, if a photon can
escape to infinity from $r_i$.

In the following section the conditions for the unbound motion of
photons are analyzed.

\subsubsection{Unbound motion of photons}

The effective potential for photons at spatial infinity is zero.
Therefore in principal any photon has the chance to escape the
holostar permanently. Usually this doesn't happen, due to the
photon's angular momentum.

Disregarding the quantum-mechanical tunnel-effect, the fate of a
photon, i.e. whether it remains bound in the gravitational field
of the holostar or whether it escapes permanently on a classical
trajectory to infinity, is determined at the angular momentum
barrier, which is situated in the exterior space-time at $r = 3
r_+ / 2$. The exterior effective potential for photons possesses a
local maximum at this position. If the effective potential for a
photon at $r = 3 r_+ / 2$ is less than 1, i.e. $V_{eff}(3 r_+/2) <
1$, the photon has a non-zero radial velocity at the maximum of
the angular momentum barrier. Escape is classically inevitable.
The condition for (classical) escape for photons is thus given by:

\begin{equation} \label{eq:Veff:1}
\frac{r_i}{r_0} \leq 3 \left(\frac{r_+}{2
r_0}\right)^{\frac{2}{3}}
\end{equation}

Any photon emitted from the region defined by equation
(\ref{eq:Veff:1}) will escape, if its motion is purely geodesic.

On the other hand, any photon whose internal turning point of the
motion lies outside the "photon-escape region" defined by equation
(\ref{eq:Veff:1}) will be turned back at the angular momentum
barrier and therefore is bound. Likewise, all massive particles
outside the photon escape region must be bound as well, because
the exterior effective potential of a massive particle is always
larger than that of a photon with the same $r_i$, i.e. the angular
momentum barrier for a massive particle is always higher than that
of a photon emitted from the same $r_i$. Therefore every particle
with interior turning point of the motion outside the "photon
escape region" will be turned back by the angular momentum
barrier.

For large holostars, the region of classical escape for photons
becomes arbitrarily small with respect to the holostar's overall
size. A holostar of the size of the sun with $r_+/r_0 \approx
10^{38}$ has an "unbound" interior region of $r_i \leq 4. 10^{25}
r_0 \approx 0.5 \, nm$. The radial extension of the "photon escape
region" is 13 orders of magnitude less than the holostar's
gravitational radius. The gravitational mass of this region is
negligible compared to the total gravitational mass of the
holostar.\footnote{Quite interestingly, for a holostar of the mass
of the universe ($r \approx 10^{61} r_{Pl}$), the temperature at
the radius of the photon-escape region is $T \approx 2.7 \cdot
10^{10} K \approx 2.3 MeV$, which is quite close to the
temperature of nucleosynthesis.}

\subsubsection{Unbound motion of massive particles}

For particles with non-zero rest-mass the analysis is very much
simplified, if the effect of the angular momentum barrier is
neglected.\footnote{For most combinations of $r_i$ with $\beta_i^2
< 1$, the angular momentum barrier hasn't a significant effect on
the question, whether a the particle is bound or unbound.}
Massive particles generally have an effective potential at spatial
infinity larger than zero. A necessary, but not sufficient
condition for a massive particle to be unbound is, that the
effective potential at spatial infinity be less than 1. This
condition translates to:

\begin{equation} \label{eq:re:1}
r_i < \frac{r_0}{1 - \beta_i^2} = r_0 \gamma_i^2
\end{equation}

According to equation (\ref{eq:re:1}) escape on a classical
geodesic trajectory for a massive particle is only possible from a
region a few Planck-lengths around the center, unless the particle
is highly relativistic. For example, massive particles with
$\beta_i^2 = 0.5$ can only be unbound, if they originate from the
region $r_i < 2 r_0$. If the region of escape for massive
particles is to be macroscopic, the proper tangential velocity
$\beta_i^2$ at the turning point of the motion must be
phenomenally close to the local speed of light. Note however that
in such a case there usually is an angular momentum barrier in the
exterior space-time (see the discussion in the previous section).

\subsection{An upper bound for the particle flux to infinity}

The lifetime of a black hole, due to Hawking evaporation, is
proportional to $M^3$. Hawking radiation is independent of the
interior structure of a black hole. It depends solely on the
exterior metric up to the event horizon. As the exterior
space-times of the holostar and a black hole are identical
(disregarding the Planck-size region between gravitational radius
and membrane) the estimated lifetime of the holostar, due to loss
of interior particles, should not significantly deviate from the
Hawking result.

An upper bound for the flux of particles from the interior of the
holostar to infinity can be derived by the following, albeit
very crude argument:

Under the - as we will later see, unrealistic - assumption, that
the effects of the negative radial pressure can be neglected, the
particles move on geodesics and the results derived in the
previous sections can be applied.

For large holostars and ignoring the pressure the particle flux to
infinity will be dominated by photons or other zero rest-mass
particles, such as neutrinos, emanating from the "photon escape
region" $r_i < C r_+^{2/3} r_0^{1/3}$ defined by equation (\ref{eq:Veff:1}).

The gravitational mass $\Delta M$ of this region, viewed by an
asymptotic observer at infinity, is proportional to $r_+^{2/3}
r_0^{1/3}$.

The exterior asymptotic time $\Delta t$ for a photon to travel
from $r_i$ to the membrane at $r_h \simeq r_+$ is given by:

$$\Delta t = \int_{r_i}^{r_+}{\sqrt{\frac{A}{B}}\frac{dr}{\beta_r}}
= \int_{r_i}^{r_+}{\frac{r}{r_0}
\frac{dr}{\sqrt{1-(\frac{r_i}{r}})^3}}$$

For a large holostar with $r_i \ll r_+$ the integral can be
approximated by:

$$\int_{r_i}^{r_+}{\frac{r}{r_0} \frac{dr}{\sqrt{1-(\frac{r_i}{r}})^3}}
\approx \int_{0}^{r_+}{\frac{r}{r_0} dr} = \frac{r_+^2}{2 r_0}$$

Note, that the time of travel from the membrane, $r_h$, to a
position $r$ well outside the gravitational radius of the holostar
is of order $r - r_h$, i.e. very much shorter than the time of
travel from the center of the holostar to the membrane, which is
proportional to $r_h^2$.

Under the assumption that the continuous particle flux to infinity
is comparable to the time average of the - rather conceived -
process, in which the whole "photon escape region" is moved in one
bunch from the center of the holostar to its surface, one finds
the following estimate for the mass-energy-flux to infinity for a
large holostar:

\begin{equation} \label{eq:flux:1}
\frac{\Delta M}{\Delta t} \propto {\left( \frac{r_0}{r_+}
\right)}^{4/3} \propto {\left( \frac{\sqrt{\hbar}}{M}
\right)}^{4/3}
\end{equation}

This flux is larger than the flux of Hawking radiation, for
which the following relation holds:

$$\frac{dM}{dt} \propto \left(\frac{\sqrt{\hbar}}{M}\right)^2$$

However, the pressure effect has not been taken into account in
equation (\ref{eq:flux:1}). As will be shown in the following
sections, the pressure reduces the photon flux two-fold: First it
reduces the local energy of the outward moving photons, so that
less energy is transported to infinity. Second, if the local
energy of the individual photons is reduced with respect to pure
geodesic motion, the chances of classical escape for a photon are
dramatically reduced, because most photons will not have enough
"energy" to escape when they finally reach the pressure-free
region beyond the membrane.

The first effect reduces the energy of the photon flux by a factor
$r_+^{-1/3}$, as can be derived from the results of the following
section. This tightens the bound given in equation
(\ref{eq:flux:1}) to $dM/dt \propto 1/M^{5/3}$.

The second effect will effectively switch off the flux of photons.
As will be shown later, the energy of an ensemble of photons
moving radially outward or inward changes in such a way, that the
ensemble's local energy density is always proportional to the
local energy density of the interior matter it encounters along
its way. Therefore an ensemble of photons "coming from the
interior", having reached the radial position $r_h$ of the
membrane, will be indistinguishable from the photons present at
the membrane. The majority of the photons at the membrane,
however, will have a turning point of the motion close to $r_h$,
meaning that escape on a classical trajectory is impossible.
Therefore, whenever photons coming from the interior have reached
the surface of the holostar, $r_h$, their energy will be so low,
that the vast majority of the photons are bound. Classically it
appears as if no photon will be able to escape from the holostar.

Massive particles which have high velocities at their interior
turning points of the motion behave like photons. Therefore the
discussion of the previous paragraph applies to those particles as
well. Highly relativistic massive particles will not be able to
carry a significant amount of mass-energy to infinity. For massive
particles escape to infinity is only possible, if the motion
starts out from a region within one (or two) fundamental lengths
of the center (see equation (\ref{eq:re:1})). But this region
contains only very few particles, if any at all.

The holostar therefore must be regarded as classically stable,
just as a black hole. Once in a while, however, a particle
undergoing random thermal motion close to the surface might
acquire sufficient energy in order to escape or tunnel through the
angular momentum barrier. Furthermore there are the tidal
forces in the exterior space-time, giving rise to "normal" Hawking
evaporation.

Taking the pressure-effects into account, the mass-energy flux to
infinity of the holographic solution will be quite comparable to
the mass-energy flux due to Hawking-evaporation of a black hole.
The exponent $x$ in the energy-flux equation $dM/dt \propto 1/M^x$
will lie somewhere between $5/3$ and $2$, presumably quite close,
if not equal to $2$.

Even the very crude upper bound of equation (\ref{eq:flux:1})
yields quite long life-times. For a holostar with the mass of the
sun, the evaporation time due to equation (\ref{eq:flux:1}) is
still much larger than the age of the universe ($T \approx 10^{44}
s$).

\subsection{Pressure effects and self-consistency}

In this section the effect of the pressure on the internal motion
of particles within the holostar is studied. I will demonstrate
that the negative, purely radial pressure, equal in magnitude to
the mass-density, is an essential property of the holostar, if it
is to be a self-consistent static solution.

Let us consider the radial movement (outward or inward) of a
spherical shell of particles with a proper thickness $\delta l =
\sqrt{r/r_0} \delta r$, situated at radial coordinate position $r$
within the holostar. The shell has a proper volume of $\delta V =
4 \pi r^2 \delta l$ and a total local energy content of $\rho
\delta V = \delta l /2$.

In this section the analysis will be restricted to zero rest-mass
particles, referred to as photons in the following discussion. In
the geometric optics approximation photons move along
null-geodesic trajectories. Note that the pressure will have an
effect on the local energy of the photon. However, as the local
speed of light is independent of the photon's energy, the pressure
will not be able to change the geometry of a photon trajectory,
i.e. the values of $r(t), \theta(t), \varphi(t)$ as determined
from the equations for a null-geodesic trajectory.

For pure radial motion there can be no cross-overs,
i.e. no particle can leave the region defined by the two
concentric boundary surfaces of the shell.

The equation of motion for a null-geodesic in the interior of the
holostar is given by:

\begin{equation} \label{eq:motion:photon}
\frac{dr}{dt} = \frac{r_0}{r} \sqrt{1-\frac{r_i^3}{r^3}}
\end{equation}

For $r \gg r_i$ the square-root factor is nearly one. Therefore
whenever the photon has reached a radial position $r \approx 10
r_i$, a negligible error is made by setting $r_i=0$, which
corresponds to pure radial motion.

Equation (\ref{eq:motion:photon}) with $r_i = 0$ has the solution:

\begin{equation} \label{eq:r(t)photon}
r(t) = \sqrt{2 r_0 t - r^2(0)}
\end{equation}

The radial distance between two photons, one travelling on the
inner boundary of the shell, starting out at $r(0) = r_i$, one
travelling on its outer boundary, starting out from $r(0) = r_i +
\delta r_i$, is given by:

\begin{equation} \label{eq:dr(t)photon}
\delta r(t) = \sqrt{2 r_0 t - (r_i + \delta r_i)^2} - \sqrt{2 r_0 t - r_i^2}
\end{equation}

If $r(t) \gg r_i$, Taylor-expansion of the square-root yields:

\begin{equation} \label{eq:dr(t)photon:appr}
\delta r \approx \delta r_i \frac{r_i}{r}
\end{equation}

In terms of the proper thickness $\delta l = \delta r \sqrt{A}$ of
the shell we find the following relation:

\begin{equation} \label{eq:dr/dl:photon}
\frac{\delta l}{\delta l_i} = \sqrt{\frac{r_i}{r}}
\end{equation}

Therefore, if the shell moves radially with the local speed of
light, its proper thickness changes according to an inverse square
root law.

Whenever the proper radial thickness changes during the movement
of the shell, work will be done against the negative radial
pressure. The rate of change in proper thickness $d (\delta l(r))$
per radial coordinate displacement $dr$ of the shell is given by:

\begin{equation} \label{eq:dl:photon}
d (\delta l) = - \delta l_i \sqrt{r_i} \frac{dr}{2 r^{3/2}}
\end{equation}

Due to the anisotropic pressure (the two tangential pressure
components are zero) any change in volume along the tangential
direction will have no effect on the total energy. The purely
radial pressure has only an effect on the energy of the shell, if
the shell expands or contracts in the radial direction. The work
done by the negative radial pressure therefore is given by:

\begin{equation} \label{eq:dE:photon}
dE = - P_r 4 \pi r^2 d (\delta l) = - \delta l_i \sqrt{r_i} \frac{dr}{4 r^{3/2}}
\end{equation}

Because the radial pressure is negative, the total energy of the
shell is reduced when the shell is compressed along the radial
direction.

The total pressure-induced local energy change of the shell, when
it is moved outward from radial coordinate position $r_i$ to
another position $r \gg r_i$ is given by the following integral:

\begin{equation} \label{eq:dE2:photon}
\Delta E = \int_{r_i}^{r}{dE} = \frac{\delta l_i}{2}
\left(\sqrt{\frac{r_i}{r}} - 1\right)
\end{equation}

But $\delta l_i/2$ is nothing else than the original total local
energy of the shell, $\delta E_i$. Therefore we find the following
expression for the final energy of the shell:

\begin{equation} \label{eq:DeltaE:photon}
\delta E = \delta E_i + \Delta E = \delta E_i \sqrt{\frac{r_i}{r}}
\end{equation}

This result could also have been obtained by assuming the ideal
gas law $\delta E \propto P_r \delta V$ with $\delta V = 4 \pi r^2
\delta l$.

The total energy density in the shell therefore changes according
to the following expression:

\begin{equation} \label{eq:rho(r):photon}
\rho(r) = \frac{\delta E}{\delta V} =
 \frac{\delta E_i}{4 \pi r^2 \delta l_i} = \frac{\delta E_i}{\delta V_i} \frac{r_i^2}{r^2}
 = \rho(r_i) \frac{r_i^2}{r^2}
\end{equation}

We have recovered the inverse square law for the mass-density. The
holographic solution has a remarkable self-consistency. Any
spherical shell carrying a fraction of the total local energy,
moving inward or outward with the local velocity of light, changes
its energy due to the negative radial pressure in such a way, that
the local energy density of the shell at any radial position $r$
always remains exactly proportional to the actual local energy
density $\rho(r)$ of the (static) holographic solution.

This feature is essential, if the holographic solution is to be a
self consistent (quasi-) static solution.\footnote{The term
quasi-static is used, because internal motion within the holostar
is possible, as long as there is no net mass-energy flux in a
particular direction. A radially directed outward flux of (a
fraction of the) interior matter is possible, if this flux is
compensated by an equivalent inward flux. Note that the outward
and inward flowing matter must not necessarily be of the same
kind. If the outward flowing matter consists of massive particles
with a finite life-time, the inward flowing matter is expected to
carry a higher fraction of the decay products, which will be
lighter, possibly zero rest-mass particles.} A holostar evidently
contains matter. A fraction of this matter will consist of
non-zero rest-mass particles. These particles will undergo random
thermal motion. Furthermore, due to Hawking evaporation or
accretion processes there might be a small outward or inward
directed net-flux of mass-energy between the interior central core
region and the boundary. Even if there is no net mass-energy flux,
the different particle species might have non-zero fluxes, as long
as all individual fluxes add up to zero. If the internal motion
(thermal or directed) takes place in such a way, that the local
mass-density of the holostar ($\rho \propto 1/r^2$) is
significantly changed from the inverse square law in a time scale
shorter than the Hawking evaporation time-scale, the holostar
cannot be considered (quasi-) static.

Movement of an interior particle at the speed of light from $r_i
\simeq r_0 $ to the membrane at $r = r_h $ takes an exterior time
$t \simeq r_h^2/(2r_0) \propto M^2$. The movement of a particle
from $r_i = r_h/2$ to the membrane is not much quicker: $t \simeq
3 r_h^2/(8 r_0)$. In any case the time to move through the
holostar's interior is much less than the Hawking evaporation
time-scale, which scales as $M^3$. Therefore a necessary condition
for the holostar to be a self-consistent, quasi-static solution
is, that the radial movement of a shell of zero rest mass
particles should not disrupt the local (static) mass-density.

More generally any local mass-energy fluxes should take place in
such a way, that the overall structure of the holostar is not
destroyed and that local mass-energy fluxes aren't magnified to an
unacceptable level at large scales.

\subsection{Number of interior particles and holographic principle}

The self-consistency argument given above leads to some
interesting predictions. Let us assume that the outmoving shell
consists of essentially non-interacting zero rest-mass particles,
for example photons or neutrinos (ideal relativistic gas
assumption). For a large holostar the mass-density in its outer
regions will become arbitrarily low. Therefore the ideal gas
approximation should be a reasonable assumption for large
holostars. Under this assumption the number of particles in the
shell should remain constant.\footnote{In general relativity the
geometric optics approximation implies the conservation of photon
number.} This allows us to determine the number density of zero
rest-mass particles within the holostar up to a constant factor:

Let $N_i$ be the number of particles in the shell at the position
$r_i$. Then the number density $n(r)$ of particles in the shell,
as it moves inward or outward, is given by the respective change
of the shell's volume:

\begin{equation} \label{eq:numberdensity}
n(r) = \frac{N_i}{\delta V} =  \frac{N_i}{4 \pi r^2 \delta l} =
\frac{N_i}{4 \pi \sqrt{r_i} \delta l_i} \frac{1}{r^{\frac{3}{2}}}
= n(r_i) \left(\frac{r_i}{r}\right)^{\frac{3}{2}}
\end{equation}

Under the assumption that the interior matter-distribution of the
holostar is (quasi-) static, and that the local composition of the
matter at any particular $r$-position doesn't change with time,
self-consistency requires that the number density of zero
rest-mass particles per proper volume should be proportional to
the number density predicted by equation (\ref{eq:numberdensity}).

The total number of zero rest-mass particles in a holostar will
then by given by the proper volume integral of the number density
(\ref{eq:numberdensity}) over the whole interior volume:

\begin{equation} \label{eq:NumberParticles}
N \propto \int_0^{r_h}{\frac{dV}{r^{3/2}}} = \int_0^{r_h}{\frac{4
\pi r^2 \sqrt{\frac{r}{r_0}}dr}{r^{3/2}}} \propto r_h^2 \propto
A_h
\end{equation}

We arrive at the remarkable result, that the total number of zero
rest-mass particles within a holostar should be proportional to
the area of its boundary surface, $A_h$. The same result holds for
any concentric sphere within the holostar's interior. This is an -
albeit still very tentative - indication, that the holographic
principle is valid in classical general relativity, at least for
large compact self gravitating objects.

Under the assumption that the region $r < r_0/2 \approx
\sqrt{\hbar}$ can contain at most one particle (see the discussion
in \cite{Petri/charge}), we find:

\begin{equation}
N = \left(\frac{r_h}{r_0/2}\right)^2 \approx \frac{r_h^2}{\hbar} = \frac{A_h}{4 \pi \hbar}
\end{equation}

\subsection{Local radiation temperature and the Hawking temperature}

Under the assumption that the mass-energy density of the holostar
is dominated by ultra-relativistic particles, the mean energy per
ultra-relativistic particle can be determined from the energy
density $\rho \propto 1/r^2$ and the number-density given in
equation (\ref{eq:numberdensity}).

\begin{equation} \label{eq:Tmean}
\overline{E}(r) = \rho(r) / n(r) \propto \frac{1}{\sqrt{r}}
\end{equation}

This relation could have been obtained directly from the
pressure-induced energy-change of a geodesically moving shell of
zero-rest mass particles: As the number of particles in the shell
remains constant, but the shell's total energy changes according
to $\sqrt{r_i/r}$, the mean energy per particle must change in the
same way as the total energy of the shell varies.

In a gas of relativistic particles in thermal equilibrium the mean
energy per relativistic particle is proportional to the local
temperature in appropriate units. This hints at a local radiation
temperature within the holostar proportional to $1 / \sqrt{r}$.
This argument in itself is not yet too convincing. It - so far -
only applies to the low-density regime in the outer regions of a
holostar, where the motion is nearly geodesic and thus interaction
free. It is questionable, if a temperature in the thermodynamic
sense can be defined under such circumstances.

However, there is another argument for a well defined local
radiation temperature with $T \propto 1 / \sqrt{r}$: At the high
pressures and densities within the central region of the holostar
most of the known particles of the Standard Model will be
ultra-relativistic and their mutual interactions are strong enough
to maintain a thermal spectrum. The energy-density of radiation in
thermal equilibrium is proportional to $T^4$. The energy density
$\rho$ of the holostar is known to be proportional to $1/r^2$.
Radiation will be the dominant contribution to the mass-energy at
high temperature, so this argument also hints at a local
temperature within the holostar's central region proportional to
$1/\sqrt{r}$.

Therefore it is reasonable to assume that the holostar has a well
defined internal local temperature of its zero-rest mass
constituent particles everywhere, i.e. not only in the hot central
region, and that this temperature follows an inverse square-root
law in $r$.

This temperature is isotropic. This statement should be
self-evident for the high temperature central region of the
holostar, where the radiation has a very short path-length and the
interaction time-scale is short. But one also finds an isotropic
temperature in the outer regions of a holostar, where the
radiation moves essentially unscattered. Because of spherical
symmetry, only radiation arriving with a radial component of the
motion at the detector need be considered. Imagine a photon
emitted from the hot inner region of the holostar with an energy
equal to the local temperature at the place of emission, $r_e$.
Due to the square-root dependence of the temperature, the local
temperature at the place of emission, $r_e$, will be higher than
the local temperature at the place of the detector, $r_a$, by the
square-root of the ratio $r_a / r_e$. However, on its way to the
detector the photon will be red-shifted due to the pressure-effect
by exactly the same (or rather inverse) square-root factor, so
that its energy, when it finally arrives at the detector, turns
out to be equal to the local temperature at the detector. The same
argument applies to a photon emitted from the low-temperature
outer region of the holostar. Due to the pressure effect this
photon will be blue-shifted when it travels towards the detector.
Generally one finds, that the pressure induced red-shift (or
blue-shift) exactly compensates the difference of the local
temperatures between the place of emission, $r_e$, and the place
of absorption $r_a$ of a zero rest mass particle.

Disregarding pressure effects, one could naively assume that an
individual photon emitted from an interior position $r_i$ would
undergo gravitational blue shift, as it moves "down" in the
effective potential towards larger values of $r$. If this were
true, a photon moving from a "hot" inner position to a "cold"
outer position would become even hotter. This result is
paradoxical. In fact, the apparent energy change due to the naive
application of the gravitational redshift-formula is exactly
opposite to the pressure-induced effect:

$$\frac{\nu}{\nu_i} = \sqrt{\frac{g_{00}(r_i)}{g_{00}(r)}} = \sqrt{\frac{r}{r_i}}$$

Therefore the naive application of the gravitational Doppler-shift
formula to the interior space-time of the holostar leads to grave
inconsistencies. In the interior of the holostar the usual
gravitational Doppler-shift formula is not
applicable.\footnote{Meaning, that the gravitational Doppler shift
is not the only effect that influences the frequency of a photon
as a function of position.} This has to do with the fact, that its
derivation requires not only a stationary space-time, but also
relies on the geodesic equations of motion, which are only the
"true" equations of motion in vacuum. Although particles move on
geodesics in the (rather unrealistic) case of a pressure-free
"dust-universe", this is not true when significant pressures are
present.\footnote{This fact can be experienced by anyone living on
the surface of the earth. None of us, except astronauts in space,
move geodesically. Geodesic motion means free fall towards the
earth's center. The pressure forces of the earth's surface prevent
us from moving on such a trajectory. From the viewpoint of general
relativity the earth's surface exerts a constant force
accelerating any object lying on its surface against the direction
of the "gravitational pull" of the earth.} Therefore one should
not expect the gravitational Doppler-shift law to be applicable in
space-time regions where the pressure is significant.

Note finally, that the frequency shift due to the interior
pressure applies to all zero rest-mass particles. Furthermore, the
pressure-induced frequency shift is insensitive to the route
travelled by the zero rest-mass particles. It solely depends on
how the volume available to an individual particle has changed,
i.e. only depends on the number-density of the particles which is
a pure function of radial position. If $r_a$ is the radial
position where the photon that was emitted from $r_e$ with
frequency $\nu_e$ is finally absorbed, one finds.

\begin{equation} \label{eq:Redshift1}
\frac{\nu_e}{\nu_a} = \sqrt{\frac{r_a}{r_e}}
\end{equation}

It is now an easy exercise to show, that any geodesically moving
shell of zero rest mass particles preserves the Planck-distribution:

The Planck-distribution is defined as:

$$ n(\nu, T) \propto \frac{\nu^2 d \nu}{e^{\frac{\nu}{T}}-1}$$

$n$ is the density of the photons, $\nu$ their individual
frequency and $T$ the temperature. The left side of the equation,
i.e. the number density of the photons with a given frequency,
scales as $1/r^{3/2}$ according to equation
(\ref{eq:numberdensity}). The right side of the equation has the
same dependence. The frequency $\nu$ of any individual photon
scales with $1/r^{1/2}$ according to equation
(\ref{eq:Redshift1}). The same shift applies to the frequency
interval $d \nu$. Therefore the factor $\nu^2 d\nu$ on the right
side of the equation also scales as $1/r^{3/2}$. The argument of
the exponential function, $\nu / T$ is constant, because both the
frequency of an individual photon $\nu(r)$, as well as the overall
temperature $T(r)$ have the same $r$-dependence.

We find that the Planck-distribution is preserved by the geodesic
motion of a non-interacting gas of radiation within the holostar.
This astonishing result directly follows from the holographic
metric, the effects of the negative radial pressure and
self-consistency.

The local temperature law for the holostar allows a quick
derivation of the Hawking temperature: The zero rest-mass
particles at the surface of the holostar will have a local
temperature proportional to $1/\sqrt{r_h}$. If this is the true
surface temperature of the holostar, one should be able to relate
this temperature to the Hawking temperature of a black hole. The
Hawking temperature is measured in the exterior space-time at
spatial infinity. As the exterior space-time is pressure free, any
particle moving out from the holostar's surface to infinity will
undergo "normal" gravitational red-shift. The red-shift factor is
given by the square-root of the time-coefficient of the metric at
the position of the membrane, i.e. $\sqrt{r_0/r_h}$. Multiplying
the local temperature with this factor gives the temperature of
the holostar at infinity. We find $T \propto 1/r_h = 1 /(r_+ +
r_0)$. Disregarding the rather small value of $r_0$ the Hawking
temperature of a black hole has exactly the same dependence on the
gravitational radius $r_+ = r_h - r_0$ as the holostar's local
surface temperature, measured at infinity. We have just derived
the Hawking temperature up to a constant factor. A simple
dimensional analysis shows that the factor is of the order unity.
In \cite{Petri/thermo} a more definite relationship will be
derived.

One can use the Hawking temperature to fix the local temperature
at the holostar's membrane. The local temperature of the membrane
is blue-shifted with respect to the Hawking temperature, due to
the exterior gravitational field of the holostar. Under the
assumption that the blue-shifted Hawking temperature, $T_H$, is
equal to the local temperature of the membrane, $T(r_h)$, we find:

\begin{equation}
T(r_h) = T_{H} * \sqrt{\frac{B(\infty)}{B(r_h)}} = \frac{\hbar}{4
\pi \sqrt{r_h r_0}}
\end{equation}

Knowing the local temperature within the holostar at one point
allows one to determine the temperature at an arbitrary internal
position:

\begin{equation} \label{eq:Tlocal}
T = \frac{\hbar}{4 \pi \sqrt{r r_0}}
\end{equation}

With the above equation for the local temperature, the unknown
length parameter $r_0$ can be estimated. Raising equation
(\ref{eq:Tlocal}) to the fourth power gives:

\begin{equation} \label{eq:T4/rho}
T^4 = \frac{\hbar^4}{2^5 \pi^3 r_0^2} \frac{1}{8 \pi r^2} =
\frac{\hbar^4}{2^5 \pi^3 r_0^2} \rho
\end{equation}

which implies:

\begin{equation} \label{eq:r0^2}
\frac{r_0^2}{\hbar} = \frac{\hbar^3}{2^5 \pi^3} \frac{\rho}{T^4}
\end{equation}

Under the assumption, that we live in a large holostar of cosmic
proportions we can plug in the temperature of the cosmic microwave
background radiation (CMBR) and the mean matter-density of the
universe into the above equation. If the recent results from WMAP
\cite{WMAP/cosmologicalParameters} are used, i.e. $T_{CMBR} =
2.725 \, K$ and $\rho \approx 0.26 \rho_{crit}$, where
$\rho_{crit} = 3 H^2 / (8 \pi)$ is determined from the
Hubble-constant which is estimated to be approximately $H \approx
71 (km/s)/Mpc$, we find:

\begin{equation} \label{eq:r0^2:est}
r_0^2 \approx 3.52 \hbar
\end{equation}

which corresponds to $r_0 \approx 1.88 \sqrt{\hbar}$. Therefore
$r_0$ is roughly twice the Planck-length, which is quite in
agreement to the theoretical prediction $r_0 \approx 1.87 r_{Pl}$
at low energies, obtained in \cite{Petri/charge}.

\subsection{\label{sec:massive:acc}Geodesic acceleration and pressure - necessary conditions for nearly geodesic motion of massive particles}

In this and the following sections the radial motion of non zero
rest-mass (massive) particles will be analyzed in somewhat greater
detail. The main purpose of this analysis is to show, that as in
the case of photons, geodesic motion of massive particles is
self-consistent within the holostar solution, i.e. preserves the
energy-density $\rho \propto 1/r^2$. It cannot be stated clearly
enough, though, that for massive particles geodesic motion is at
best an approximation to the true motion of the particles within
the pressurized fluid consisting of massive particles and photons
alike. The radial pressure of the space-time will always exert an
acceleration on a massive particle, so that massive particles can
never move truly geodesically. Furthermore, the higher
relativistic the motion of a massive particle becomes, the higher
its proper acceleration in its own frame will be, because the
proper acceleration grows with $\gamma^3$. Therefore the results
derived in this and the following sections might only be
interpretable as "thought-experiments", giving an answer to the
question of what would happen, if the motion were geodesic in
full. Such a thought-experiment is nonetheless worthwhile, because
the "true" motion of the massive particles will lie somewhere "in
between" the motion of photons and the geodesic motion of the
massive particles. If both, the motion of photons and the geodesic
motion of massive particles can be shown to be compatible with the
holostar's internal energy density, it is quite likely that any
other motion, geodesic or not, will be compatible as well.

The motion of a massive particle in the holostar is
subject to two effects: Geodesic proper acceleration and
acceleration due to the pressure forces.

The geodesic (proper) acceleration $g$ for a massive particle at
its turning point of the motion, i.e. at the position where it is
momentarily at rest with respect to the $(t, r, \theta, \varphi)$
coordinate system, is given by the following expression:

\begin{equation} \label{eq:g}
g = \frac{d \beta_r}{d \tau} = \frac{1}{2} \sqrt{\frac{r_0}{r}}
\frac{1}{r}
\end{equation}

In the interior of the holostar, the geodesic acceleration is
always radially outward directed, whereas it is inward-directed in
the exterior space-time. We find that the geodesic acceleration is
always directed towards the membrane. In a certain sense the
membrane can be considered as the true source of the gravitational
attraction.\footnote{Note, that the sum $\rho + P_r + 2 P_\theta$,
which when integrated over the whole space-time gives the so
called Tolman-mass (sometimes also referred to as the "active
gravitational mass"), is zero everywhere except at the membrane.}

Due to the negative radial pressure, an interior (massive)
particle will also be subject to a radially inward directed proper
acceleration resulting from the "pressure force".

Under the rather bold assumption, that the negative radial
pressure in the holostar is produced in the conventional sense,
i.e. by some yet to be found "pressure-particles" moving radially
inward, which once in a while collide with the massive particles
moving outward, the momentum transfer in the collision process
will result in a "net-force" acting on the massive particles. The
acceleration by the pressure force, $a_P$, can then be estimated
as follows for a particle momentarily at rest in the $(t, r,
\theta, \varphi)$-coordinate system:

\begin{equation} \label{eq:aP}
\frac{dp}{d\tau} = m a_P = P_r \sigma = - \frac{\sigma}{8 \pi r^2}
\end{equation}

$m$ is the (special relativistic) mass-energy of the particle and
$\sigma$ its cross-sectional area with respect to the
"pressure-particles". The cross-sectional area will depend on the
characteristics of the field (or particles) that generates the
radial pressure. $\sigma$ might also depend on the typical
interaction energy at a particular $r$-value. If the pressure is
gravitational in origin, one would expect the cross-sectional area
to be roughly equal to the Planck-area.

The geodesic acceleration $g$ has a $1/r^{3/2}$-dependence,
whereas the pressure-induced acceleration $a_P$ follows an inverse
square law ($a_P \propto 1/r^2$), whenever $\sigma$ and $m$ can be
considered constant. For ordinary matter and the strong and
electro-weak forces this condition is fulfilled whenever the
particles are non-relativistic\footnote{Note that the way equation
(\ref{eq:aP}) is set up, $m$ is not the rest-mass of the particle,
but rather the "relativistic mass". For relativistic particles $m$
has to be replaced by $m \rightarrow E = m \gamma$.}, which will
be true in the outer regions of the holostar. For large $r$ it
should be possible to neglect the pressure-induced acceleration
with respect to the geodesic acceleration. In this case the motion
of a massive particle will become geodesic to a high degree of
precision. The region of "almost" geodesic motion is characterized
by $g \gg a_P$, which leads to:

\begin{equation} \label{eq:rgeodesic}
r \gg \left(\frac{\sigma}{4 \pi}\frac{1}{m}\right)^2 \frac{1}{r_0}
\end{equation}

If cross-sectional areas typical for the strong force ($\sigma_{S}
\approx 36 \pi \hbar^2/m_p^2 \approx 40 mb$) are used, the radial
coordinate position $r_{eq}$ where geodesic and pressure-induced
acceleration become equal is given by:

\begin{equation}
r_{eq} \approx \left(\frac{9 \hbar^2}{{m_p}^3}\right)^2
\frac{1}{r_0} \approx 10^{83} cm
\end{equation}

This is roughly a factor of $10^{55}$ larger than the radius of
the observable universe.

For a cross-sectional area of roughly the Planck-area ($\sigma
\approx r_0^2$) and for a particle with the mass of a nucleon ($m
\approx 10^{-19} r_0$) we find $r_{eq} \approx 10^{36} r_0 \approx
17 m$, roughly $0.5$ percent of the gravitational radius of the
sun. According to equation (\ref{eq:Tlocal}) the local temperature
of the holostar at this point is roughly $T_{eq} \approx 10^{13}
K$. This corresponds to a thermal energy of roughly $1 GeV$, i.e.
the rest mass of the nucleon. For an electron the radial position
of equal geodesic and pressure induced acceleration will be larger
by the squared ratio of the proton-mass to the electron mass:
$r_{eq} \approx 5.8 \cdot 10^4 km$. At this point the local
temperature is roughly $T_{eq} \approx 5 \cdot 10^9 K$,
corresponding to an energy of $500 keV$, i.e. the rest mass of the
electron.

Using equation (\ref{eq:Tlocal}) the inequality in equation
(\ref{eq:rgeodesic}) can be expressed as:

\begin{equation} \label{eq:acc:equal}
m > \frac{\sigma}{4 \pi \sqrt{r_{eq} r_0}} = \frac{\sigma}{\hbar}
T
\end{equation}

In the following discussion I make the assumption, that the
cross-sectional area $\sigma$ of the pressure-particles is
comparable to the Planck-area, i.e. $\sigma \approx A_{Pl} \approx
\hbar$. From equation (\ref{eq:acc:equal}) we find, that whenever
the local temperature of one of the constituent particles of the
holostar falls below its rest-mass, the outward directed geodesic
acceleration becomes larger than the inward directed acceleration
due to the pressure, allowing the particle to move outward on a
trajectory which becomes more and more geodesical.

What happens in the region of the holostar, where the temperature
is higher than the rest-mass of the particle? In this case, the
mass $m$ in equation (\ref{eq:acc:equal}) must be replaced by the
total mass-energy of the particle, $m \rightarrow E = \sqrt{m^2 +
p^2}$. Equation (\ref{eq:acc:equal}), which is the condition of
zero net acceleration, then reflects the trivial condition $E
\simeq T$. This condition is fulfilled, at least approximately,
whenever the local temperature is higher than the rest mass of the
(massive) particles. Therefore we arrive at the remarkable
conclusion, that whenever the interior particles become
ultra-relativistic, their net-acceleration in the frame of the
observer at rest in the ($t, r, \theta, \varphi$) coordinate
system is nearly zero, i.e. the holostar is nearly static in this
regime.

In \cite{Petri/thermo} it is shown, that $E = s T$ for
ultra-relativistic particles within the holostar. $s$ is the
entropy per relativistic particle, which is nearly constant and
lies in the range between $3.15$ and $3.5$ for reasonable
assumptions concerning the ratio of bosonic to fermionic degrees
of freedom.\footnote{$s = 3.37$ is the preferred value for the
entropy per particle. As shown in \cite{Petri/thermo} the mean
entropy per particle, $s$, attains this value for the
supersymmetric high temperature phase, where all particles are
ultra-relativistic, the fermionic and bosonic degrees of freedom
are equal and the chemical potentials of fermions and bosons are
opposite to each other. Keep in mind, that $s$ is the mean entropy
per particle. In the holostar bosons and fermions have different
entropy. Therefore the relation $E = s T$ doesn't apply to bosons
or fermions individually, but rather to the average of both
species. In order to keep things simple, I will not make this
necessary distinction here.} The condition for zero
net-acceleration in the radiation dominated central region of the
holostar will be fulfilled exactly, if

\begin{equation} \label{eq:acc:equal:sigma}
\frac{E}{T} = \frac{\sigma}{\hbar} = s
\end{equation}

The temperature $T_i$, where a particle with rest mass $m$ starts
to move outward, is given by:

\begin{equation} \label{eq:T:out}
T_i = \frac{m}{s} \approx \frac{m}{3}
\end{equation}

It is not altogether clear, whether zero net-acceleration is
exactly achievable in the hot interior region of the holostar. For
$\sigma / \hbar < s$ the geodesic outward directed acceleration is
larger than the pressure-induced deceleration. If the holostar
solution is combined with the results of quantum gravity and the
Immirzi parameter is fixed at $\gamma = s/(\pi \sqrt{3})$ (see the
discussion in \cite{Petri/charge}) then $\sigma/\hbar = s$, at
least at the Planck energy. At lower energies the mean
cross-sectional area of the particles is expected to lie in the
range $\pi \sqrt{3/4} < \sigma / \hbar < s $ (for a fundamental
spin-1/2 particle). Therefore for spin-1/2 fermions we expect the
mean cross-sectional area $\sigma$ (with respect to the pressure
forces) to be always less than the mean entropy per particle $s$,
except at the central region $r \approx r_0$.

\subsection{A possible origin of the negative radial pressure}

Note that this section is highly speculative and trods into
uncharted territory without the appropriate guide. On the other
hand, the issues addressed here must be solved in one way or the
other, if the holostar is to be a truly self-consistent model, not
only for black holes comparable to the mass of the sun, but for
holostars of arbitrary size, up to and exceeding the observable
radius of the universe. It might well be that the solution to the
problems addressed in this section, if there is any, will turn out
to be completely different from what is presented here.

It is an astonishing coincidence, that the local temperature of
the holostar at the radial position $r_{eq}$, i.e. where the net
acceleration (geodesic and pressure-induced) becomes nearly zero
for a particular particle, is roughly equal to the rest mass of
the particle. At least this is the case, if the result $\sigma
\approx s \hbar$, which was obtained independently in
\cite{Petri/charge}, is used. This can be considered as
indication, that the negative radial pressure within the holostar
is not just a curious mathematical feature of the solution, but
could be a real, measurable physical effect, and that the
assumption that the pressure is gravitational in origin, with a
cross-sectional area $\sigma$ comparable to the Planck-area, is
not too far off the track.

In fact, if the pressure were produced by a continuous flow of
particles moving radially inward, and that interact only very
weakly with the outflowing "ordinary matter", this could explain
the purely radial nature of the pressure. If the inflowing
"pressure-particles" carry a mass-energy equivalent to that of the
outflowing ordinary matter, this could at the same time explain
the mystery, that the holostar is a static solution which requires
that any outflow of mass-energy must be accompanied by an
equivalent inflow.\footnote{\label{fn:ord:matter}There is no
mystery, as long as the outmoving particles reverse their outward
directed motion in the exterior space-time and thus "swing" back
and forth between exterior and interior space-time (without
friction), as suggested by the holostar equations of motion for a
stable massive particle. The motion of the massive particles would
be time-symmetric and any outmoving particle would be confronted
by a highly relativistic flux of particles moving inward. The
mystery arises, when we identify the holostar with the observable
universe. It will be shown later, that any outmoving massive
particle experiences an isotropic radially outward directed
Hubble-flow in its co-moving frame. This is exactly what we
ourselves observe. However, our situation in the expanding
universe appears time-asymmetric: We haven't yet noticed a flow of
particles hitting us head-on in our "outward" directed motion
(although we might already have noticed the effects of the
inflowing matter in the form of a positive cosmological constant,
i.e. a negative pressure). This must not altogether be a
contradiction. First, if the outflowing and inflowing matter moves
highly relativistically, as is expected from the holostar
equations of motion for a single massive particle, the
cross-sectional area for a collision between an outmoving "matter
particle" and an in-moving "pressure particle" can become almost
arbitrarily small, as $\sigma \propto \hbar^2/E^2$ for ordinary
matter. If the collision energy in the center of mass frame
(=coordinate frame) is of order of the Planck energy, the
cross-sectional area will be comparable to the Planck area. For
larger collision energies a black hole will be formed, so that the
cross-sectional area is expected to rise as the surface area of a
black hole, which is proportional to $E^2$. All in all we expect a
cross-sectional area of $\sigma \propto \hbar^2/E^2 + E^2$.
Secondly, the outflowing particles and the inflowing particles
must not necessarily be the same. The inflowing particles could be
decay products of the outmoving particles or altogether different
species. Any weakly interacting "dark matter" particle would
probably qualify. If the inflowing particles interact only weakly
with the outflowing "ordinary" matter, they could deliver an
inward-directed energy flow comparable to the outward directed
flow and a time-symmetric situation with respect to the net energy
flow could be restored.} Therefore we arrive at two conditions
that should be met by the "pressure particles", whenever "ordinary
matter" and "pressure-particles" must be regarded as distinct:

\begin{itemize}
\item The pressure particles should interact very weakly with
ordinary matter (at least at energies below roughly 1 MeV) \item
the mass-energy density of the pressure particles should be
equivalent to the mass-energy density of ordinary matter
\end{itemize}

The first condition can - in principle - be fulfilled by several
particles. The graviton, the supposed messenger particle for the
gravitational force, looks like the most suitable candidate. The
second condition suggests that supersymmetric matter might be the
preferred candidate for the pressure-particles, if they exist: It
seems quite improbable that massless gravitons, with only two
degrees of freedom, can deliver an energy flow exactly equal to
the energy flow of ordinary matter at any arbitrary radial
position within the holostar. But exact supersymmetry predicts
equal numbers\footnote{With equal numbers I don't mean equal
number densities per proper volume, but rather equal numbers of
degrees of freedom, i.e. equal number of particle species. The
number densities of bosons and fermions in thermodynamic
equilibrium in the holostar will be different, even if the number
of fermionic and bosonic species are equal (for a detailed
discussion see \cite{Petri/charge}).} and masses of the
supersymmetric particles. If the "pressure particles" were the
supersymmetric partners of ordinary matter, it might be possible
to fulfil the second condition quite trivially.\footnote{If
supersymmetry is broken, the masses of the superparticles cannot
be much higher than the $W$- or $Z$-mass. In this case the
negative radial pressure could still be generated by an
appreciable, but not too high number-density of the lightest
superparticle (LSP).} Furthermore, for interaction energies below
the electro-weak unification scale supersymmetric matter and
ordinary matter are effectively decoupled from each other.
Supersymmetric matter therefore fits both conditions well. Note
also, that the holostar solution assumes that the cosmological
constant is exactly zero. Exact supersymmetry guarantees just
this. Therefore - from a purely theoretical point of view - we
would get a much more consistent description of the phenomena, if
supersymmetry were realized exactly even in the low density / low
energy regions of a large holostar.

If the principal agents of the negative radial pressure consist of
supersymmetric matter, this requires an efficient mechanism which
converts the outflowing ordinary matter into the supersymmetric
"pressure-particles". This process is expected to take place in
the membrane. The membrane is in many respects similar to a
two-dimensional domain wall. Therefore the membrane could induce
interactions similar to the conversion processes that are believed
to take place when ordinary matter crosses a two-dimensional
domain wall. On the other hand, if the lightest supersymmetric
particle were light enough, ordinary matter could just decay into
the lightest superparticle anywhere on its way
outward.\footnote{The time of decay would have to be roughly equal
to the radius of the holostar. As the proton's lifetime is several
orders of magnitude larger than the current age of the universe,
this would require a rather large holostar/universe.}

Supersymmetry might also provide the solution to the problem, why
the "pressure particles" (bosons) preferably move inward, whereas
ordinary matter (fermions) moves outward.\footnote{As has been
shown in the previous section, the problem only exists in the
low-density (dynamic) regions of the holostar, where massive
particles move geodesically. The high temperature central region
of the holostar poses no such problem: Due to the approximate
balance between pressure-induced and geodesic acceleration, the
holostar is expected to be nearly static and time-symmetric in
this region.} Such behavior would be easier to understand if one
could think up a mechanism that explains this time-asymmetric
situation: If geodesic movement in the holostar were fully
$T$-symmetric, the time reversed process would be equally likely,
in which case the "pressure-particles" should be ordinary
matter.\footnote{In fact, if the "pressure-particles" move fast
enough, they could well be ordinary matter (see footnote
\ref{fn:ord:matter}). An interaction energy between outflowing
ordinary matter and inflowing pressure particles of order of the
Planck energy doesn't seem too far fetched, if the discussion in
sections \ref{sec:delta:rho} and \ref{sec:phi} is taken
seriously.} In the following argument I make use of the fact, that
the membrane appears as the primary source of gravitational
attraction in the low density region of the holostar, at least for
ordinary matter in nearly geodesic motion. Could the membrane
expel the "pressure particles", after the interactions in the
membrane have converted ordinary matter into supersymmetric
"pressure particles"? The gravitational force is always
attractive, unless we were able to find some form of matter with
$M^2 < 0$.

This is the point where supersymmetry might come up with an
answer: The Higgs-field, which is expected to give mass to the
particles of the Standard Model, is characterized by a quantity
$M^2$, which is usually identified with the mass squared of the
field. Whenever $M^2 > 0$, all particles of the Standard model are
massless. Whenever $M^2$ falls below zero, the Higgs-mechanism
kicks in. In the supersymmetric extension of the Standard Model
the dependence of $M^2$ on the energy/distance-scale can be
calculated. It turns out that $M^2$ is positive at the
Planck-scale and becomes negative close to the energy scale of the
Standard-Model. Therefore the condition $M^2 < 0$ will be
fulfilled in the problematic low-density region of the holostar,
i.e. whenever $T < M_{Higgs}$. The peculiar property $M^2 < 0$ of
the Higgs-field at low energies therefore might provide the
mechanism, by which supersymmetric matter is rather expelled from
the membrane, whereas ordinary matter is attracted. If
supersymmetry can actually provide such a mechanism, we could
understand why the holostar is a static solution, not only for
small holostars\footnote{For small holostars, where $M_H^2 > 0$ in
the whole interior, all of its constituent particles should be
massless. For massless particles the condition $T \approx E$ is
trivially fulfilled. If $\sigma / \hbar = s = E / T$ the
pressure-induced inward directed acceleration and the geodesic
outward directed acceleration are equal throughout the whole
interior. In such a case the holostar is truly static,
time-symmetric and in thermal equilibrium. Note also, that an
extended static central region provides an excellent of
"protection" against continued gravitational contraction of the
whole space-time to a point-singularity.}, but also for an
arbitrarily large holostar, where a large fraction of the interior
matter is situated in the regime where the Higgs-field(s) have
$M^2 < 0$ and "ordinary matter" moves outward.

\subsection{Nearly geodesic motion of massive particles}

In the following we are interested in the low density regions of
the holostar, where the motion of massive particles should become
more and more geodesic, i.e. for $T \ll m$. The geodesic equation
of pure radial motion for a massive particle, starting out at rest
from $r=r_i$, is given by equation (\ref{eq:beta:r2:m:radial}):

\begin{equation} \label{eq:motion:m}
\beta_r = \frac{dr}{dt} \frac{r}{r_0} = \sqrt{1- \frac{r_i}{r}}
\end{equation}

Integration of the above equation gives:

\begin{equation} \label{eq:motion:m:t}
2 r_0 t = \sqrt{1-\frac{r_i}{r}}(r^2 + \frac{3 r_i}{2} r) +
\frac{3}{4} r_i^2 \ln{\left(\frac{2
r}{r_i}(\sqrt{1-\frac{r_i}{r}}+1)-1\right)}
\end{equation}

For $r \gg r_i$ the logarithm can be neglected and the
square-root can be Taylor-expanded to first order:

\begin{equation} \label{eq:motion:m:t2}
2 r_0 t \approx r^2 + r_i r
\end{equation}

or

\begin{equation} \label{eq:motion:m:r}
r \approx -\frac{r_i}{2} + \frac{1}{2} \sqrt{r_i^2 + 8 r_0 t}
\end{equation}

Two massive particles separated initially at $r_i$ by a
radial coordinate separation $\delta r_i$ and moving geodesically
outward, will have a radial coordinate separation $\delta r(r)$
that tends to the following constant value when $r \gg r_i$

\begin{equation} \label{eq:delta_r:m}
\delta r(r) \rightarrow -\frac{\delta r_i}{2}
\end{equation}

The minus sign in the above equation is due to a "cross-over" of
the massive particles, which takes place at a very early stage of
the motion, i.e. where $r \approx r_i$. After the cross-over the
radial coordinate separation of the massive particles quickly
approaches the value $\delta r_i/2$ and remains effectively constant
henceforth.

Constant coordinate separation means that the proper radial
separation viewed by an observer at rest in the ($t, r, \theta,
\varphi$)-coordinate system, $\delta l$, develops according to:

\begin{equation} \label{eq:delta_l:m}
\delta l(r) = \delta l_i \sqrt{\frac{r}{r_i}}
\end{equation}

Therefore, quite contrary to the movement of the zero-rest mass
particles, an outmoving shell of massive particles expands along
the radial direction.

Due to the negative radial pressure the energy in the shell
increases as the shell expands radially. A similar calculation as
in the zero rest mass case gives the following total energy change
in the shell:

\begin{equation} \label{eq:E:massive}
\delta E(r) = \delta E_i \sqrt{\frac{r}{r_i}}
\end{equation}

The proper volume of the shell, as measured from an observer at
rest in the chosen coordinate system, changes according to:

\begin{equation} \label{eq:dB:massive}
\delta V(r) = \delta V_i \left(\frac{r}{r_i}\right)^{\frac{5}{2}}
\end{equation}

A factor proportional to $r^2$ comes from the proper surface area
of the shell, a factor of $r^{1/2}$ from the proper expansion of
the shell's radial dimension.

Note that exactly as in the case of zero rest-mass particles the
mass-energy density of the expanding shell, viewed from an observer at
rest, follows an inverse square law.

\begin{equation} \label{eq:rho:massive:shell}
\rho(r) = \frac{\delta E(r)}{\delta V} = \frac{\delta E_i
\sqrt{\frac{r}{r_i}}}{\delta V_i (\frac{r}{r_i})^{\frac{5}{2}}} =
\rho_i \left(\frac{r_i}{r}\right)^2
\end{equation}

Therefore the geodesic motion of a shell of massive particles also
is self-consistent with the static holographic mass-energy density. In
fact, this self-consistency is independent of the path of the
motion and the nature of the particle. It only depends on the
radial pressure which guarantees, that for any conceivable motion,
geodesic or not, the energy of the particles will change such that
the energy-density law $\rho \propto 1/r^2$ within the holostar is
preserved.

Under the assumption that no particles are created or destroyed in
the shell, the number-density of massive particles develops
according to the inverse proper volume of the shell, as no massive
particles can leave the shell in the regime, where $\delta r_i =
const$.

\begin{equation} \label{eq:n:massive}
n_m(r) = n_i (\frac{r_i}{r})^{\frac{5}{2}}
\end{equation}

In section \ref{sec:frames} it will be shown that expansion
against the negative pressure has the effect to produce new
particles in the co-moving frame. Therefore the assumption that
the particle-number remains constant in the shell, is not entirely
correct. On the other hand, genuine particle-production in the
shell must obey the relevant conservation laws. Some of those,
such as conservation of lepton- and baryon-number, are empiric.
Those "empiric" conservation laws can be violated without severe
consequences with respect to our established physical theories.
There are only few conservation laws that are linked to first
principles, such as local gauge-symmetries. One of these
conservation laws is charge-conservation. Therefore particle
creation in the shell must observe charge conservation. As charge
is quantized in units of the electron charge the difference
between positively and negatively charged elementary particles in
the shell must remain constant, so that the net number-density of
charged particles scales with $1/r^{5/2}$.

Even if particle production in the co-moving shell is taken into
account, the number-density of the massive particles declines much
faster than the number-density of the zero-rest mass particles.
This is a consequence of the fact, that a geodesically moving
shell of photons is compressed in the radial direction, whereas a
geodesically moving shell of massive particles expands. If no
particles are created or destroyed, the respective number
densities behave differently.

The ratio of zero-rest mass particles to massive particles should
be independent of the local Lorentz frame of the observer (see
however the discussion in section \ref{sec:frames}). Therefore,
whenever the zero-rest mass particles are chemically and
kinematically decoupled from the non-zero rest-mass particles,
their ratio can be used as a "clock" by an observer co-moving with
the massive particles. However, any net momentum transfer between
massive and zero rest-mass particles will distort this relation.
Therefore this particular "clock" cannot be considered as highly
accurate.

According to equation (\ref{eq:E:massive}) the energy of the shell
increases with $\sqrt{r/r_i}$. If we assume no particle creation
in the shell, each massive particle must acquire an increasingly
larger energy $E = m_0 \sqrt{r/r_i}$. Where does this energy come
from?

The energy is nothing else than the energy of the motion of a
massive particle, as viewed by an asymptotic observer at rest with
respect to the $(t, r, \theta, \varphi)$ coordinate system.
According to the equations of motion the (almost) radial velocity,
measured by an observer at rest in the $(t, r, \theta, \varphi)$
coordinate system is given by:

\begin{equation} \label{eq:beta2}
\beta^2(r) \simeq \beta_r^2(r) = 1-\frac{r_i}{r}
\end{equation}

which implies:

\begin{equation} \label{eq:gamma}
\gamma^2(r) = \frac{r}{r_i}
\end{equation}

The total energy $E(r)$ of a massive particle with rest mass $m$,
viewed by an observer at rest in the coordinate system, will then
be given by:

\begin{equation} \label{eq:E(r):massive}
E(r) = \gamma(r) m_0 = m_0 \sqrt{\frac{r}{r_i}}
\end{equation}

This is just the energy-increase per particle, which has been
derived from the pressure-induced increase due to the radial
expansion of the shell against the negative pressure.

Therefore, from the perspective of an exterior observer, energy is
conserved not only locally, but globally in the holostar
space-time. This is remarkable, because global energy conservation
is not mandatory in general relativity. In fact, there exist only a limited class of space-times (such as asymptotically flat space-times) in which a global concept of energy can be rigorously defined. Except for a a small class of symmetric space-times it is generally impossible to define a locally meaningful concept of gravitational energy.\footnote{So far no realistic cosmological
space-time has been found, in which global energy-conservation
holds. In the standard cosmological models global
energy-conservation is heavily violated in the radiation dominated
era. If the holostar turns out to be a realistic alternative to
the standard cosmological models, and energy is conserved globally
in the holostar space-time, one could use global energy
conservation as a selection principle to chose among various
possible solutions.}

\subsection{Entropy area law}

For a large holostar the total number of non-relativistic
(massive) particles ($N_m \propto r_h^{3/2}$) can be neglected
with respect to the number of ultra-relativistic (zero rest-mass)
particles ($N \propto r_h^2$). According to the $T \propto
1/\sqrt{r}$-law the main contribution to the mass of a large
holostar comes from its outer low temperature regions. But
whenever the temperature becomes lower than the rest mass of the
particle, the number density of the non-relativistic massive
particles is thinned out with respect to the number density of the
yet relativistic particles. On the other hand, for a small
holostar the internal local temperature is so high, that the
majority of massive particles will become ultra-relativistic, in
fact massless whenever the Higgs-mechanism fails to function,
because $M_H^2 > 0$. Therefore the dominant particle species of a
holostar, large or small, will be ultra-relativistic or zero rest
mass particles.

Under the reasonable assumption that the entropy of the holostar
is proportional to the number of its particles, one recovers the
Hawking entropy-area law for black holes up to a constant
factor.\footnote{with a presumably small correction for the massive
particles} A dimensional analysis shows, that the factor is of
order unity. We therefore find:

\begin{equation}
S \propto N \propto \frac{A}{\hbar}
\end{equation}

A more definite relationship will be derived in \cite{Petri/thermo}.

\subsection{Motion of massive particles in their own proper time}

In this section I will examine the equations of motion from the
viewpoint of a moving massive particle, i.e. from the viewpoint of
the co-moving material observer who moves geodesically. The
geodesic equation of radial motion for a massive particle,
expressed in terms of its own proper time, is given by:

\begin{equation} \label{eq:dr/dtau}
\frac{dr}{d\tau} = \sqrt{\frac{r_0}{r_i}}\sqrt{1-\frac{r_i}{r}}
\end{equation}

The radial coordinate velocity $dr / d\tau$ is nearly constant for
$r \gg r_i$. Integration of the above equation gives:

\begin{equation} \label{eq:tau2}
\tau = \sqrt{\frac{r_i}{r_0}}\left( r \sqrt{1-\frac{r_i}{r}} +
\frac{r_i}{2}
\ln{\left(\frac{2r}{r_i}\big(\sqrt{1-\frac{r_i}{r}}+1\big)-1\right)}\right)
\end{equation}

For large $r \gg r_i$ this can be simplified:

\begin{equation} \label{eq:tau}
\tau \cong \sqrt{\frac{r_i}{r_0}} \left(r + \frac{r_i}{2}
\ln{\frac{4 r}{r_i}}\right) \cong \sqrt{\frac{r_i}{r_0}} r
\end{equation}

The proper time it takes a material observer to move along a
radial geodesic trajectory through the holostar space-time is
proportional to $r$. This is very much different from what the
external asymptotic observer sees. The time measured by an
exterior clock at infinity has been shown to be proportional to
$r^2$.

Under the assumption that we live in a large holostar formula
(\ref{eq:tau}) is quite consistent with the age of the universe,
unless $r_i \ll r_0$.\footnote{As has been noted before, it is quite
unlikely, that any particle can be emitted from $r_i \ll r_0$.} As
can be seen from equation (\ref{eq:tau}), it takes a material
observer roughly the current age of the universe in order to
travel geodesically from the Planck-density region at the
holostar's center (i.e. at $r_i \approx r_0$) to the low density
region at $r \approx 10^{61} r_0$, where the density is comparable
to the density of the universe observed today: For $r_i = r_0$ we
find $\tau = r \approx 1.6 \cdot 10^{10} y$, if $r\approx 10^{61}
r_{Pl}$. The proper time of travel could be much longer: If a
massive particle is emitted (with zero velocity) from $r_i > r_0$,
the proper time of travel to radial position $r$ is larger than
the former value by the square root ratio of $r_i$ to $r_0$.
Therefore the holostar solution is compatible with the age of the
oldest objects in our universe ($\approx 1.3 - 1.9 \,  \cdot
10^{10} y$), but would also allow a much older age.

Note, that the holostar solution has no need for inflation.
According to the equations governing the geodesic motion of a
material observer in the holostar, the whole observable universe
has started out from a space-time region in thermal equilibrium,
which was causally connected. The "scale factor" $r$ develops
proportional to $\tau$. The "expansion", defined by the local
Hubble-radius, also develops proportional to $\tau$. Therefore any
causally connected region remains causally connected during the
"expansion". The causal horizon and the particle-horizon remain
always proportional. Furthermore the number-density law $n_m
\propto 1/r^{5/2}$ for massive particles indicates, that very
massive particles that have decoupled kinematically from the
radiation at an early epoch, such as magnetic monopoles, become
very much thinned out with respect to the radiation or the lighter
particles, such as baryons, which decouple much later.

\subsection{A linear and a quadratic redshift distance relation}

From equation (\ref{eq:Redshift1}) a linear redshift-distance
relation can be derived, which is in some sense similar to the
redshift-distance relations of the standard Robertson Walker
models of the universe.

Imagine a concentric shell of material observers moving radially
outward through the holostar. Place two observers in galaxies at
the inner and outer surfaces of the shell and another observer in
a galaxy midway between the two outer observers. When the observer
in the middle reaches radial coordinate position $r_e$, the two
other observers are instructed to emit a photon with frequency
$\nu_e$ in direction of the middle observer.\footnote{All
observers could (at least in principle) synchronize their clocks
via the microwave background radiation or the total matter
density.} At this moment the proper radial thickness of the shell,
i.e. the proper distance between the two outer galaxies, shall be
$\delta l_e$. Let the three galaxies travel geodesically outward.
Some million years later (depending on how large $\delta l_e$ has
been chosen), the photons from the edge-galaxies will finally
reach the observer in the middle galaxy. The observer in the
middle determines his radial coordinate position at the time of
absorption, $r_a$. According to equation (\ref{eq:Redshift1}) the
photons will have been red-shifted by the squareroot of the ratio
of $r_e / r_a$. In order to derive the redshift-distance relation
we only need to know, how the proper distance between the galaxies
has changed as a function of $r$. According to equation
(\ref{eq:delta_l:m}) the proper radial distance grows proportional
to the square-root of the radial coordinate value, whenever the
galaxies move geodesically:

\begin{equation} \label{eq:RedshiftDistance}
1+z = \frac{\nu_e}{\nu_a} = \sqrt{\frac{r_a}{r_e}} = \frac{\delta
l_a}{\delta l_e}
\end{equation}

The final result is, that the light emitted from the distant
galaxies is red-shifted by the ratio of the proper distances of
the galaxies at the time of emission to the time of absorption.

However, this is the result that an observer at rest in the ($t,
r, \theta, \varphi$)-coordinate system would see. The co-moving
observer will find a different relation (see also the discussion
in the following section and in section \ref{sec:frames}). For the
co-moving observer the proper radial distance has to be multiplied
with his special relativistic $\gamma$-factor. If we denote the
proper separation between the galaxies in the system of the
co-moving observer with an overline, we find:

\begin{equation} \label{eq:RedshiftDistance:co1}
\frac{\overline{\delta l_a}}{\overline{\delta l_e}} =
\frac{r_a}{r_e}
\end{equation}

If we insert this into equation (\ref{eq:RedshiftDistance}) the
result is:

\begin{equation} \label{eq:RedshiftDistance:co}
(1+z)^2 = \left(\frac{\nu_e}{\nu_a}\right)^2 =
\frac{\overline{\delta l_a}}{\overline{\delta l_e}}
\end{equation}

This result might seem paradoxical. However, this is exactly what
the co-moving observer must see: The stress-energy tensor of the
holostar space-time is radially boost invariant. Therefore any
radial boost should not affect the local physics. This means that
in the co-moving frame, as well as in the coordinate frame, the
frequency of the photons should be proportional to the local
radiation temperature, i.e. $\nu \propto T$. The local radiation
temperature, however, depends on the inverse square-root of the
radial coordinate value: $T \propto 1/\sqrt{r}$. Due to Lorentz
contraction (or rather Lorentz-elongation in the co-moving frame)
the proper distance $\overline{\delta l}$ in the radial direction
between two geodesically moving massive particles develops
proportional to $r$ (see also the next section). Putting all this
together gives: $\overline{\delta l} \propto r \propto 1/T^2
\propto 1 /\nu^2 \propto 1/(1+z)^2$.

The quadratic redshift dependence should show up in the
measurement of the Hubble-constant at high redshifts. Its effect
should be similar to that of a cosmological constant.

\subsection{\label{sec:Hubble}An isotropic Hubble flow of massive particles}

With the equations of motion for massive particles one can show,
that an observer co-moving with the massive particles within the
outmoving shell will see an isotropic Hubble-type expansion of the
massive particles with respect to his point of view.

First let us calculate the Hubble-flow viewed by the co-moving
observer in the tangential direction, by analyzing how the proper
distance between the radially moving observer and a neighboring
radially moving particle develops. I.e. observer and particle
always have the same $r$-coordinate value. Because of spherical
symmetry the coordinate system can be chosen such, that both move
in the plane $\theta = \pi/2$. The observer moves along the radial
trajectory $\varphi = 0$ and the neighboring particle along
$\varphi = \varphi_0$. The proper distance between observer and
particle is given by:

$$l = r \varphi_0$$

After a time $d\tau$ the distance will have changed due to the
radial motion of both particles:

$$\frac{dl}{d\tau} = \frac{dr}{d\tau} \varphi_0$$

The "speed" by which the particle at $\varphi_0$ moves away from the
observer is given by:

$$v = \frac{dl}{d\tau} = \frac{dr}{d\tau} \varphi_0 = \frac{dr}{d\tau} \frac{l}{r}$$

Therefore the Hubble-parameter in the tangential direction is
given by:

$$H_{\perp} = \frac{v}{l} = \frac{dr}{d\tau} \cdot \frac{1}{r}$$

Note that special relativistic effects due to the highly
relativistic motion of the co-moving observer don't have to be
taken into account, because the distances and velocities measured
are perpendicular to the direction of the motion.

For the derivation of the Hubble parameter in the radial direction
the Lorentz-contraction due to the relativistic motion has to be
taken into account. The proper radial separation of two particles
in geodesic motion, as seen by the observer at rest in the ($t$,
$r$, $\theta$, $\varphi$) coordinate system, develops as:

$$ l = l_i \sqrt{\frac{r}{r_i}}$$

This formula has to be corrected. Due to the relative motion of
the co-moving observer, the observer at rest in the ($t$, $r$,
$\theta$, $\varphi$) coordinate system will see a proper length,
which has been Lorentz-contracted. Therefore the co-moving
observer must measure a proper length which is larger by the
special relativistic $\gamma$-factor. In the system of the
co-moving observer the formula for the proper length (denoted by
barred quantities) is then given by:

$$\overline{l} = \overline{l_i} \frac{r}{r_i}$$

This gives:

\begin{equation}
H_r = \frac{\frac{d\overline{l}}{d\tau}}{\overline{l}} =
\frac{dr}{d\tau} \cdot \frac{1}{r} = H_{\perp}
\end{equation}

The radial and the tangential local Hubble-values are equal. They
just depend on $r$ and the "proper radial coordinate velocity"
$dr/d\tau$. Its value is given by equation (\ref{eq:dr/dtau}). It
is nearly constant for $r \gg r_i$. Therefore the isotropic
Hubble-parameter can finally be expressed as:

\begin{equation} \label{eq:Hubble}
H(r) = \frac{1}{r} \sqrt{\frac{r_0}{r_i}-\frac{r_0}{r}} \simeq
\frac{1}{r} \sqrt{\frac{r_0}{r_i}} \simeq \frac{1}{\tau}
\end{equation}

\subsection{\label{sec:delta:rho}Density perturbations and their evolution in the holostar universe}

The Hubble law given in equation (\ref{eq:Hubble}) contains an
unknown parameter, the nearly constant proper radial coordinate
velocity $dr/d\tau \approx \sqrt{r_0/r_i}$. If the holostar is an
appropriate model for the universe, $dr/d\tau$ can be determined
by comparing the measured Hubble constant with our current radial
coordinate position $r$, which can be estimated from the
mass-density. This will be done in section \ref{sec:measurement}.
However, it would be quite helpful if $dr / d\tau$ could be
determined by some independent method. The evolution of the
density-fluctuations from the time of decoupling to the structures
we find today might provide such a means:

The fluctuations in the microwave background radiation have been
determined to be roughly equal to $\delta T / T= 10^{-5}$. These
small temperature fluctuations are interpreted as the relative
density fluctuations at the time of decoupling, i.e. at the time
when the microwave background radiation temperature was believed
to be roughly 1000 times higher than today. In the adiabatic
approach (instantaneous decoupling) the respective fluctuations in
the baryon-density at the time of decoupling are $\delta = \delta
\rho_B / \rho_B \approx 3 \delta T / T$.\footnote{This estimate is
based on the assumption, that decoupling happened very fast,
almost instantly. Today's refined estimates take into account,
that the decoupling took longer. In this case the ratio $\delta
\rho_B / \rho_B$ at decoupling is lower than in the simple
adiabatic model up to a factor of 10 (see for example
\cite{Silk}).} From these fluctuations the large scale
distribution of galaxies, as we see them today, should have
formed. The density fluctuation in the distributions of galaxies
on a large scale is roughly of order unity today: $\delta_{today}
= \delta \rho / \rho \simeq 1$. In the standard cosmological
models this evolution of the density fluctuations is quite
difficult to explain. The problem is, that in the dust
approximation (no significant pressure, velocity of sound nearly
zero), which is assumed to be an accurate description of the
universe after decoupling, the fluctuations grow with $\delta
\propto t^{2/3}$. After decoupling the universe is matter
dominated. In the matter dominated era $r \propto t^{2/3}$. The
temperature $T$ and the scale factor $r$ are related in the
standard cosmological models as $T \propto 1/r$. Combining these
dependencies we get:

\begin{equation}
\delta \propto t^{2/3} \propto r \propto \frac{1}{T}
\end{equation}

The above formula predicts $\delta_{today} \approx 10^{-2}$, which
is roughly two orders of magnitude less than the observed value.

In order to explain the rather large density fluctuations today,
the standard cosmological model is usually extended to encompass
cold dark matter.

In the holostar universe the evolution of the density-fluctuations
can be explained quite easily. Only one parameter, $dr/d\tau$,
which appears in equation (\ref{eq:Hubble}) for the Hubble value
need to be adjusted.

Let us first consider the dust case, i.e. the evolution of the
density fluctuations in the holostar by deliberately ignoring the
effects of the pressure. After decoupling the expansion in the
holostar universe (in the co-moving frame) is very similar to the
expansion in the standard Robertson Walker models. The expansion
is isotropic and the density is matter-dominated. Therefore the
usual standard model formula for the evolution of the density
fluctuations should be applicable. In a matter-dominated Robertson
Walker universe with no pressure, the density fluctuations evolve
according to the following differential equation (see for example
\cite{Peacock}):

\begin{equation} \label{eq:delta}
\ddot{\delta} + 2 \frac{\dot{r}}{r} \dot{\delta} - 4 \pi \rho
\delta = 0
\end{equation}

For the holostar we find:

\begin{equation}
r = \sqrt{\frac{r_0}{r_i}} \tau
\end{equation}

\begin{equation}
\frac{\dot{r}}{r} = \frac{1}{\tau}
\end{equation}

\begin{equation}
4 \pi \rho = \frac{1}{2r^2} = \frac{r_i}{2 r_0}\frac{1}{\tau^2}
\end{equation}

With these relations equation (\ref{eq:delta}) reduces to the
following differential equation:

\begin{equation} \label{eq:delta:2}
\ddot{\delta} + \frac{2}{\tau} \dot{\delta} - \frac{r_i}{2 r_0} \frac{1}{\tau^2} \delta = 0
\end{equation}

The equation can be solved by the following ansatz:

\begin{equation}
\delta = \tau^n
\end{equation}

This gives a quadratic equation for $n$

\begin{equation}
n^2 + n -\frac{r_i}{2 r_0} = 0
\end{equation}

which can be solved for $n$:

\begin{equation}
n = -\frac{1}{2} \pm \sqrt{\frac{1}{4} + \frac{r_i}{2 r_0}}
\end{equation}

In the holographic universe we have the following dependencies:

$$r \propto \tau \propto \frac{1}{T^2}$$

Therefore we can express $\delta$ as a function of temperature
$T$:

\begin{equation} \label{eq:delta:T}
\delta \propto \tau^n \propto \frac{1}{T^{2n}} \propto
\frac{1}{T^{\epsilon}}
\end{equation}

with

\begin{equation} \label{eq:exponent:delta}
\epsilon = -1 \pm \sqrt{1+\frac{2 r_i}{r_0}}
\end{equation}

The exponent $\epsilon$ in the above equation can be estimated
from the known ratio of the density contrast $\delta_{dec} /
\delta_{today}$ and the ratio of the decoupling temperature to the
CMBR-temperature.

In the holostar the temperature at decoupling will be larger than
in the standard cosmological model, because the number ratio of
baryons to photons doesn't remain constant. If radiation and
matter do not interact after decoupling (no re-ionization!) and
the baryon number remains constant in the co-moving volume, the
baryon-density will scale as $1/r^{3} \propto T^6$ in the frame of
the geodesically moving material observer. However, the discussion
in section \ref{sec:frames} indicates, that the expansion against
the negative pressure creates new particles. The simplest
assumption compatible with the experimental data is that the
baryon-density will scale as $1/r^2 \propto T^4$. Under these
circumstances the Saha-equation yields a temperature at decoupling
of roughly $T_{dec}\approx 4 900 K$, when no dark matter component
is assumed and the baryon-density today is set to the total
matter-density as determined by WMAP
\cite{WMAP/cosmologicalParameters}, i.e. $\rho_B = \rho_m \approx
2.5 \cdot 10^{-27} kg / m^3$.\footnote{This matter density
corresponds to $\Omega_m \approx 0.26$ or roughly $1.5$ nucleons
per cubic meter at a critical density $\rho_c$ determined from a
Hubble value of $H \simeq 71 km/s / Mpc$.} A decoupling
temperature of $4 900 K$ corresponds to a red-shift $z_{dec}
\simeq 1790$. In order to achieve $\delta_{today} \approx 1$ from
$\delta_{dec} \approx 3 \cdot 10^{-5}$ at $z_{dec} = 1790$, the
exponent $\epsilon$ in equation (\ref{eq:delta:T}) must be roughly
equal to $1.48$. For more realistic scenarios in which decoupling
doesn't happen instantaneously $\delta_{dec}$ is estimated to be
lower, up to a factor of 10 \cite{Silk}. For a very slow
decoupling scenario with $\delta_{dec} \approx 0.3 \cdot 10^{-5}$
we require $\epsilon \approx 1.7$.

For $r_i = 4 r_0$ we get $\delta \propto 1/T^2$, which quite
certainly is too high. For $r_i = 2 r_0$ we find $\epsilon =
\sqrt{5}-1 \simeq 1.24$, which is too low. A good compromise is
given by $r_i \approx 3 r_0$, which provides a fairly good fit for
moderately slow decoupling:

\begin{equation}
r_i = 3 r_0 \,  \rightarrow \,
\delta \propto \frac{1}{T^{\sqrt{7}-1}} \simeq \frac{1}{T^{1.65}}
\end{equation}

The value $r_i = 3 r_0$ is interesting, because for this value the
critical mass-energy density in the standard cosmological model,
which is given by $\rho_c = 3 H^2 / (8 \pi)$ is exactly equal to
the energy density of the holostar, which is given by $\rho = 1 /
(8 \pi r^2)$.

We find that the evolution of the density fluctuations requires
$r_i \approx r_0$ and thus $H \approx 1/r$. However, we cannot
expect highly accurate results from the above treatment. In the
holostar universe it is not possible to neglect the effects of the
pressure. In order to incorporate pressure into the treatment
above, we need to know how the anisotropic pressure will manifest
itself in the co-moving frame. In \cite{McManus/Coley} universes
with anisotropic pressure were studied. The authors found, that an
anisotropic pressure will appear as an isotropic pressure for the
observer co-moving with the cosmic fluid. The isotropic pressure
in the co-moving frame, $P$, is related to the anisotropic
pressure components as $P = (P_r + 2 P_\theta)/3$. If this
relation is applied to the holostar, we find $P = -1/(24 \pi r^2)
= -\rho/3$. Except for the sign this is the pressure of normal
radiation. In \cite{Petri/thermo} $P = 1/(24 \pi r^2)$ was found
for the holostar in thermodynamic equilibrium. This discrepancy
clearly demonstrates, that there is an open problem with respect
to the right sign of the pressure in the holostar solution with
respect to the co-moving frame. Nonetheless, the sign in the
pressure shouldn't influence the square of the velocity of sound,
which is the relevant parameter for the density evolution in a
Robertson Walker universe including pressure. Therefore I make the
ansatz, that the density fluctuations in the holostar after
decoupling evolve just as the density fluctuations of a radiation
dominated universe. The evolution equation (\ref{eq:delta}) then
has to be replaced by:

\begin{equation} \label{eq:delta:p}
\ddot{\delta} + 2 \frac{\dot{r}}{r} \dot{\delta} - 4 \pi \rho \frac{8}{3}
\delta = 0
\end{equation}

If the above equation is expressed in terms of $r_i/r_0$, the only
difference to the dust case in equation (\ref{eq:delta:2}) is to
replace $r_i/r_0 \rightarrow (8 r_i)/(3 r_0)$, so that the
exponent in equation (\ref{eq:delta:T}) is given by:

\begin{equation} \label{eq:exponent:delta:p}
\epsilon = -1 \pm \sqrt{1+\frac{16 r_i}{3 r_0}}
\end{equation}

In order to achieve an exponent of roughly $1.65$ we need $1+16
r_i/(3 r_0) \approx 7 $, which requires

\begin{equation}
\frac{r_i}{r_0} \approx \frac{9}{8}
\end{equation}

Note, that the above value is very close to unity. For $r_i = r_0$
we find

\begin{equation}
\delta \propto \frac{1}{T^{\sqrt{19/3}-1}} \simeq \frac{1}{T^{1.52}}
\end{equation}

which is a good compromise between models assuming instantaneous
decoupling ($\epsilon \approx 1.4$) or very slow decoupling
($\epsilon \approx 1.7$).

Equation (\ref{eq:exponent:delta:p}) is quite sensitive to the
value of $r_i / r_0$. It seems quite clear that $r_i$ cannot
possibly lie outside the range $0.5 r_0 < r_i < 2 r_0$. If the
temperature at decoupling was $4 900 K$ and decoupling took place
very fast, the best fit would be $r_i/r_0 \simeq 0.9$. For
non-geodesic motion (after decoupling) one expects that the
decoupling temperature will be lower, somewhere in the range
between $3 500$ and $4 900 K$. A lower decoupling temperature and
non-geodesic motion both require a higher exponent in equation
(\ref{eq:delta:T}), so that $r_i \simeq r_0$. For non-adiabatic
(i.e. not instantaneous) decoupling the baryonic density
fluctuations at decoupling have been estimated to be lower (see
for example \cite{Silk}), placing $\delta_{dec}$ in the range $0.3
\cdot 10^{-5} < \delta_{dec} < 3 \cdot 10^{-5}$, in which case
$r_i/r_0$ should be very close to unity.

Whatever the exact value of $r_i/r_0$ might be, we are drawn to
the conclusion that the massive particles making up the matter in
our universe must have moved nearly geodesically from $r_i \approx
r_0$. This is very difficult to believe without further
evidence.\footnote{\label{fn:ri}Note also that $r_i \approx r_0$
indicates, that the particles must have had extremely high radial
velocities at decoupling: With the - incorrect - assumption that
the number of particles remains constant during the expansion the
order of magnitude of the relativistic $\gamma$-factor at
decoupling can be roughly estimated: $\gamma_{dec} \approx
\sqrt{r_{dec}/r_i} \approx 1.7 \cdot 10^{27}$. Such high
velocities at decoupling are only conceivable, if the massive
particles that constitute the matter of the universe today have
truly originated from $r_i \approx r_0$. This possibility is
discussed in section \ref{sec:phi}.}

An accurate estimate for the exponent in equation
(\ref{eq:delta:T}) and therefore an accurate prediction for the
Hubble-value can only be made if the true equations of motion of
massive particles in the holostar space-time, including pressure,
are used. Furthermore at distances $r \approx r_0$ the
discreteness of the geometry and quantum effects will come into
play. A detailed analysis of the motion of particles in the
holostar, including the region close to its center, must be left
to future research.

\subsection{\label{sec:phi}On the angular correlation of the microwave background radiation}

As has already hinted in footnote \ref{fn:ri} the experimentally
determined ratio $r_i/r_0 \approx 1$, that appears as an
independent parameter in equation (\ref{eq:Hubble}) for the Hubble
value only makes sense, if the particles in the universe as we see
it today have originated from a few Planck-distances from the
holostar's center. The most probable scenario is that a
(presumably massive) precursor particle was created close to the
center and then moved out on a nearly geodesic trajectory
henceforth. Nearly geodesic motion through the hot interior region
of the holostar is only conceivable for a massive particle of
roughly Planck mass, which only interacts gravitationally.
Somewhere on its trajectory, not too far from the radial
coordinate value where the temperature has dropped below the
electro-weak unification scale, the particle then must have
decayed into ordinary matter, endowing the final products of the
decay process (electrons, protons, neutrons and neutrinos - or
more generally quark/leptons) with the high momentum gathered on
its outward track. This scenario is quite an extravagant claim,
for which it would be quite helpful to have another - independent
- verification. The missing angular correlation of the microwave
background radiation at large angles provides such a means.

The recent WMAP data \cite{WMAP/cosmologicalParameters} have
revealed, that there is practically no correlation between the
fluctuations in the microwave background radiation at an angular
separation larger than approximately $60^\circ$, which corresponds
to roughly $1$ radian. This feature can be quite effortlessly
explained by the motion of massless and massive particles in the
holostar:

As can be seen from equation (\ref{eq:Nrev}) particles that were
emitted just a few Planck distances from the center of a holostar
have a limited angular spread. According to the scenario proposed
beforehand let us assume, that a massive, uncharged particle with
a fairly long lifetime is created with not too high a tangential
velocity $\beta_i$ close to the center of the holostar. Its
maximum angular spread can be calculated:

\begin{equation} \label{eq:implicit:beta}
\varphi_{max} = \beta_i \sqrt{\frac{r_i}{r_0}}
\int_{0}^1{\frac{dx}{ \sqrt{1 - x \left(1 -
\beta_i^2(1-x^2)\right)}}}  =  \beta_i
\sqrt{\frac{r_i}{r_0}} \, \xi(\beta_i^2)
\end{equation}

$\xi = \xi(\beta_i^2)$ is the value of the definite integral in
the above equation for the value of $\beta_i$ at the turning point
of the motion. As remarked in section \ref{sec:eq:motion} its
value lies between $1.4 < \xi \leq 2$. The above equation gives an
implicit relation for $\beta_i^2$ , which is the mean velocity
$\beta_i$ of the particle at the radial coordinate position of its
turning point of the motion (or rather the radial coordinate
position $r_i$ where the motion of the particle has become nearly
geodesical). Equation (\ref{eq:implicit:beta}) can be solved
iteratively for $\beta_i^2$, whenever the maximum angular spread
$\varphi_{max}$ and the ratio $r_i/r_0$ is known.

$\varphi_{max}$ and $r_i/r_0$ can be determined experimentally.
However, there is a subtlety involved in the experimental
determination of $r_i/r_0$ from the characteristics of the
expansion. In the derivation of equation (\ref{eq:Hubble}) for the
Hubble-value I have assumed, that the motion started out from
$r_i$ at rest. This is unrealistic. At $r_i \approx r_0$ there is
an extremely high temperature, so that even a particle of nearly
Planck mass will have an appreciable velocity at its (true)
turning point of the motion, $r_i$. Therefore the ratio $r_i/r_0$
in the Hubble equation (\ref{eq:Hubble}) doesn't refer to the true
turning point of the motion $r_i$, but rather to a fictitious
"zero-velocity" turning point $\widetilde{r_i}$, which describes
the radial part of motion far away from the turning point. Both
values are related: In section \ref{sec:eq:motion} it has been
shown, that the radial part of the motion of a particle with an
appreciable tangential velocity at its turning point is nearly
identical to the motion of a particle that started out from a
somewhat smaller radial coordinate value, whenever $r \gg r_i$.
The relation between the true turning point of the motion and the
apparent turning point of the motion is given by:

\begin{equation}
r_i = \gamma_i^2 \widetilde{r_i}
\end{equation}

where $\gamma_i$ is the special relativistic $\gamma$-factor of
the particle at its true turning point of the motion $r_i$.

The experimentally determined ratio $r_i/r_0$ in the Hubble
equation (\ref{eq:Hubble}) or in the equation for the density
evolution (\ref{eq:delta:p}) therefore rather refers to
$\widetilde{r_i}/r_0$, whereas $r_i$ in equation
(\ref{eq:implicit:beta}) rather refers to the true turning point
of the motion. In section \ref{sec:delta:rho} this ratio has been
estimated as $\widetilde{r_i}/r_0 \approx 1$, if the density
perturbations found in the microwave background radiation $\delta
\approx 10^{-5}$ evolve according to equation (\ref{eq:delta:T})
in combination with equation (\ref{eq:exponent:delta:p}) to the
value observed today. Let us denote the ratio
$\widetilde{r_i}/r_0$ with $\kappa$:

\begin{equation}
\kappa = \frac{\widetilde{r_i}}{r_0} = \frac{1}{\gamma_i^2}
\frac{r_i}{r_0}
\end{equation}

The equation for the maximum angular spread then reads:

\begin{equation}
\varphi_{max}^2 = \beta_i^2 \xi^2 \frac{r_i}{r_0} = \beta_i^2
\gamma_i^2 \xi^2 \kappa
\end{equation}

so that:

\begin{equation} \label{eq:determine:betai}
\beta_i^2 \gamma_i^2  = \frac{\beta_i^2}{1-\beta_i^2} \simeq
\frac{\varphi_{max}^2}{\xi^2 \kappa}
\end{equation}

Equation (\ref{eq:determine:betai}) has to be solved for
$\beta_i^2$ (or alternatively for $\beta_i^2 \gamma_i^2$). Once
$\beta_i \gamma_i$ is known, the true turning point of the motion
can be determined:

\begin{equation}
r_i = \gamma_i^2 \widetilde{r_i} = (1 + \beta_i^2 \gamma_i^2)
\kappa r_0
\end{equation}

where the relation $\gamma^2 = 1 + \beta^2 \gamma^2$ was used.

For $\varphi_{max} = 60^\circ = \pi /3 \approx 1$ the maximum
angular spread is nearly unity (in radians). For $\kappa = 1$ we
find $\xi \approx 1.77$. With these values

\begin{equation}
\beta_i^2 \gamma_i^2 \simeq 0.319
\end{equation}

so that

\begin{equation}
\beta_i^2 = \frac{\beta_i^2 \gamma_i^2}{1+ \beta_i^2 \gamma_i^2}
\simeq 0.242
\end{equation}

and

\begin{equation} \label{eq:ri:true}
\frac{r_i}{r_0} \simeq 1.319
\end{equation}

If we know $\beta \gamma$ of a massive particle at the radial
position $r_i$, where the particle's motion has become nearly
geodesical (i.e. the particle is effectively decoupled from the
other particles), we can determine the mass of the particle. The
momentum of any massive particle with mass $m_0$ is given by

\begin{equation}
p = \beta \gamma m_0
\end{equation}

The massive particle will be immersed in a very hot radiation bath
of ultra-relativistic or zero rest mass particles. Most likely it
was created at the hottest possible spot of the holostar, i.e.
somewhere in the region $r_0/2 < r <
r_0$.\footnote{\label{fn:r0}There is good theoretical reason to
believe, that $r_0/2$ provides a universal cut-off for the region
that can by occupied by any one particle. No particle will be able
to enter the region bounded by the smallest possible area quant of
quantum gravity which turns out to be roughly equal to $4 \pi
(r_0/2)^2$ (see \cite{Petri/charge}). Therefore the smallest
"separation" between two particles should be roughly equal to
$r_0$, as far as classical reasoning can still be trusted at the
Planck scale. In \cite{Petri/thermo} the number of particles
within a spherical concentric region has been determined as $N =
\pi / s r^2/\hbar \simeq r^2 / \hbar$, as $s \approx \pi$. With
the experimental estimate $r_0 \approx 2 \sqrt{\hbar}$ from
equation (\ref{eq:r0^2:est}) we find, that $N \simeq 1$ for $r =
r_{Pl} \approx r_0/2$. Note also, that in \cite{Petri/charge} is
has been shown, that $r_0/2$ is the radius of the membrane of an
elementary extremal holostar with zero (or negligible) mass, i.e.
the smallest holostar possible.} During its "movement" from the
"point" of its creation to its "point" of decoupling, $r_i \approx
r_0$, the radiation will have imprinted its momentum on the
massive particle $m_0$. In the holostar the mean momenta
$\overline{p}_\gamma$ of the ultra-relativistic particles are
proportional to the interior radiation temperature, given by
equation (\ref{eq:Tlocal}).

\begin{equation}
\overline{p}_\gamma = s T_\gamma = \frac{s}{4 \pi} \frac{\hbar}{\sqrt{r_i r_0}}
\end{equation}

In \cite{Petri/thermo} it is shown, that the factor of
proportionality, $s$, is equal to the entropy per particle (which
is slightly larger than $\pi$ for a great variety of
circumstances). Putting all equations together and using the
relation $r_0^2 \approx 4s / \pi$ at the Planck-energy proposed in
\cite{Petri/charge} we find:

\begin{equation}
m_0 \simeq \frac{1}{8 } \sqrt{\frac{s}{\kappa \pi}}
\frac{\sqrt{\hbar}}{\beta_i \gamma_i^2}
\end{equation}

With $\kappa = 1$ and $\varphi_{max} = 1$ the mass of the
"precursor" particle $m_0$ is given by:

\begin{equation}
m_0 \simeq 0.201 \, m_{Pl}
\end{equation}

if $s = 3.37174$ is used. This value is the mean entropy per
ultra-relativistic particle in the supersymmetric phase expected
to be present at the center of the holostar, where the fermionic
and bosonic degrees of freedom are equal and the chemical
potentials of bosons and fermions are proportional to the
temperature and opposite to each other (for a more detailed
discussion see \cite{Petri/thermo}).

The particle has a mass roughly a fifth of the Planck mass. Due to
$\overline{p}_\gamma = s T_\gamma = \sqrt{s / \pi}/4 \approx 0.561
\, m_0 \approx 0.113 \, m_{Pl}$ the production of this particle
will be somewhat inhibited at $r=r_i$, as an energy of $E =
\sqrt{\overline{p_\gamma}^2 + m_0^2} \approx 0.231 \, m_{Pl}$ is
required. However, at $r = r_0/2 \approx r_i/2$ there is just
enough energy available to create the "precursor" particles in (in
pairs) in substantial numbers and with the right momenta, either
by the collision of any two massive or massless particles of the
surrounding radiation bath, or - slightly more efficiently - by
Unruh radiation: At a radial coordinate position $r = r_0/2
\approx r_i/2.6$ the local radiation temperature $T_\gamma$ will
be up by a factor of $\sqrt{2.6}$ with respect to the temperature
at $r_i$, so that the mean momentum of the radiation will be given
by $\overline{p}_\gamma \approx 0.183 \, m_{Pl}$. The mean energy
of the massive particle at $r_0/2$ will be higher as well, but not
with the same factor as the radiation, because of its
non-negligible rest-mass. We find: $E_0 \approx
\sqrt{\overline{p_\gamma}^2 + m_0^2} \approx 0.272 \, m_{Pl}$. The
mean momentum of the radiation quanta doesn't yet suffice to
produce the precursor particles efficiently in pairs, at least not
with the right momentum. However, the mean energy of the two
radiation quanta is almost large enough to produce two precursor
particles with zero momentum. Presumably the most efficient
production of the precursor particles is via Unruh radiation
between $r_0/2 < r < r_0$. The Unruh-temperature at $r = r_0/2$
will be twice the radiation temperature (see section
\ref{sec:Unruh}). The energy of a particle produced by Unruh
radiation therefore is $E \approx 0.366 \, m_{Pl} \approx 1.35 \,
E_0$. The production of the precursor particles by Unruh radiation
should be quite efficient, even if the high chemical potential of
the particles, $\mu = 1.353 \, T_\gamma$, in the supersymmetric
phase is taken into account.\footnote{For the relevance of
chemical potentials and the thermodynamics of highly relativistic
matter states see \cite{Petri/thermo}.}

The mass of the "precursor" particle $m_0$ only depends moderately
on $\kappa$. If $\kappa$ lies in the range $0.5 < \kappa < 4$ we
find the following mass-range for $m_0$:

\begin{equation} \label{eq:preon:mass:range}
0.164 < m_0 < 0.243
\end{equation}

in Planck units.

Below a table for the masses and the tangential velocity $\beta_i$
of the precursor particle is given for several values of $\kappa$
with $\varphi_{max} = 56.8^\circ$:
\\
\begin{tabular} {c||c|c|c|c} \label{tab:m0}
$\kappa$ & $\beta_i$ & $\xi$ & $m_0$ \\ \hline
1/2 & 0.647 & 1.65 & 0.165 \\
3/4 & 0.553 & 1.72 & 0.188 \\
1   & 0.490 & 1.77 & 0.201 \\
3/2 & 0.405 & 1.83 & 0.218 \\
2   & 0.353 & 1.87 & 0.228 \\
5/2 & 0.316 & 1.89 & 0.234 \\
3   & 0.288 & 1.91 & 0.238 \\
4   & 0.250 & 1.93 & 0.243 \\
\end{tabular}
\linebreak
\\

If equation (\ref{eq:delta:p}) for the density evolution (with
pressure) is correct, $\widetilde{r_i}$ should lie in the range
$3/4 < \widetilde{r_i}/r_0 < 5/4$, so that the bound for the mass
will be roughly $0.19 < m < 0.21$ in Planck units. A value of
$\widetilde{r_i}$ that lies this range also fits better with the
current measurements of the Hubble-constant (see section
\ref{sec:measurement}).

We find, that if the current expansion rate of the universe and
the evolution of the density perturbations after decoupling can be
(approximately) described by the holostar solution, one requires a
new particle. Lets call it the "preon".\footnote{The "pre" stands
for "precursor". Unfortunately the term "geon", which might be
considered even more appropriate (the syllable "ge" then could
refer to "genesis" as well as "geometric") has already been
taken.} Its properties can be quite accurately inferred from the
above discussion: Its mass should be less than one quarter of the
Planck mass. Its exact value shouldn't lie too far outside the
mass range given by equation (\ref{eq:preon:mass:range}). The
particle should interact only gravitationally and should be
relatively long lived ($\tau \approx 10^{-6} ... 10^{-2} s$), so
that it can survive up to electro-weak transition (or slightly
longer).

It is possible to estimate the mass-energy of the preon by
another, independent argument. According to the discussion in
footnote \ref{fn:r0} the proper volume occupied by a single
particle at the holostar's center is given by

\begin{equation}
V_1 = \int_0^{r_0/2}{dV} = \frac{8 \pi}{7 \sqrt{2}}
\left(\frac{r_0}{2}\right)^3 \approx \frac{8 \pi}{7 \sqrt{2}}
V_{Pl}
\end{equation}

On its way outward the volume available to the preon will become
larger, which enables it to decay into lighter particles, such as
neutrons, protons and electrons. The decay is expected to conserve
mass-energy locally. What will the energy-density of the stable
decay products be at large $r$-values? To simplify the
calculations it is convenient to consider a thought-experiment, in
which the proper length of the expanding volume remains constant
in the radial direction. This thought-experiment simplifies the
calculations as we don't have to take into account any changes of
the internal energy in the expanding volume due to the radial
pressure.\footnote{Keep in mind that the actual motion of the
preon is different from the thought-experiment. However, the
thought-experiment allows us to neglect the effects of the
negative radial pressure without changing the physical results
obtained from a more realistic treatment. If the pressure-effects
are included in the total energy-balance, the actual (geodesic)
motion of the preon yields the same results as the method of
"constant internal energy" in the thought-experiment.} Any motion
that leaves the internal energy of a spherically outmoving volume
unaffected requires that the volume develops as the proper surface
area of an expanding spherical shell with constant proper
thickness, i.e. $V \propto r^2$.

At the radial position $r = 9.18 \cdot 10^{60} \, r_{Pl}$, which
corresponds to the radius of the observable universe today, the
volume $V_1$ will have expanded to:

\begin{equation} \label{eq:V:today}
V_{today} = \left(\frac{r}{r_{Pl}}\right)^2 V_1 \simeq 1.0 \cdot 10^{18} m^3
\end{equation}

Assuming local mass-energy conservation and assuming that the
mass-energy of the preon ends up predominantly in neutrons (i.e.
no significant dark matter component; energy-contributions of
electrons, neutrinos and photons negligible with respect to the
baryons), the total number of neutrons within this volume will be
given by:

\begin{equation} \label{eq:N:proton:from:preon}
N_{n} = \frac{E_{preon}}{m_p} \approx 1.68 \cdot 10^{18}
\end{equation}

if the mass-energy of the preon is assumed to be $E_{preon} =
p_\gamma(r_0) \simeq \sqrt{s/\pi} \, m_{Pl}/8 = 0.13 \, m_{Pl}$.
From equations (\ref{eq:V:today}, \ref{eq:N:proton:from:preon})
the number-density of nucleons in the universe today can be
estimated as:

\begin{equation}
n_{n} = 1.68 \frac{1}{m^3}
\end{equation}

amounting to roughly $1.7$ nucleons per cubic meter. This is very
close to the number-density of nucleons in the universe derived
from the total matter-density determined by WMAP assuming no
significant dark matter component, $n_n = 1.48 / m^3$ (see section
\ref{sec:measurement}). Therefore a preon-mass in the range
between $0.1$ to $0.2$ Planck masses is quite consistent with the
findings in the observable universe today.

The assumption that the preon eventually decays into nucleons at a
temperature slightly below the nucleon rest-mass, enabled us to
give a - very crude - estimate for the absolute value of the
density contrast $\delta$ at the time of baryogenesis.
Astoundingly this crude estimate fits quite well with the
experimentally determined values of the density contrast today
$\delta_{today} \approx 1$ and at the time of decoupling
$\delta_{dec}\approx 10^{-5}$:

The red-shift $z_b$ where the mean energy of the radiation is
equal to the nucleon rest-mass is given as:

\begin{equation}
z_{b} \approx \frac{m_p/3.37}{T_{CMBR}} \approx 1.19 \cdot 10^{12}
\end{equation}

At this redshift the number-density of the nucleons $n_b$ in the
holostar will be higher than today by a factor of $z_b^4$, i.e.

\begin{equation}
n_b \approx \frac{1.5}{m^3} z_b^4 \approx \frac{3 \cdot
10^{48}}{m^3}
\end{equation}

which corresponds to a matter-density of $\rho_b \approx 5 \cdot
10^{21} kg/m^3$. This value is roughly a factor of thousand higher
than the typical neutron star density and four orders of magnitude
higher than the typical density of stable nuclei.

If the nucleons in our universe originate from the preon at
roughly this redshift, one would expect a density-contrast
$\delta_b$ on the order of the nucleon to preon mass at this time.
With a preon mass $m_{preon} \approx 0.15 m_{Pl}$ we find

\begin{equation}
\delta_b \approx \frac{m_p}{m_{preon}} \approx 5 \cdot 10^{-19}
\end{equation}

As has been shown in section \ref{sec:delta:rho} the density
contrast evolves as a power-law with respect to the redshift. For
$r_i = r_0$ the exponent is given by $\epsilon = \sqrt{19/3}-1
\simeq 1.517$, so that:

\begin{equation}
\delta \propto \frac{1}{z^{1.517}}
\end{equation}

Therefore we can "predict" the density-contrast at any redshift $z
< z_b$ from the density-contrast at the time of baryogenesis. We
find:

\begin{equation}
\delta_{dec} = \delta_b \left(\frac{z_b}{z_{dec}}\right)^{1.517}
\approx 1.2 \cdot 10^{-5}
\end{equation}

with $z_{dec} \approx 1800$ and

\begin{equation}
\delta_{today} = \delta_b \left(\frac{z_b}{1}\right)^{1.517}
\approx 1.05
\end{equation}

So far the maximum angular correlation distance of the microwave
background radiation ($\varphi_{max} \approx 60^\circ$) has been
put in by hand, as determined from the observations. It would be
nice, if this value could be derived by first principles. There
appears to be a way to do this. For this purpose let us consider
the angular spread of zero mass particles in the holostar. The
maximum angular spread for a photon emitted from $r_i$ is given
by:

\begin{equation} \label{eq:max:angle:photon}
\varphi_{max} = \frac{\sqrt{\pi}}{3}
\frac{\Gamma(\frac{1}{3})}{\Gamma(\frac{5}{6})}
\sqrt{\frac{r_i}{r_0}}\simeq 1.4022 \sqrt{\frac{r_i}{r_0}}
\end{equation}

Let us consider photon-pair production (or the production of any
other massless particle in pairs) by Unruh radiation. In order to
produce a photon pair with a mean momentum equal to the local
radiation temperature $T_\gamma$, the Unruh-temperature $T_U$
should be twice the radiation temperature. In section
\ref{sec:Unruh} the following relation between radiation
temperature and Unruh-temperature will be derived:

\begin{equation}
\frac{T_U}{T_\gamma} \simeq \frac{r_0}{r}
\end{equation}

In order for the Unruh temperature to be twice as high as the
radiation temperature we need $r_i = r_0/2$. Note that according
to \cite{Petri/charge} $r_i=r_0/2 \approx r_{Pl}$ is the position
of the membrane of an "elementary" extreme holostar, which appears
to be the smallest possible "size" for a (classical) holostar.
There are some good reasons to identify an "elementary" extreme
holostar (with nearly zero mass) with an elementary particle. As a
fundamental particle cannot be composite, i.e. there cannot be any
particle "within" a fundamental particle, and as $4 \pi (r0/2)^2$
is very close to the smallest area quantum of quantum gravity, it
is very probable that the region $r < r_0/2$ is not accessible to
any particle, probably not even well defined. Therefore Unruh
creation of photon pairs occurs efficiently just at the most
central conceivable location that is available for a "particle"
(or more generally: a discrete geometric entity) within a large
holostar. If we insert the ratio $r_i/r_0 = 1/2$ into equation
(\ref{eq:max:angle:photon}) we get:

\begin{equation} \label{eq:max:angle:photon:2}
\varphi_{max} = \frac{\Gamma(\frac{1}{3})}{\Gamma(\frac{5}{6})}
\sqrt{\frac{\pi}{18}} \simeq 0.9915 = 56.81^\circ
\end{equation}

Voil$\grave{a}$.

Finally it should be noted, that $\varphi_{max} \approx 1$ makes
the "expansion" of particles that move radially outward in the
holostar nearly equal for the radial and the tangential
directions: $\delta l_\perp = r \varphi_{max}$, whereas $\delta
l_r \approx r$.

\subsection{\label{sec:measurement}Estimating cosmological parameters from the radiation temperature}

In this section the total local mass-energy density, the local
Hubble value, the radial coordinate value $r$ and the proper time
$\tau$ in a "holostar universe" will be determined from the
temperature of the microwave background radiation.

The total energy density in the holostar universe can be
determined from the radiation temperature, whenever $r_0^2$ is
known. It is given by equation (\ref{eq:T4/rho}):

\begin{equation}
\rho = \frac{2^5 \pi^3 r_0^2}{\hbar^4} T^4
\end{equation}

There is some significant theoretical evidence for $r_0^2 \approx
4 \sqrt{3/4} \hbar$ at the low energy scale. However $r_0^2 = 4
\hbar$ or a value a few percent higher than $4 \hbar$ might also
be possible (for a somewhat more detailed discussion see
\cite{Petri/charge}). With $r_0^2 = 4 \sqrt{3/4} \hbar$ and
$T_{CMBR} = 2.725 K$ we find:

\begin{equation}
\rho = 2.425 \cdot 10^{-27} \frac{kg}{m^3} = 1.450 \cdot
\frac{m_p}{m^3} =  4.702 \cdot 10^{-124} \rho_{Pl}
\end{equation}

This is almost equal to the total matter-density of the universe
determined by WMAP \cite{WMAP/cosmologicalParameters}:

\begin{equation}
\rho_{WMAP} = 2.462 \cdot 10^{-27} \frac{kg}{m^3}
\end{equation}

From the matter density the radial coordinate position $r$ within
the holostar can be determined.

\begin{equation}
r = \frac{1}{\sqrt{8 \pi \rho}} = 9.199 \cdot 10^{60} r_{Pl} =
1.575 \cdot 10^{10} ly
\end{equation}

This value is quite close the radius of the observable universe.

The local Hubble-constant can be determined from the matter
density via equation (\ref{eq:Hubble}):

\begin{equation}
H = \sqrt{8 \pi \rho \frac{r_0}{r_i}}
\end{equation}

With $r_i/r_0 = 1$ and the matter density determined beforehand we
find:

\begin{equation} \label{eq:H:r0qg}
H = 2.021 \cdot 10^{-18} \, [\frac{1}{s}] = 62.36 \,
[\frac{km/s}{Mpc}]
\end{equation}

The Hubble-constant comes out quite close to the value that is
used in the concordance model by the WMAP-group
\cite{WMAP/cosmologicalParameters} with $H = 71 \, km/s \, / Mpc$.
This is an encouraging result. Note, however, that the value of
the Hubble constant is model-dependent. It is possible to relate
the Hubble value of the various standard cosmological models to
the Hubble value of the holostar solution via the mass-density. In
the standard cosmological models we find:

\begin{equation}
\rho_m = \Omega_m \rho_c = \Omega_m \frac{3 H_s^2}{8 \pi}
\end{equation}

For the holostar the mass-density is given by

\begin{equation}
\rho_m = \frac{r_i}{r_0} \frac{H_h^2}{8 \pi}
\end{equation}

Setting the mass densities equal, the local Hubble-values can be
related:

\begin{equation} \label{eq:Hubble:standard:holo}
H_h^2 = \Omega_m \frac{3 r_0}{r_i} H_s^2
\end{equation}

Both values are equal, if $\Omega_m = r_i/(3 r_0)$. This is almost
the case for $\Omega_m \approx 0.26 \approx 1/4$ according to WMAP
and $r_i/(3 r_0) \approx 1/3$ as determined in section
\ref{sec:delta:rho}. This result is quite robust. If we determine
$H_h$ from $H_s$ via equation (\ref{eq:Hubble:standard:holo}) for
various combinations of $\Omega_m$ and $H_s$ that have been used
in the past\footnote{Some years ago $\Omega_m \approx 0.3$ and
$H_s \approx 65 km/s / Mpc$ was a common estimate.} we get the
result of equation (\ref{eq:H:r0qg}) with an error on the order of
a few percent.

From the local Hubble-value determined in (\ref{eq:H:r0qg}) the
proper time $\tau$ can be derived:

\begin{equation}
\tau = \frac{1}{H} = 9.180 \cdot 10^{60} t_{Pl} = 1.57 \cdot
10^{10} y
\end{equation}

This is somewhat larger than the result recently announced by the
WMAP group, $t = 1.37 \cdot 10^{10} y$.

If we set $r_i/r_0 = 3/4$ we find an almost perfect agreement of
the holostar's local Hubble value $H_h = 71.85$ (km/s)/Mpc and the
current proper time $\tau = 13.6 \, Gy$ with respect to the values
determined by WMAP \cite{WMAP/cosmologicalParameters} ($H = 71$
(km/s)/Mpc and $t = 13.7 \, Gy$).

The holostar solution is quite compatible with the recent findings
concerning the large scale structure and dynamics of the universe.
The recent WMAP results are reproduced best by setting $r_i/r_0 =
3/4$. From the evolution of the density perturbations (see section
\ref{sec:delta:rho}) one would rather expect $r_i \simeq r_0$. It
is quite clear that $r_i/r_0$ cannot lie very much outside the
range $3/4 < r_i/r_0 < 1$, which corresponds to an age of the
universe in the range between $13-16 \, Gy$.

Note that the above age comes close to the ages of the oldest
globular clusters. If $r_i/r_0$ is chosen larger than unity, the
proper time $\tau$ in the holostar universe will be larger. For
$r_i/r_0 \approx 2$ the holostar universe can easily accommodate
even the old estimates for the ages of the globular clusters,
which not too far ago have been thought to lie between $13$ to
$19$ billion years. The problem with such an assignment is, that
the Hubble value comes out far too low. For $r_i = 2 r_0$ we find
$H \approx 44$ (km/s)/Mpc, which appears incompatible with today's
experimental measurements, even if the large systematic errors in
calibrating the cosmological distance scale are taken into
account.

From a theoretical point of view (see the discussion in section
\ref{sec:Unruh}, where the CMBR-temperature is interpreted as
Unruh-temperature) I find $r_i = r_0$ the most preferable choice.
However, this choice is based on equation (\ref{eq:TU/T:2}), which
assumes $(T_i/m) \, (\sigma/\hbar) = 1$. This relation is based on
classical reasoning (at the Planck-energy scale) so that one
should expect some moderate adjustment due the quantum nature of
space-time at this scale. Therefore it is too early to make a
definite numerical prediction for the ratio $r_i/r_0$. If
$r_i/r_0$ can be pinned down theoretically we are in the very much
desirable position to make a precise prediction for the Hubble
value, which should be an order of magnitude better than today's
measurement capabilities. Therefore any significant advances in
the understanding of the motion of particles in the holostar
universe could be of high practical value for the development of a
"high precision cosmology" in the near future.

\subsection{Local matter distribution and self-similarity}

For a very large holostar of the size of the universe, its outer
regions will consist of low-density matter, which should be quite
comparable in density and/or distribution to the matter we find in
our universe. Such matter can be expected form local hierarchical
sub-structures comparable to those found in our universe, as long
as the scale of the sub-structures remains compatible with the
overall mass-energy distribution $\rho \propto 1/r^2$ of the
holostar.\footnote{The largest scale for significant local
deviations from the mass-density is given by the proper
circumference at radial position r, i.e. roughly the Hubble
length.}

On not too large a scale some local regions might collapse,
leaving voids, others might expand, giving rise to filamental
structures. However, any local redistribution of mass-energy has
to conserve the total gravitational mass of the holostar, its
total angular momentum and - quite likely - its entropy, i.e. its
total number of particles. These exterior constraints require that
regions of high matter-density must be accompanied by voids.

Furthermore one can expect, that the local distribution of matter
within a holostar will exhibit some sort of self-similarity, i.e.
the partly collapsed regions should follow a $1/r^2$-law for the
local mass-density. This expected behavior is quite in agreement
with the observations concerning the mass-distribution in our
universe: The flat rotation curves of galaxies, as well as the
velocity dispersion in galaxies and clusters of galaxies hint
strongly, that the matter distribution of the local matter in
galaxies and clusters follows an $1/r^2$-law. As far as I know,
there has been no truly convincing explanation for this apparently
universal scaling law, so far.

\subsection{\label{sec:frames}Some remarks about the frames of the asymptotic and the co-moving observer}

Most of the discussion about the properties of the holostar has
been in the frame of the asymptotic observer, who is at rest in
the ($t, r, \theta, \varphi$)-coordinate system.

If the holostar is to serve as a model for an expanding universe,
one must interpret the phenomena from the frame of the co-moving
observer, who moves nearly geodesically on an almost radial
trajectory through the low density outer regions of the holostar.

The frames of the co-moving and the asymptotic observer are
related by a Lorentz boost in the radial direction. Due to the
small tidal acceleration in the holostar's outer regions, the
extension of the local Lorentz frames can be fairly large. The
proper acceleration experienced by the co-moving observer, if
there is any, should be very low in the regions which have a
density comparable to the density that is observed in the universe
at the present time.

The holostar's interior space-time is boost-invariant in the
radial direction, i.e. the stress-energy tensor is unaffected by a
radial boost. The co-moving observer moves nearly radially for $r
\gg r_i$. His radial $\gamma$-factor grows as the square root of
his radial coordinate value, whereas his tangential velocity goes
rapidly to zero with $1/r^3$. The radially boosted co-moving
observer therefore will see exactly the same total stress-energy
tensor, and thus the same total energy density as the observer at
rest in the ($t, r, \theta, \varphi$)-coordinate system. The above
statement, however, only refers to the total energy density. It is
not a priori clear if the individual contributions to the
mass-energy, i.e. massive particles and photons, have the same
$r$-dependence as the total energy density.

Let us first consider the case of massive particles. The observer
at rest in the coordinate system measures an energy density $\rho$
of the massive particles, which is proportional to $1/r^2$. A
factor of $1/r^{5/2}$ comes from the number density given in
equation (\ref{eq:n:massive}), a factor of $r^{1/2}$ from the
special relativistic $\gamma$-factor that must be applied to the
rest mass of the particles according to equation (\ref{eq:E(r):massive}).

From a naive perspective (neglecting the effects of the pressure)
it appears as if the co-moving observer and the observer at rest
in the ($t, r, \theta, \varphi$)-coordinate system disagree on how
the energy and number densities of massive particles develop with
$r$. Because of the highly relativistic motion of the co-moving
observer, the observer at rest in the ($t, r, \theta,
\varphi$)-coordinate system will find that the proper volume of
any observer co-moving with the massive particles is
Lorentz-contracted in the radial direction. Therefore the
co-moving observer will measure a larger proper volume, enlarged
by the radial $\gamma$-factor, which is proportional to
$\sqrt{r}$. If we denote the volume in the frame of the co-moving
observer by an overline, we find $\overline{V} \propto \gamma
r^{5/2} \propto r^3$. As long as the massive particles aren't
created or destroyed, the number-density of the massive particles
in the co-moving frame therefore must scale as $1/r^3$.
Furthermore, for the co-moving observer the neighboring massive
particles are at rest to a very good approximation.\footnote{The
tidal acceleration is negligible; the extension of the local
Lorentz frame of the observer is nearly equal to the local Hubble
length, i.e. of the order the current "radius" of the universe,
$r$.} Therefore the mass-energy density in the co-moving frame
should be nothing else than the (presumably) constant rest-mass of
the particles multiplied by their number-density. From this it
follows, that the mass-energy density of the massive particles
should scale as $1/r^3$ as well.

This naive conclusion, however, is false. It does not take into
account the energy change in the co-moving volume due to the
radial pressure. Any radial expansion in the co-moving frame
affects the internal energy. Due to Lorentz-elongation in the
co-moving volume the radial thickness $\overline{l_r}$ of the
expanding shell develops proportional to $r$ in the co-moving
frame. From this the internal energy-change $\delta E$ in the
shell can be calculated for a small radial displacement $\delta
r$. We find:

\begin{equation} \label{eq:dE:massive:co}
\overline{\delta E} = - \overline{P_r} A \, \overline{\delta l_r}
\propto \frac{\delta r}{2}
\end{equation}

with $\overline{P_r} = P_r = -1 / (8 \pi r^2)$ and $\overline{A} =
4 \pi r^2$. Note that the radial pressure in the co-moving frame
is exactly equal to the radial pressure in the coordinate frame
due to the boost-invariance of the stress-energy tensor in the
radial direction.

From equation (\ref{eq:dE:massive:co}) the radial dependence of
the internal total energy in the co-moving frame follows:

$$\overline{E} \propto r$$

We have already seen that the co-moving volume develops as
$\overline{V} \propto r^3$, so that the energy density of the
massive particles in the co-moving frame, taking the pressure into
account, develops exactly as in the coordinate frame, i.e.
$\overline{\rho} \propto \overline{E} / \overline{V} \propto
1/r^2$.

It is quite obvious that such a dependence is not compatible with
the assumption that both the rest-mass and the number of the
massive particles in the co-moving volume remain constant. Either
the rest-mass of the massive particles must increase during the
expansion or new particles (massive or radiation) have to be
created by the negative pressure. There is no indication
whatsoever that the rest mass of the nucleon or the electron have
changed considerably during the evolution of the universe, at
least for temperatures at and below the time of nucleosynthesis.
Therefore we cannot avoid the conclusion that the negative
pressure has the effect to create new particles. Particle creation
in an expanding universe is not new. It is one of the basic
assumptions of the venerable steady-state model of the universe.
Furthermore particle production via expansion against a negative
pressure is a well known phenomenon from the inflational equation
of state. There are differences. Whereas the isotropic negative
pressure of the inflational phase keeps the energy-density in the
expanding universe constant during the expansion, the
energy-density in the holostar develops as $1/r^2$, because the
negative pressure only acts in one of three spatial directions. As
a result the particle creation rate in the holostar is quite low
at the present time: Roughly one neutron per cubic kilometer every
10 years is required.

The above analysis indicates that the energy density of the
massive particles in the co-moving frame should be proportional to
$1/r^2$, exactly as in the coordinate frame. A radial boost not
only leaves the total energy density unaffected, but also the
respective energy densities of the different particle
species.\footnote{This argument is not water tight. If the
particles produced by the negative pressure are different from the
(massive) particle species that catalyze their production, there
might be a redistribution of mass-energy between the different
species during the expansion.} With the reasonable assumption that
the rest mass of the massive particles is constant (at least for
temperatures well below the rest mass of the particles) the number
densities must also evolve according to $1/r^2$.

Let us now discuss the number- and energy densities of the zero
rest-mass particles in the two frames. This problem is closely
related to the question whether the co-moving observer experiences
a different radiation temperature than the observer at rest in the
($t, r, \theta, \varphi$)-coordinate system. The standard argument
for the red-shift of radiation in an expanding Robertson-Walker
universe is, that the wavelength of the radiation is stretched
proportional to the expansion.\footnote{A more sophisticated
derivation is based on the existence of a Killing vector field in
a Robertson-Walker universe (see for example
\cite[p.101-104]{Wald/GR}). The derivation makes use of the well
known fact, that the scalar product of the photon wave-vector with
the Killing-vector is constant for geodesic motion of photons.
This argument, however, requires the geodesic equations of motion,
and therefore is only water-tight for a dust universe without
significant pressure.} This gives the known $T \propto
1/r$-dependence for the radiation temperature in the standard
cosmological models. If this argument is applied to the holostar,
one would expect that the $T \propto 1 / \sqrt{r}$-law (in the
frame of the observer at rest) is transformed to a $\overline{T}
\propto 1/r$-law in the frame of the co-moving observer, due to
Lorentz-contraction (or rather elongation) of the photon
wavelength.

But a $\overline{T} \propto 1/r$-law wouldn't be consistent for a
small holostar, where the energy density is expected to be
dominated by radiation in true thermal equilibrium. The energy
density of thermalized radiation is proportional to $T^4$. If the
only contributor to the total energy density is radiation, the
radiation energy density will transform exactly as the total
energy density in a radial boost. However, the total energy
density is radially boost-invariant. Therefore in the radiation
dominated era the energy density of the radiation in any radially
boosted frame should scale as $1/r^2$. This however implies the
$\overline{T} \propto 1/\sqrt{r}$-dependence, at least if the
radiation is in thermal equilibrium in the boosted
frame.\footnote{If the holostar were truly static in the high
temperature regime, i.e. geodesic and pressure induced
acceleration cancel exactly, there is no problem. There would be
no directed motion and therefore $\overline{T} = T \propto
1/\sqrt{r}$ trivially.}

In a radially boost-invariant space-time one would expect on more
general grounds, that it is - in principle - impossible to
determine the radial velocity of the motion, at least by direct
measurements performed by the observer co-moving with the
matter-flow. A $\overline{T} \propto 1/r$-law would imply a
radiation temperature incompatible with the energy-density of the
radiation, which would allow the co-moving observer to determine
his radial velocity. This can be seen as follows:

Naively one would assume, that the number-density of photons in
the co-moving frame is given by $\overline{n_\gamma} = n_\gamma /
\gamma \propto 1/r^2$, due to Lorentz elongation of the co-moving
volume with respect to the observer at rest. However, at a closer
look one has to take into account the measurement process. Photons
cannot be counted just by putting them in a box and then taking
out each individual particle. As photons always move with the
local speed of light, such a procedure, which would be good for
massive particles, doesn't work. The right way to count photons is
to place a small ideal (spherical) absorber somewhere in the
space-time, count all the hits per proper time interval and relate
the obtained number to the volume (or surface area) of the
absorber. But this procedure requires, that the time-delay due to
the highly relativistic motion of the co-moving observer has to be
taken into account. The co-moving observer will count many more
photons in a given standard interval of his proper time, than the
observer at rest would count in the same interval. The
time-dilation introduces another $\gamma$-factor that exactly
cancels the $\gamma$-factor from the Lorentz contraction of the
proper volume. Therefore we arrive at the remarkable result, that
the number-density of photons should be the same for both
observers, i.e. independent of a radial boost.\footnote{This
result could also have been obtained by calculating the
pressure-induced energy-change in the co-moving frame. Quite
interestingly this change is zero for photons, because the proper
radial extension of a geodesically moving shell of photons remains
constant in the co-moving frame. Therefore the energy-density of
the photons in the co-moving frame evolves inverse proportional to
the proper surface area of the shell, i.e. $\overline{\rho_\gamma}
\propto 1/r^2$. Assuming that the photon number in the shell
remains constant in the co-moving frame (which is equivalent to
assuming a thermal spectrum) we then find $\overline{n_\gamma}
\propto 1/r^{3/2}$.}

If we had $\overline{T} \propto 1/r$ in the co-moving frame, we
would get $\overline{\rho_\gamma} \propto \overline{n_\gamma}
\overline{T} \propto 1/r^{5/2}$, which implies
$\overline{\rho_\gamma} \propto \overline{T}^{5/2}$ in the
co-moving frame. However, in thermal equilibrium
$\overline{\rho_\gamma} \propto \overline{T}^4$. Even if the
argument given above for the number-density of photons were
incorrect, i.e. only the volume were Lorentz contracted and time
dilation would play no role, we would have $\overline{\rho_\gamma}
\propto \overline{n_\gamma} \overline{T} \propto 1/r^3$, implying
$\overline{\rho_\gamma} \propto \overline{T}^3$ in the co-moving
frame, which wouldn't work either.

Therefore it seems reasonable to postulate, that a radial boost
from the $(t, r, \theta, \varphi)$-coordinate system to the system
of the co-moving observer should not only leave the total energy
density unaffected, but also other physically important
characteristics such as the thermodynamic state of the system,
i.e. whether or not the radiation can be characterized as thermal.
This then implies that the thermodynamic relation between
energy-density and temperature for an ultra-relativistic gas
should be valid in the radially boosted frame of the co-moving
observer, i.e. $\overline{\rho} \propto \overline{T}^4$, which
requires the $\overline{T} \propto 1/\sqrt{r}$ law.

The reader may object, that any radial boost will produce a large
anisotropy in the radiation temperature, as measured in the
boosted frame. I.e. the radiation will be blue-shifted for photons
travelling opposite to the motion of the co-moving observer and
red-shifted for photons travelling in the same direction, due to
"normal" Doppler shift. However, this again doesn't take into
account the subtle effects of the negative radial pressure. The
radially boosted observer will find that the volume that lies in
front of him is Lorentz-contracted in the radial direction. The
photons coming from the front side therefore come from a radially
"squeezed" volume. But any volume contraction in the radial
direction will reduce the energy of the photons because of the
negative radial pressure. The blue-shift of the photons due to the
(kinematical) Doppler-shift will be exactly compensated by the
red-shift originating from the pressure-induced Lorentz
contraction. A similar effect occurs for the photons coming from
the rear, i.e. moving in the same direction as the observer.

Although perhaps some new insights are required in order to
resolve the problem of relating the observations in the different
frames in a satisfactory fashion, nature appears to have taken a
definite point of view: If we live in a large holostar, we clearly
are in the position of the co-moving observer. Except for a small
dipole anisotropy, which can be explained by the small relative
motion of our local group with respect to the isotropic
Hubble-flow, the CMBR is isotropic. Furthermore, in Planck-units
the CMBR-temperature is $T_{CMBR} \simeq 2 \cdot 10^{-32} T_{Pl}$,
whereas the radius of the observable universe is roughly $r
\simeq 9 \cdot 10^{60} r_{Pl}$ and the mass-density $\rho
\simeq 5 \cdot 10^{-124} \rho_{Pl}$. These simple figures strongly
suggest, that the $T \propto 1/\sqrt{r}$ and $\rho \propto T^4$
laws are realized in our universe in the system of the co-moving
observer.

\subsection{On the baryon to photon ratio and nucleosynthesis}

The discussion of the previous section has paved the way to
address the problem of nucleosynthesis in the holostar universe. A
quite remarkable by-product of the discussion in this section is a
surprisingly simple explanation for the baryon to photon ratio in
the universe.

We have seen in the previous section that the number density of
massive particles $\overline{n_m}$ and photons
$\overline{n_\gamma}$ in the co-moving frame develop as:

\begin{equation}
\overline{n_m} \propto \frac{1}{r^2} \propto T^4
\end{equation}

and

\begin{equation}
\overline{n_\gamma} \propto \frac{1}{r^{\frac{3}{2}}} \propto T^3
\end{equation}

As consequence the ratio of the energy-densities per proper volume
of massive particles to photons remains constant in the co-moving
frame in the holostar universe\footnote{Note that $\rho_\gamma
\propto \rho_m \propto 1/r^2$ has already been established in the
coordinate frame.}, i.e:

\begin{equation} \label{eq:rho:T}
\overline{\rho_m} \propto \overline{\rho_\gamma} \propto
\frac{1}{r^2} \propto T^4
\end{equation}

Comparing the energy density of electrons $\overline{\rho_e}$ and
photons $\overline{\rho_\gamma}$ at the present time, we find that
they are almost equal. In fact, assuming the chemical potential of
the photons to be zero, the respective energy densities turn out
as

\begin{equation}
\overline{\rho_\gamma} = 8.99 \cdot 10^{-128} \, \rho_{Pl}
\end{equation}

and

\begin{equation}
\overline{\rho_{e}} = 2.23 \cdot 10^{-127} \, \rho_{Pl} = 2.52
\overline{\rho_\gamma}
\end{equation}

if we assume an electrically uncharged universe, a proton to
nucleon ratio of $7/8$ and a universe consisting predominantly out
of nucleons (no significant dark matter component). $\rho_{Pl}$ is
the Planck-density, $m_{Pl}/r_{Pl}^3$.

In the context of the standard cosmological model this fact
appears as a very lucky coincidence, which happens just at the
current age of the universe and won't last long: The energy
density of radiation and matter evolve differently in the standard
cosmological model. Basically $\rho_m \propto T^3$, whereas
$\rho_\gamma \propto T^4$, so that the energy density of the
radiation falls off with $T$ compared to the energy density of the
massive particles.\footnote{The standard cosmological model
assumes that the number ratio of baryons to photons $\eta$ remains
constant in the expanding universe. With the recent WMAP data this
ratio is now estimated (at the time of nucleosynthesis) as $\eta
\simeq 6.5 \cdot 10^{-10}$. The postulate $\eta = const$ is quite
different to the evolution of the different particle species in
the holostar, where rather the energy densities, and not the
number-densities of the fundamental particle species remain
constant during the evolution.}

The particular value of the baryon to photon ratio $\eta$ is a
free parameter in the standard cosmological model. The most recent
experimental determination of $\eta$ via primordial
nucleosynthesis is given by \cite{WMAP/cosmologicalParameters} as
$\eta \approx 6.5 \cdot 10^{-10}$ (at the time of
nucleosynthesis). Unfortunately we still lack an established
theoretical framework by which this value could be calculated or
even roughly estimated from first principles. In the holostar the
value of $\eta$ is linked to the nearly constant energy densities
of the different fundamental particle species. Whenever we know
what the ratio of the energy densities should be, we can estimate
the present value of $\eta$.

Can the ratio between the energy densities of photons and
electrons in the holostar universe be predicted by first
principles? In order to answer this question let us turn our
clocks backward until the radiation temperature in the holostar
registers somewhere above the electron rest mass, but well below
the mass of the muon or pi-meson (for example $T \approx 1-10 \,
MeV$). The rather fast electromagnetic reactions at the high
temperatures and densities ensure that the energy is distributed
nearly equally among the relativistic degrees of freedom of
electrons/positrons and photons, respectively.\footnote{I never
found the basic assumption of the standard cosmological model
utterly convincing, according to which the photon to baryon ratio
should be frozen at a constant value that is postulated to have
remained nearly constant from the time of baryogenesis ($T \approx
1 GeV$) to the time of nucleosynthesis ($T \approx 0.1 \, MeV$)
and forever thereon. There is no problem with such an assumption
after the time where photons and matter became chemically
decoupled, i.e. after nucleosynthesis ended. It is not difficult
to verify that in a homogeneously expanding universe with
negligible pressure and negligible particle-changing interactions
(i.e. no particle creation/destruction) the particle numbers of
the different stable species in any volume co-moving with the
expansion is conserved. Note, however, that this feature stands
and falls with the assumption of a homogenous universe with a
universal cosmological time, i.e. a universe which looks exactly
the same at any spatial position at any fixed value of the
cosmological time. In such a universe the particles (massive or
mass-less alike) have no other way to "go" than to move with the
geodesically expanding volume. In the holostar universe the
situation is different, as the particles move through a static
space-time and can "go anywhere they wan't", as long as they obey
the local equations of motion (which are not equivalent with the
geodesic equations of motion!). But even in a homogeneously
expanding universe it requires quite a bit of "fine-tuning" with
respect to the physics if the photon to electron ratio should be
vastly different from unity at or slightly below the temperature
where photons and electrons chemically decouple, i.e. at $T <
m_e/3 \approx 0.15 MeV$. Taking into account the strength of the
electro-weak and the strong interactions in the energy range from
$0.2-1000 \, MeV$ and taking into account the expansion rate of
the universe during this period, the mutual interactions are fast
enough to maintain a thermal spectrum in the whole range $T
\approx 0.2 \ldots 1000 MeV$ in the standard cosmological model.
But in near thermal equilibrium the energy density of the
ultra-relativistic electrons (and positrons) must be comparable to
that of the photons. This conclusion remains true even if there is
an appreciable net baryon lepton asymmetry and is independent from
the ratio of photons to baryons that might have been produced by
the mutual annihilation of baryon/anti-baryon pairs at the time of
baryogenesis: In the cosmological time period between baryogenesis
and nucleosynthesis, when the temperature falls from $1 GeV$ to
$0.2 MeV$, the Hubble-expansion is not fast enough to shut off the
electro-magnetic interactions. Therefore, when the temperature has
reached the electron mass threshold, the energy density of photons
and electrons will have become comparable to each other, equal
within one order of magnitude if the possibly non-zero chemical
potentials of the particles are taken into account. In fact, the
same argument applies to baryogenesis. It is not understandable
that an asymmetry on the order of $10^9$ photons to baryons could
have arisen in the first place: At the time where the annihilation
of baryons/antibaryons starts to set in, the energy densities of
baryons and photons are nearly equal. Even if there were a ratio
of $10^9+1$ baryons to $10^9$ anti-baryons at $T \approx 1
GeV/2.7$, the annihilation of baryon/anti-baryon pairs wouldn't
happen instantaneously. It rather has the effect that energy is
transferred to the photons (and the other relativistic particles),
keeping the temperature constant. Yet the universe doesn't halt
its expansion. Rather the continued expansion against the
(positive) radiation pressure will gradually reduce the energy
density of all particle species at a nearly constant temperature.
There is no conservation law for photons which would forbid
photons to be destroyed in favor of maintaining a constant
energy/temperature. Furthermore, at the nearly constant
temperature of the phase transition the photons are still in close
thermal contact with the baryons, so that the photon and baryon
energy densities remain nearly equal. As the temperature remains
constant (with the mean momentum of the photons equalling the
baryon rest mass), the number-densities of photons and baryons
(and of the yet highly relativistic electrons, muons and
neutrinos, Pi-mesons etc.) will remain nearly equal as well. The
process stops, when a baryon density slightly above the final
density of the relic baryons is reached, for example two baryons
to one anti-baryon. Only then will the temperature drop below the
threshold for the creation of baryon/anti-baryon pairs,
effectively shutting off the reactions that maintained thermal
equilibrium up to this point. A similar sequence of events happens
at the electron-mass threshold, so that it is quite inconceivable
that we should encounter the vast number of $10^9$ photons per
relic electron (or baryon) just below the electron mass threshold.
Only after photons and electrons have decoupled chemically, i.e.
well below the electron mass threshold, the photon to electron
ratio in an expanding universe develops independently from a
thermal distribution.} When the temperature falls below the
threshold defined by the electron-mass, the last (few!) remaining
positrons are quickly dispatched, so that the number of photons
should be nearly equal to the number of the relic electrons. The
same will be true for the respective energy densities, so that as
a very rough estimate we can assume, that the energy densities of
photons and electrons are equal at decoupling.\footnote{In the
simple analysis here I neglect the difference in energy-densities
of photons and electrons due to the fact, that photons are bosons
whereas electrons are fermions. Likewise the non-negligible
effects of any non-zero chemical potentials of photons and
electrons are neglected.}

At temperatures below nucleosynthesis, i.e. $T < 0.1 MeV$, the
energy densities of photons and electrons in the holostar evolve
nearly independently. Even when the reactions that have maintained
thermal equilibrium before have ceased, the energy-densities of
the two particles species will be maintained at a nearly constant
ratio, whose exact value is determined by the physics at the
time-period where electrons and photons have decoupled
chemically.\footnote{It is quite remarkable, that the
energy-densities of the different fundamental particle species in
the holostar evolve with a constant ratio throughout the whole
holostar: At high temperature the nearly equal energy densities
are maintained by the reactions establishing thermal equilibrium.
At low temperatures the geodesic motion in combination with the
negative radial pressure ensures, that the massive and massless
particles maintain a constant energy ratio during their motion.}
An exact determination of this ratio is beyond the scope of this
paper. Quite likely chemical potentials will play an important
role (see \cite{Petri/thermo}). However, it is quite encouraging
that the very rough estimate $\rho_e \approx \rho_\gamma$ at the
time of chemical decoupling is within a factor of three of the
experimental value $\rho_e \approx 2.5 \rho_\gamma$ determined at
the present time.

The discussion above allows us to make a rough estimate for the
value of $\eta$ at the present time. Assuming that photons and
electrons decoupled chemically at $T = m_e/3$, assuming equal
energy-densities of electrons and photons at this time we find:

\begin{equation} \label{eq:eta}
\eta_{today} \approx \frac{3 T_{CMBR}}{m_e} = 1.38 \cdot 10^{-9}
\end{equation}

Remarkably, this very rough estimate is only a factor of 2 higher
than the WMAP-result $\eta_{WMAP} \approx 6.5 \cdot 10^{-10}$.

We can also compare the value in equation (\ref{eq:eta}) to the
maximum possible value of the baryon to photon ratio in the
universe today. Under the assumption that there is no significant
dark matter-component $\eta_{max}$ can be estimated from the
cosmological number densities of photons and
electrons.\footnote{The photon number density is determined by the
Planck formula, assuming the chemical potential of the photons to
be zero. The maximum value for the electron number density is
determined from the total matter-density of the universe according
to WMAP, assuming no significant dark matter component, a proton
to nucleon ratio of 7/8 and an uncharged universe with no
significant antimatter.} We find $\eta_{max} \approx 3.14 \cdot
10^{-9}$, which is a factor of two higher than the estimate of
equation (\ref{eq:eta}).

We are now in the position to discuss nucleosynthesis in the
holostar universe. The primordial synthesis of the light elements
proceeds somewhat differently as in the standard cosmological
model. The two key parameters governing the respective reaction
rates, the number density of the nucleons (baryons) $n_b$ at the
temperature of nucleosynthesis and the (Hubble) expansion rate,
turn out to be significantly different in both models.

In the standard cosmological model the number-density of the
nucleons depends on the cube of the temperature, i.e. $n_b \propto
1/R^3 \propto T^3$, whereas according to the discussion above the
co-moving observer in the holostar finds that the number density
of non-relativistic particles scales as the fourth power of the
temperature, $\overline{n_b} \propto 1/r^2 \propto T^4$. Therefore
the number-density of the nucleons at the temperature of
nucleosynthesis will be roughly nine orders of magnitude higher
than in the standard model.

This won't have a large effect on the ratio of primordial Helium-4
to Hydrogen. The numerical value of the He/H ratio is mainly due
to the n/p ratio in thermal equilibrium at $T \approx 0.8 \, MeV$,
i.e. the temperature where the weak reactions are shut
off.\footnote{Neutron decay during the time where the temperature
drops to roughly $0.1 MeV$, i.e. the temperature where deuterons
are produced in significant numbers (and quickly end up in the
more stable $He^4$), also plays a non-negligible role in the
standard cosmological model. Note that the shut-off temperature of
the weak interactions, which is $0.8 MeV$ in the standard model,
depends on the Hubble rate, and thus will be (slightly) different
in the holostar.} However, the higher number-density will have a
considerable effect on the amount of Deuterium, Helium-3 and
Lithium-7 produced in the first few seconds of the universe.

The relative abundance of these elements depends quite sensitively
on the nuclear reaction rates, which are proportional to the
number-densities of the reacting species. In order to accurately
calculate the relative number densities of all the other elements,
excluding H and He-4, one has to consider the dynamic process in
which the nuclear reactions compete against the cosmic expansion.
Eventually the nuclear reaction rates will fall below the
expansion rate, ending the period of nucleosynthesis. Therefore
the second key parameter in primordial nucleosynthesis, besides
the nucleon number-densities, is the cosmic expansion rate, which
is proportional to the Hubble-value.

In the matter dominated era the standard model predicts $H \propto
1/t \propto 1/R^{3/2} \propto T^{3/2}$, whereas in the radiation
dominated era $H \propto T^2$. In the radiation dominated era the
dependence of the expansion rate on the temperature is equal in
both models. Taking the different dependencies in the
matter-dominated era into account, one can relate the
Hubble-constant $H_{h}$ in the holostar universe and the
Hubble-constant $H_s$ in the standard model at the time of
nucleosynthesis. We find:

\begin{equation}
H_h \approx \sqrt{\frac{T_{eq}}{T_{CMBR}}} H_s = \sqrt{z_{eq}} H_s \simeq 60 H_s
\end{equation}

where $z_{eq} \approx 3450$ is the red-shift at which the standard
model becomes radiation dominated, according to the recent WMAP
findings. The above result is quite consistent with the fact, that
the "age" of the universe at $T = 0.1 \, MeV$  is roughly $\tau
\approx 7 s$ in the holostar, whereas in the standard cosmological
model we find $t \approx 200 s$ at the same temperature.

It should be evident, that due to the differences in the two key
parameters nucleosynthesis in the holostar universe will proceed
quite differently from the standard model. Without actually
solving the rate equations it is difficult to make quantitative
predictions. In general the higher number density of nucleons in
the holostar will lead to a more effective conversion of Deuterons
to the stable Helium-4 nucleus, reducing the relative abundance of
the few Deuterium nuclei that survive their conversion to He-4. On
the other hand, the significantly larger expansion rate enhances
the chance, that the less stable nuclei such as Deuterium and He-3
will survive through the much shorter time period of primordial
nucleosynthesis. The nuclear reactions are shut-off faster,
providing some amount of "compensation" for the faster reaction
rates due to the higher number densities. Yet it would be quite a
coincidence, if nucleosynthesis in the holostar would lead to the
same abundances of Deuterium, Helium-3 and Lithium-7 as in the
standard cosmological model. Whether the holostar model of the
universe can explain the experimentally determined abundances of
the light elements in a satisfactory fashion, therefore must be
considered as an open question.

\subsection{\label{sec:Unruh}The Unruh temperature}

A massive particle at rest in the ($r, \theta, \varphi, t$)
coordinate system is subject to a geodesic acceleration given by
equation (\ref{eq:g}). With this acceleration the following
Unruh-temperature can be associated

\begin{equation} \label{eq:TUnruh}
T_{U} = \frac{g \hbar}{2 \pi} = \frac{\hbar}{4 \pi r}
\sqrt{\frac{r_0}{r}} =  T_\gamma \frac{r_0}{r}
\end{equation}

where $T_\gamma$ is the local radiation temperature given by
equation (\ref{eq:Tlocal}).

At $r \approx r_0$ the Unruh temperature is comparable to the
Planck-temperature. Therefore particles with masses up to the
Planck-mass can be created out of vacuum in the vicinity of a
roughly Planck-size region close to the center of the holostar.

If we multiply the Unruh-temperature with the local radiation
temperature of equation (\ref{eq:Tlocal}) we find the following
interesting relation, which doesn't include the - unknown -
parameter $r_0$:

\begin{equation}
T_{U} T_\gamma = \frac{\hbar^2}{16 \pi^2 r^2} = \frac{\hbar^2}{2
\pi} \rho
\end{equation}

The Unruh temperature, as derived in equation (\ref{eq:TUnruh}),
only applies to an observer at rest in the holostar. It cannot be
experienced by an observer moving geodesically, because such an
observer is by definition unaccelerated in his own frame of
reference.

However, pure geodesic motion is not possible within a holostar,
due to the pressure. Even in the low-density regions of a holostar
where the motion is predominantly geodesic, the pressure will
provide a small deceleration, which can be estimated by boosting
the pressure induced acceleration, $a_P$, given by

\begin{equation}
a_P = \frac{\sigma}{m} P = \frac{\sigma}{m} \frac{1}{8 \pi r^2}
\end{equation}

to the frame of the co-moving observer. If we denote the frame of
the co-moving observer with an overline, we get $\overline{a_P} =
\gamma^3 a_P$, because the deceleration $a_P$ is parallel to the
(radial) boost. $\gamma$ is given by equation (\ref{eq:gamma}).
Therefore the acceleration (or rather deceleration) due to the
pressure in the frame of the co-moving observer is given by:

\begin{equation}
\overline{a_P} = \frac{\sigma}{m} \frac{1}{8 \pi r^2}
\left(\frac{r}{r_i}\right)^{\frac{3}{2}} = \frac{\sigma}{2 m}
\frac{1}{4 \pi \sqrt{r r_i^3}}
\end{equation}

This acceleration can be associated with an Unruh temperature,
which the co-moving observer should be able to measure in
principle. It is quite remarkable, that the Unruh-temperature in
the frame of the co-moving observer turns out to be exactly
proportional to the local radiation temperature $T_\gamma$:

\begin{equation}
\overline{T_U}(r) = \frac{\hbar \overline{a_P}}{2 \pi} =
\frac{\sigma}{4 \pi m r_i} \sqrt{\frac{r_0}{r_i}} T_\gamma(r)
\end{equation}

For $r_i \gg r_0$ the Unruh-temperature will be much lower than
the radiation temperature. Whenever the particles move
geodesically, the ratio of their Unruh-temperature to the local
radiation temperature is nearly constant and given by:

\begin{equation} \label{eq:TU/T}
\frac{\overline{T_U}}{T_\gamma} = \frac{\sigma}{\hbar}
\frac{T_i}{m} \frac{r_0}{r_i}
\end{equation}

$T_i = T_\gamma(r_i)$ is the local radiation temperature at the
turning point of the motion of the geodesically moving particle
(or rather the temperature at the radial coordinate position
$r_i$, where the motion has become geodesical). According to the
discussion in \ref{sec:massive:acc} and the results obtained in
\cite{Petri/charge} $\sigma / \hbar \approx s = m / T_i$,
therefore the factor in front of $r_0/r_i$ in equation
(\ref{eq:TU/T}) is quite close if not equal to one, so that

\begin{equation} \label{eq:TU/T:2}
\frac{\overline{T_U}}{T_\gamma} \simeq \frac{r_0}{r_i}
\end{equation}

For a particle of nearly Planck-mass emitted from the central
region of the holostar, the Unruh-temperature will be comparable
to the radiation temperature, if the particle can be considered to
move geodesically from the beginning. This opens up the
possibility to explain the isotropy of the radiation temperature
in the co-moving frame by the Unruh-effect, since the
Unruh-temperature is always isotropic in the accelerated frame. In
sections \ref{sec:delta:rho}  and \ref{sec:measurement} it has
been shown, that the Hubble value and the evolution of the density
fluctuations in a holostar-universe are compatible with the
experimentally determined values of our universe, if
$\widetilde{r_i}/r_0 = 3/4$, where $\widetilde{r_i}$ is the
"fictitious" zero-velocity turning point of the motion. In fact,
for more realistic scenarios (non-instantaneous decoupling) we
find $r_i \simeq r_0$. Note also, that in section \ref{sec:phi} it
has been shown that the true turning point of the motion $r_i$
(for $\widetilde{r_i}/r_0 = 3/4$) should lie roughly at $r_i
\approx 1.08 \, r_0$ (equation (\ref{eq:ri:true})). This strongly
suggests, that the microwave-background temperature is in fact
equal to the Unruh temperature, i.e. $r_i = r_0$, so that
$T_\gamma = \overline{T_U}$ and thus $H = 1/r = 1 /\tau$.

However, in order to attain an Unruh-temperature comparable to the
temperature of the CMBR, the proper acceleration in the co-moving
frame would have to be enormous: $a = 2 \pi \overline{T_U} / \hbar
\approx 10^{20} m/s^2$ for $\overline{T_U} \simeq 2.73 K$. If the
microwave background radiation is supposed to be nothing else than
Unruh radiation, why then don't we notice this enormous
acceleration? This answer to this open question might be found in
the pressure\footnote{A diver in water can "compensate" the
(constant) gravitational field of the earth if his mean density is
equal to the density of the pressurized fluid (water) in which he
floats. Maybe the negative pressure, which has the same energy
density as the "ordinary matter", provides a similar compensation.
This is even more suggested by the fact, that in general
relativity the "true" coordinate independent gravitational effect
is not a constant acceleration field (which can be transformed
away), but rather the tidal acceleration, which is quite low in
the outer regions of the holostar.} or - possibly - in the
rotation of the holostar.

\section{The holostar as a unified description for the fundamental building blocks of nature?}

As has been demonstrated in the previous chapter, the holostar has
properties which are in many respects similar to the properties of
black holes and the universe.

It is quite remarkable, that the fairly simple model of the
holostar appears to have the potential to explain the properties
of objects, that so far have been treated as distinct building
blocks of nature, in one uniform description. Black holes and the
universe didn't appear to have much in common, although both are
strongly gravitating systems. The holostar solution points at a
deeper connection between self gravitating systems of any size.
The holostar solution might even be of some relevance for
elementary particles.

This chapter is dedicated to a first preliminary discussion of the
question, whether the holostar can serve as an alternative -
possibly unified - description for black holes, the universe and
elementary particles.

\subsection{The holostar as alternative to black holes?}

The holostar possesses properties very similar to properties
attributed to a black hole. It has an entropy and temperature
proportional, if not equal to the Hawking entropy and temperature.
Its exterior space-time metric is equal to that of a black hole,
disregarding the Planck coordinate distance between the membrane and
the gravitational radius. Therefore, as viewed from an exterior
observer, the holostar is indistinguishable from a black hole.

With respect to the interior space-time, i.e. the space-time
within the event horizon (or within the membrane), there are
profound differences:

A black hole has no interior matter, except - presumably - a point
mass $M$ at the position of its central singularity.\footnote{A
rotating Kerr-black hole has a ring-singularity.} The number and
the nature of the interior particles of a black hole, or rather
the particles that have gone into the black hole, is undefined.
Any concentric sphere within the event horizon is a trapped
surface. The holostar has a continuous interior matter
distribution, no singularity at its center, no trapped surfaces
and - as will be shown in \cite{Petri/thermo} - a definite number
of interior particles proportional to its boundary area. The
absence of trapped surfaces and of an accompanying singularity is
a desirable feature in its own right. Furthermore, the holostar
appears to be the most radical fulfillment of the holographic
principle: The number of the holostar's interior particles is
exactly proportional to its proper surface area. Many researchers
believe the holographic principle to be one of the basic
ingredients from which a possible future universal theory of
quantum gravitation can be formed.

A black hole has an event horizon. The unique properties of the
event horizon, i.e. its constant surface gravity and its
(classically) non-decreasing area, determine the thermodynamic
properties of a black hole, i.e. its Hawking temperature and
entropy. The holostar's thermodynamic properties are determined by
its interior matter configuration. Therefore the holostar solution
"needs" no event horizon. In fact, it possesses no event horizon,
because the metric coefficients $B$ and $1/A$ never become zero in
the whole space-time. Note, however, that for a large holostar the
minimum value of $B$ becomes arbitrarily close to zero. $B$
reaches it's lowest value at the boundary $r_h = r_+ + r_0$:

\begin{equation}
B_{min} = B(r_h) = \frac{r_0}{r_h} = \frac{1}{1 + \frac{r_+}
{r_0}}
\end{equation}

For a holostar with a gravitational radius of $n$ Planck lengths,
i.e. $r_+ = n r_0$, one gets $B_{min} = 1/(n+1)$ and $A_{max} =
n+1$. A holostar with the mass of the sun has $n \approx 2 \cdot
10^{38}$ and therefore $B_{min} \approx 10^{-38}$.

Instead of an event horizon the holostar possesses a two
dimensional membrane of tangential pressure, who's properties are
very similar to the properties of the event horizon of a black
hole as expressed by the membrane paradigm \cite{Price/Thorne/mem,
Thorne/mem}.

The absence of an event horizon is a desirable feature of the
holostar. If there is no event horizon, there is no information
paradox. Unitary evolution of states is possible throughout the
whole space-time.

From the viewpoint of an exterior observer the holostar appears as
a viable alternative to a black hole. It is in most respects
similar to a black hole but doesn't appear to be plagued with the
undesirable consequences of an event horizon (information paradox)
and of a central singularity (breakdown of causality and/or
predictability).

Furthermore the holostar's interior is truly Machian. What matters
are the relative positions and motions of the interior particles
with respect to the whole object. The holostar's spherical
membrane serves as the common reference, not the ill-defined
notion of empty asymptotic Minkowski space.

Both the holostar and a black hole are genuine solutions to the
field equation of general relativity. From the viewpoint of the
theory both solutions have a good chance to be physically realized
in the real world. At the time this paper is written the holostar
solution looks like an attractive alternative to the black hole
solutions. However, only future research - theoretical and
experimental - will be able to answer the question, what
alternative, if any, will provide a better description of the
phenomena of the real world. Some more evidence in favor of the
holostar solution will be presented in \cite{Petri/thermo,
Petri/charge}.

\subsection{The holostar as a self-consistent model for the
universe?}

A large holostar has some of the properties, which can be found in
the universe in its actual state today: At any interior position
there is an isotropic radiation background with a definite
temperature proportional to $1/\sqrt{r}$. Geodesic motion of
photons preserves the Planck-distribution. Massive particles
acquire a radially directed outward motion, during which the
matter-density decreases over proper time. Likewise the
temperature of the background radiation decreases. The temperature
and mass density within the holostar are related by the following
formula, $\rho = 2^5 \pi^3 r_0^2 T^4$, which yields results
consistent with the observations, when the mass density and
microwave-background temperature of the universe are used as input
and $r_0$ is set to roughly twice the Planck-length.

The theoretical description of the universe's evolution is in many
aspects similar to the standard cosmological model. In the
standard model the energy density is related to the cosmological
time as $\rho \propto 1/t^2$. This relation is valid in the matter
dominated as well as in the radiation dominated epoch. It is also
valid in the holostar universe, if $t$ is interpreted as the
proper time of the co-moving observer. The standard cosmological
model furthermore has the following dependence between the total
energy density and the radiation temperature: $\rho \propto T^4$.
In the holostar universe the same relation is valid.

On the other hand, the holostar has some properties, which might
not be compatible with our universe: When massive particles and
photons are completely decoupled, the ratio of the number-density
of photons with respect to the number density of massive particles
is predicted to increase over time in the holostar-universe ($n_m
\propto 1/r^2$, whereas $n_\gamma \propto 1/r^{3/2}$). The
standard cosmological models assumes that this ratio remains
nearly constant up to high redshift. Bigbang nucleosynthesis in
the standard cosmological models wouldn't be compatible with the
observations, if the photon to baryon ratio at high redshift would
be very different from today.

Yet the holostar appears to have the potential to explain some
phenomena, which are unexplained in the standard cosmological
models:

The standard model has the $\Omega$- or flatness-problem. Why is
$\Omega$ today so close to 1? If $\Omega$ is not exactly 1, it
must have been close to 1 in the Planck-era to an accuracy of
better than $10^{-60}$, i.e. the ratio of one Planck length to the
radius of the observable universe today. Such a finetuning is
highly improbable. One would expect $\Omega = 1$, exactly.
However, in the standard cosmological models $\Omega$ is a free
parameter. There is no necessity to set it equal to one. The
holostar "solves" this problem in that there is no free parameter.
The total energy density is completely fixed. Any other total
energy density would result in a very different metric, for which
most of the results presented in this paper, as well as in
\cite{Petri/thermo, Petri/charge}, would not hold.

The standard model has the cosmological constant problem: The
recent supernova-experiments
\cite{Perlmutter/Supernovae_Lambda0.7, Riess, Perlmutter/Schmidt}
indicate that the universe is accelerating. In the standard
cosmological models such an acceleration is "explained" by a
positive cosmological constant, $\Lambda > 0$. However, the
natural value of the cosmological constant would be equal to the
Planck-density, whereas the cosmological constant today is roughly
a factor of $10^{124}$ smaller than its "natural" value. Such a
huge discrepancy suggests, that the cosmological constant should
be exactly zero. The supernova red-shift surveys, however, demand
a cosmological constant which amounts to roughly $0.7$ of the
critical mass-density today. Why does $\Lambda$ have this particular
value? Furthermore, $\Lambda$ and the mass-density scale
differently with $\tau$ (or $r$). Why then do we just happen to
live in an epoch where both values are so close?\footnote{Some
cosmological models therefore assume a time-varying cosmological
constant, which however is quite difficult to incorporate into the
the context of general relativity.} The holostar solution solves
the problem of an accelerating universe without need for a
cosmological constant. The holostar solution in fact requires a
cosmological constant which is exactly zero, which in consequence
can be interpreted as a strong indication that supersymmetry might
well be an exact symmetry of nature.

The standard model has the so called horizon problem. This problem
can be traced to the fact, that in the standard cosmological
models the scale factor of the universe varies as $r \propto
t^{2/3}$ in the matter-dominated era and as $r \propto t^{1/2}$ in
the radiation dominated era, whereas the Hubble-length varies
proportional to $r_H \propto t$. If the scale factor varies slower
than the Hubble length, the region that can be seen by an observer
today will have originated from causally unconnected regions in
the past. For example, at red-shift $z_{eq} \approx 1000$, i.e.
when matter and radiation have decoupled according to the standard
model, the radius of the observable universe consisted of roughly
$d_{e} / d_{c} \approx \sqrt{z_{eq}} \approx 30$ causally
disconnected radial patches, or $30^3$ regions. $d_e$ is the
distance scale of the expansion, $d_c$ of the causally connected
regions. The horizon problem becomes even worse in the radiation
dominated era.\footnote{In the radiation dominated era one finds
$d_{e} / d_{c} \approx z$.} The horizon problem makes it difficult
to explain why the CMBR is isotropic to such high degree.
Inflation is a possible solution to this problem. However, it is
far from clear how and why inflation started or ended.
Furthermore, the inflational scenarios need quite a bit of
finetuning and there appears to be no theoretical framework that
could effectively restrain the different scenarios and/or their
parameters. The holostar space-time solves the horizon problem in
a quite elegant way. The Hubble length $r$ is always proportional
to the scale-factor in the frame of the co-moving observer, as $r
\propto \tau$ and $H \propto 1/r$, therefore everything that is
visible to the co-moving observer today, was causally connected to
him in the past.

Related to the horizon problem is the problem of the scale factor.
With the $T \propto 1/r$-law, in the standard cosmological models
the scale factor is $r \approx 10^{30} r_{Pl}$ at the
Planck-temperature. Why would nature choose just this number?
Inflation can solve this problem, albeit not in a truly convincing
way. The holostar universe with $T \propto 1 / \sqrt{r}$ and $r
\propto \tau$ elegantly gets rid of this problem. The scale factor
evolves proportional (and within the errors equal!) to the
Hubble-length. At the Planck-temperature, both the scale-factor
and the Hubble-length are nearly equal to the Planck-length.

Inflation was originally introduced in order to explain the so
called monopole-problem. The standard cosmological model without
inflation predicts far too many monopoles. Inflation, if it sets
in at the right time, thins out the number of monopoles to a
number consistent with the observations. It might be, that the
holostar universe has an alternative solution to the problem of
monopoles or other heavy particles: Heavy particles will acquire
geodesic motion very early in the holostar, i.e. at small
$r$-coordinate value. But pure geodesic motion in the holostar
universe tends to thin out the non-relativistic particles with
respect to the still relativistic particles.

If we actually live in a large holostar, we should be able to
determine our current radial position $r$ by the formulae given in
the previous sections. In principal it safest to determine $r$ by
the total matter-density via $\rho = 1 / (8 \pi r^2)$, as no
unknown parameters enter into this determination. Using the recent
WMAP-data we find:

$$r = \frac{1}{\sqrt{8 \pi \rho}} = 9.129 \cdot 10^{60} \, r_{Pl} = 1.560 \cdot 10^{10} ly$$

This is quite close to the value $r \approx 1.5 \cdot 10^{10} ly
\approx 8.8 \cdot 10^{60} r_{Pl}$, which was thought to be the
radius of the observable universe for a rather long period of
time.

However, the (total) matter density is difficult to determine
experimentally. Although other estimates of the total mass-energy
density, for example by examining the rotation curves of galaxies
and clusters, all lie within the same range, the accuracy of the
measurements (including systematic errors) is most likely not
better than 5 \%.

A much better accuracy can be achieved through the precise
measurements of the microwave background radiation, whose
temperature is known to roughly $0.1\%$. The determination of $r$
by the $CMBR$-temperature, however, requires the measurement or
theoretical determination of the "fundamental area" $r_0^2$. Its
value can be determined experimentally from both the
matter-density and the radiation temperature, or theoretically as
suggested in \cite{Petri/charge}. If we use the theoretical value
($r_0^2 \simeq 2 \sqrt{3} \hbar$) at low energies we find:

$$r = \frac{\hbar^2}{16 \pi^2 T^2 r_0} = 9.180 \cdot 10^{60} \, r_{Pl} = 1.569 \cdot 10^{10} ly$$

In a large holostar the momentary value of the $r$-coordinate can
also be calculated from the local Hubble-value. In order to do
this, the starting point of the motion $r_i$ and the fundamental
length parameter $r_0$, or rather their ratio, have to be known.
Theoretically one would expect $r_i \simeq r_0$ (see for example
section \ref{sec:Unruh}), whereas experiments point to a somewhat
lower value. If we use $r_i/r_0 \approx 1$ and $H = 71 (km/s) /
Mpc$ we find:

$$r = \sqrt{\frac{r_0}{r_i}}\frac{1}{H} = 8.062 \cdot 10^{60} \, r_{Pl} = 1.378 \cdot 10^{10} ly$$

Very remarkably, all three different methods for the determination
of $r$ give quite consistent results, which agree by 15\%. A
rather large deviation comes from the Hubble-value, which is not
quite unexpected taking into account the difficulty of matching
the different cosmological length scales. Note however, that the
rather good agreement depends on the assumption $r_0 \simeq 1.87
\, r_{Pl}$ and $r_i/r_0 \simeq 1$. These assumptions, especially
the second, could well prove wrong by a substantial factor.

From a theoretical point of view $r_i = r_0$ is interesting
because it allows us to interpret the microwave background
radiation as Unruh radiation. If this turns out to be the right
ansatz, the Hubble constant can be predicted. Its value should be:

\begin{equation}
H = 62.36 \, \frac{km/s}{Mpc}
\end{equation}

This value is well within the measurement errors, which are mostly
due to the calibration of the "intermediate" ladder of the
cosmological distance scale via the Cepheid variables.

Without strong theoretical support for a definite value of $r_0$
the matter-density seems to be the best way to determine $r$. If
the fundamental length parameter $r_0$ can be pinned down
theoretically, such as suggested in \cite{Petri/charge}, the
extremely precise measurements of the microwave background
temperature will give the best estimate for $r$.

The simplicity of the holostar solution and the fact that it has
no adjustable parameters\footnote{Possibly an overall charge Q or
angular momentum could be included. $r_0$ is not regarded as an
adjustable parameter: The analysis here, as well as in
\cite{Petri/thermo, Petri/charge} strongly suggest, $r_0^2 \simeq
4 \sqrt{3/4} r_{Pl}$ at low energies. As long as the theoretical
"prediction" $r_i = r_0$ hasn't been confirmed independently $r_i$
should be regarded as an adjustable parameter. It is encouraging,
however, that $r_i/r_0$, as determined from the relative growth of
the density perturbations after radiation and matter became
kinematically decoupled, is quite close to the value determined
from the measurements of the Hubble-constant.} makes the holostar
solution quite attractive as an alternate model for the universe,
a model that can be quite easily verified or falsified.

At the time this paper is written it not clear, whether the
holostar will provide a serious alternative to the standard
cosmological model. It has the potential to solve many of the
known problems of the standard cosmological model. But the
standard cosmological model has been very successful so far. One
of its great achievements is the remarkably accurate explanation
for the distribution of the light elements, produced by bigbang
nucleosynthesis. Although the holostar universe is in many
respects similar to the standard cosmological models, it is
doubtable that the synthesis of the light elements will proceed in
exactly the same way as in the standard model. It would be quite a
coincidence if the holostar could give a similarly accurate
agreement between theory and observation.\footnote{Note, that
according to \cite{Petri/thermo} chemical potentials and
supersymmetry appear to play an important role in the holostar.
Supersymmetry within the holostar requires that the ratio of the
energy-densities of fermions to bosons isn't given by the usual
factor $7/8$, but rather each fermionic degree of freedom has an
energy density which is higher than that of a bosonic degree of
freedom by roughly a factor of $12.5$. If the energy density in
the radiation (neutrinos, electrons and photons) at the time of
nucleosynthesis is normalized to the photon energy density,
assuming a factor of $7/8$ for the electrons and neutrinos, one
underestimates the total energy density by a factor of roughly
$2.9$ (if the three neutrinos come only in one helicity-state) or
$4.6$ (three neutrinos with 2 helicity states). If the
energy-density is underestimated, the expansion rate comes out too
low. The "true" expansion rate at the time of nucleosynthesis will
have been higher. A higher expansion rate, however, requires a
higher number density of the baryons, in order to get the same
relative abundances of the light elements. This might partly
explain the missing factor of 6, assuming $\Omega_m \approx 0.26$
and $\Omega_b \approx 0.04$}

Furthermore, it is not altogether clear how the "true" motion of
the massive particles within the cosmic fluid will turn out.
Therefore some results from the simple analysis in this paper
might have to be revised. There also is the problem of relating
the observations in the frames of the co-moving observer and the
observer at rest, which was discussed briefly in section
\ref{sec:frames}, but which quite certainly requires further more
sophisticated consideration.

On the other hand, it is difficult to believe that the remarkably
consistent determination of $r$ by three independent methods is
just a numerical coincidence with no deeper physical meaning.

Therefore the question whether the holostar can serve as an
alternative description of the universe should be regarded as
open, hopefully decidable by research in the near future. Even if
the holostar doesn't qualify as an accurate description of the
universe, its particularly simple metric should make it a valuable
theoretical laboratory to study the so far only poorly understood
physical effects that arise in a universe with significant
(anisotropic) pressure.

\subsection{The holostar as a simple model for a self
gravitating particle with zero gravitational mass ?}

Another unexpected feature of the holographic solution is, that
one can choose $r_+ = 2 M = 0$ and still get a genuine solution with
"structure", i.e. an interior "source region" with non-zero
interior matter-distribution bounded by a membrane of roughly
Planck-area ($r_+=0$ implies $r_h = r_0$). Note that the interior
"source region" $r < r_0$ should not be considered as physically
accessible for measurements. Neither should the interior matter
distribution be considered as a meaningful description of the
"interior structure". In fact, the holostar solution with $r_h = r_0$ and
$M = 0$ should be regarded to have no physically relevant
interior sub-structure. The only physically relevant quantity is
its finite boundary area.

The $r_+ = 0$ solution therefore might serve as a very simple, in
fact the simplest possible model for a particle of Planck-size. It
describes a (spin-0) particle with a gravitating mass which is
classically zero, as can be seen from the properties of the metric
outside the "source region" (i.e. $A=B=1$):

Although it quite unlikely that an extrapolation from the
experimentally verified regime of general relativity right down to
the Planck scale can be trusted quantitatively, the holographic
solution hints at the possibility, that an elementary particle
might be - at least approximately - describable as a
self-gravitating system.

There are remarkable similarities between the properties of black
holes and elementary particles. As has already been noted by
Carter \cite{Carter/electronAsBh} the Kerr-Newmann solution has a gyromagnetic
ratio of 2, i.e. its g-factor is equal to that of the electron
(disregarding radiative corrections). The (three) quadrupole
moments of the Kerr-Newman solution are $2/3$ and $-1/3$ in
respective units, an interesting analogue to the fractional
charges of the three quarks confined within the nucleon?!

It is not altogether folly to identify elementary particles with
highly gravitating systems. For energies approaching the
Planck-scale gravity becomes comparable to the other forces.
Therefore an elementary particle will quite certainly become a
strongly gravitating system at high energies. However, the
identification of an elementary particle with a solution to the
classical field equations of general relativity at the low end of
the energy scale has so far met serious obstacles. It is difficult
to explain why the masses of the light elementary particles are
about 20 orders of magnitude smaller than the Planck-mass.

The common expectation, that the Planck-mass will be the minimum
mass of a compact self gravitating object can be traced to the
fact, that the only "natural" unit of mass which can be
constructed from the three fundamental constants of nature $c, \,
G$ and $\hbar$ is the Planck mass. Thus it seems evident, that any
"fundamental" theory of nature which incorporates the three
constants, must have fundamental particles of roughly
Planck-mass.\footnote{Supersymmetry suggests another possibility:
Although the fundamental mass-scale in any supersymmetric theory
is the Planck-mass, the near zero masses of the elementary
particles are explained from a very precise cancellation of
positive and negative contributions of bosons and fermions, due to
the (hidden) supersymmetry of nature.} This quite certainly would
be true, if mass always remains a "fundamental" parameter of
nature, even at the most extreme energy scales.

However, there might be a slight misconception about the
significance of mass in high gravitational fields. Without doubt,
"mass" is a fundamental quantity both in Newton's theory of
gravitation and in quantum field theories such as QED or QCD.
These theories have very heavily influenced the conception of mass
as a fundamental quantity of physics. In a geometric theory, such
as Einstein's theory of general relativity, mass comes not as a
natural property of a space-time-manifold. From a general
relativistic viewpoint mass isn't a geometric property at all and
enters into the equations of general relativity in a rather crude
way, as stated by Einstein himself.\footnote{Einstein once
described his theory as a building, whose one side (the left,
geometric side of the field equations) is "constructed of marble",
whereas the right side (matter-side) is "constructed from low
grade wood".} Furthermore, mass has dimensions of length. From the
viewpoint of quantum gravity length - in contrast to area (or
volume) - is a concept which is difficult to define.\footnote{See
for example \cite{Thiemann/length}} Therefore it is questionable
if mass will remain a fundamental parameter\footnote{"fundamental"
is used in the current context as "optimally adapted" for the
description of the phenomena.} in situations of high gravity,
where the geometry cannot be considered static but becomes dynamic
itself. In these situations it seems more "natural" to describe
the interacting (geometric) objects not by their mass, but rather
by the area of their boundaries. Mass appears rather as a
perturbation. Surfaces as basis for the "natural" description of
systems with high gravitational fields are not only motivated by
recent research results, such as the holographic principle
\cite{Hooft/hol, Susskind/hol}, the study of light-sheets
\cite{Bousso/lightsheets} and isolated horizons
\cite{Ashtekar/IsolatedHorizons, Ashtekar/IsolatedHorizons2}, or
M-theory, but also by the simple fact, that in natural units
$c=G=1$ area has dimension of action (or angular momentum), from
which we know that it is quantized in units of $\hbar / 2$.

Therefore the "elementary" holostar with zero gravitational mass,
but non-zero boundary area might serve as a very crude classical
approximation for the most simple elementary particle with zero
angular momentum and charge. Unfortunately none of the known
particles can be identified with the "elementary" uncharged and
spherically symmetric holostar-solution with $r_h = r_0$. Even if
a spin-zero uncharged and zero mass particle
exists\footnote{According to quantum gravity, a spin-network state
with a single spin-0 link has zero area, and thus zero entropy.
There are reasons to believe, that a spin-network state with a
single spin can be identified with an elementary particle (see
\cite{Petri/charge}). However, a zero area, zero-entropy particle
is quite inconceivable. Therefore it is likely, that a fundamental
(i.e. not composite) particle with zero spin and charge doesn't
exist. Any uncharged spin-zero particle therefore should be
composite. For all the elementary particles that are known so far
this is actually the case.}, it is unlikely that more than its
cross-sectional area can be "predicted" by the holostar solution.
Still it will be interesting to compare the properties of a
charged and/or rotating holostar to the properties of the known
elementary particles in order to see how far the equations of
classical general relativity allow us to go. Some encouraging
results, which point to a deep connection between the holostar
solution and quantum gravity spin-network states, are presented in
\cite{Petri/charge}.

\section{Discussion}

The holostar solution appears as an attractive alternative for a
black hole. It has a surface temperature, which - measured at
spatial infinity - is proportional to the Hawking temperature. Its
total number of interior particles is proportional to its proper
surface area, which can be interpreted as evidence that the
Hawking entropy is of microscopic-statistical origin and the
holographic principle is valid in classical general relativity for
self gravitating objects of arbitrary size. In contrast to a black
hole, the holostar has no event horizon, no trapped surfaces and
no central singularity, so there is no information paradox and no
breakdown of predictability.

Furthermore, the holostar solution has some potential to serve as
an alternative model for a universe with anisotropic negative
pressure, without need for a cosmological constant. It also admits
microscopic self-gravitating objects with a surface area of
roughly the Planck-area and zero gravitating mass.

The remarkable properties of the holostar solution and its high
degree of self-consistency should make it an object of
considerable interest for future research.

A field of research which presents itself immediately is the the
generalization of the holostar solution to the rotating and / or
charged case. The derivation of the charged holostar solution is
straight forward and will be discussed in a follow-up paper
\cite{Petri/charge}. A generalization to the case of a rotating
body, including spin (and charge), will be a challenging topic of
future research.

An accurate description of the thermodynamic properties of the
holostar solution is of considerable interest. In a parallel paper
\cite{Petri/thermo} the entropy/area law and the temperature-law
for the holostar will be put on a somewhat sounder foundation. The
"Hawking temperature" of the universe will be "measured", verifying
Hawking's prediction to an accuracy better than 1\%.

Another valuable field  of research will be the examination,
whether the holostar solution can serve as an alternative model
for the universe. The holostar solution appears to have the
potential to solve many problems of the standard cosmological
models, such as the horizon-problem, the cosmological constant
problem, the problem of structure formation from the small density
perturbations in the CMBR, etc. . It provides an explanation for
the numerical value of the baryon to photon ratio $\eta$. Even
some of the more recent results, such as the missing angular
correlation of the CMBR-fluctuations at angular separations larger
than $60^\circ$ or the missing quadrupole moment, which hints at a
(small) positive curvature of the universe, are explainable in the
context of the holostar space-time. On the other hand, it is far
from clear whether the holostar solution will ever be able to
explain the observed abundances of the light elements, especially
Deuterium and Lithium, in a convincing fashion, such as the
standard cosmological model can. Nucleosynthesis in a holostar
universe will be a demanding challenge. If the holostar can pass
this test, it should open up a considerable field of new and
interesting research in the field of cosmology. Quite likely
chemical potentials and supersymmetry will play an important role
in the holostar universe, not only at high temperatures above the
electro-weak unification scale, but also at the time of
nucleosynthesis or even at later times.

Lastly, the properties of the membrane, how it is formed, how it
is composed and how it manages to maintain its two-dimensional
structure will be an interesting research topic, if the holostar
turns out to be a realistic alternative for a black hole. The
study of anisotropic matter states in high gravitational fields
should provide fruitful as well.


\end{document}